\documentclass[acmsmall,authorversion,preprint,fleqn]{acmart}

\acmConference[Submitted]{Submitted}{2020}{}
\acmYear{2020}
\copyrightyear{2020}
\setcopyright{none}

\setcopyright{none}

\newcommand{\ponders}[3]{\ignorespaces}
\newcommand{\TODO}[1]{}

\newcommand{\nw}[1]{\ponders{NW}{orange}{#1}}
\newcommand{\matt}[1]{\ponders{MP}{green}{#1}} 
\newcommand{\al}[1]{\ponders{AL}{purple}{#1}}

\newcommand\spacecorrectsep{%
  \spacecorrectnosep
  \vskip\blanklineskip\relax
}
\newcommand\spacecorrectnosep{%
  \vskip-\belowdisplayskip\relax
  \vskip-\abovedisplayskip\relax
}

\usepackage[silent]{polytable}
\usepackage{doubleequals}
\usepackage{fancyvrb}
\usepackage{mathpartir}

\DefineVerbatimEnvironment%
  {core}{Verbatim}
  {xleftmargin=\mathindent}

\usepackage{placeins}

\usepackage{subcaption}
\usepackage{caption}
\usepackage{tikz}
\usepackage{tikz-cd}
\usepackage{csvsimple,booktabs, siunitx, array}

\newcommand{\ghl}[1]{\highlight{gray!25}{#1}}

\makeatletter
\@ifundefined{lhs2tex.lhs2tex.sty.read}%
  {\@namedef{lhs2tex.lhs2tex.sty.read}{}%
   \newcommand\SkipToFmtEnd{}%
   \newcommand\EndFmtInput{}%
   \long\def\SkipToFmtEnd#1\EndFmtInput{}%
  }\SkipToFmtEnd

\newcommand\ReadOnlyOnce[1]{\@ifundefined{#1}{\@namedef{#1}{}}\SkipToFmtEnd}
\usepackage{amstext}
\usepackage{amssymb}
\usepackage{stmaryrd}
\DeclareFontFamily{OT1}{cmtex}{}
\DeclareFontShape{OT1}{cmtex}{m}{n}
  {<5><6><7><8>cmtex8
   <9>cmtex9
   <10><10.95><12><14.4><17.28><20.74><24.88>cmtex10}{}
\DeclareFontShape{OT1}{cmtex}{m}{it}
  {<-> ssub * cmtt/m/it}{}

\DeclareFontShape{OT1}{cmtt}{bx}{n}
  {<5><6><7><8>cmtt8
   <9>cmbtt9
   <10><10.95><12><14.4><17.28><20.74><24.88>cmbtt10}{}
\DeclareFontShape{OT1}{cmtex}{bx}{n}
  {<-> ssub * cmtt/bx/n}{}

\newcommand{\Conid}[1]{\mathit{#1}}
\newcommand{\Varid}[1]{\mathit{#1}}
\newcommand{\anonymous}{\kern0.06em \vbox{\hrule\@width.5em}}

\usepackage{polytable}

\@ifundefined{mathindent}%
  {\newdimen\mathindent\mathindent\leftmargini}%
  {}%

\def\resethooks{%
  \global\let\SaveRestoreHook\empty
  \global\let\ColumnHook\empty}
\newcommand*{\savecolumns}[1][default]%
  {\g@addto@macro\SaveRestoreHook{\savecolumns[#1]}}
\newcommand*{\restorecolumns}[1][default]%
  {\g@addto@macro\SaveRestoreHook{\restorecolumns[#1]}}
\newcommand*{\aligncolumn}[2]%
  {\g@addto@macro\ColumnHook{\column{#1}{#2}}}

\resethooks

\newcommand{\onelinecommentchars}{\quad-{}- }
\newcommand{\commentbeginchars}{\enskip\{-}
\newcommand{\commentendchars}{-\}\enskip}

\newcommand{\visiblecomments}{%
  \let\onelinecomment=\onelinecommentchars
  \let\commentbegin=\commentbeginchars
  \let\commentend=\commentendchars}

\newcommand{\invisiblecomments}{%
  \let\onelinecomment=\empty
  \let\commentbegin=\empty
  \let\commentend=\empty}

\visiblecomments

\newlength{\blanklineskip}
\newcommand{\hsindent}[1]{\quad}
\let\hspre\empty
\let\hspost\empty

\EndFmtInput
\makeatother
\ReadOnlyOnce{polycode.fmt}%
\makeatletter

\newcommand{\hsnewpar}[1]%
  {{\parskip=0pt\parindent=0pt\par\vskip #1\noindent}}

\newcommand{\hscodestyle}{}

\newcommand{\sethscode}[1]%
  {\expandafter\let\expandafter\hscode\csname #1\endcsname
   \expandafter\let\expandafter\endhscode\csname end#1\endcsname}

  {\par\noindent
   \advance\leftskip\mathindent
   \hscodestyle
   \let\\=\@normalcr
   \let\hspre\(\let\hspost\)%
   \pboxed}%
  {\endpboxed\)%
   \par\noindent
   \ignorespacesafterend}

  {\hsnewpar\abovedisplayskip
   \advance\leftskip\mathindent
   \hscodestyle
   \let\hspre\(\let\hspost\)%
   \pboxed}%
  {\endpboxed%
   \hsnewpar\belowdisplayskip
   \ignorespacesafterend}

  {\hsnewpar\abovedisplayskip
   \advance\leftskip\mathindent
   \hscodestyle
   \let\\=\@normalcr
   \(\pboxed}%
  {\endpboxed\)%
   \hsnewpar\belowdisplayskip
   \ignorespacesafterend}

\newcommand{\plainhs}{\sethscode{plainhscode}}

\plainhs

  {\hsnewpar\abovedisplayskip
   \advance\leftskip\mathindent
   \hscodestyle
   \let\\=\@normalcr
   \(\parray}%
  {\endparray\)%
   \hsnewpar\belowdisplayskip
   \ignorespacesafterend}

  {\parray}{\endparray}

  {\(\parray}{\endparray\)}

\def\codeframewidth{\arrayrulewidth}
\RequirePackage{calc}

  {\parskip=\abovedisplayskip\par\noindent
   \hscodestyle
   \arrayrulewidth=\codeframewidth
   \tabular{@{}|p{\linewidth-2\arraycolsep-2\arrayrulewidth-2pt}|@{}}%
   \hline\framedhslinecorrect\\{-1.5ex}%
   \let\endoflinesave=\\
   \let\\=\@normalcr
   \(\pboxed}%
  {\endpboxed\)%
   \framedhslinecorrect\endoflinesave{.5ex}\hline
   \endtabular
   \parskip=\belowdisplayskip\par\noindent
   \ignorespacesafterend}

\newcommand{\framedhslinecorrect}[2]%
  {#1[#2]}

  {\(\def\column##1##2{}%
   \let\>\undefined\let\<\undefined\let\\\undefined
   \newcommand\>[1][]{}\newcommand\<[1][]{}\newcommand\\[1][]{}%
   \def\fromto##1##2##3{##3}%
   }{\) }%

  {\let\orighscode=\hscode
   \let\origendhscode=\endhscode
   \def\endhscode{\def\hscode{\endgroup\def\@currenvir{hscode}\\}\begingroup}
   \orighscode\def\hscode{\endgroup\def\@currenvir{hscode}}}%
  {\origendhscode
   \global\let\hscode=\orighscode
   \global\let\endhscode=\origendhscode}%

\makeatother
\EndFmtInput
\ReadOnlyOnce{forall.fmt}%
\makeatletter

\let\HaskellResetHook\empty
\newcommand*{\AtHaskellReset}[1]{%
  \g@addto@macro\HaskellResetHook{#1}}
\newcommand*{\HaskellReset}{\HaskellResetHook}

\newcommand\hsforall{\global\let\hsdot=\hsperiodonce}
\newcommand*\hsperiodonce[2]{#2\global\let\hsdot=\hscompose}
\newcommand*\hscompose[2]{#1}

\AtHaskellReset{\global\let\hsdot=\hscompose}

\HaskellReset

\makeatother
\EndFmtInput

\newcommand{\keyword}[1]{\mathbf{#1}}

  {\def\column##1##2{}%
   \let\>\undefined\let\<\undefined\let\\\undefined
   \newcommand\>[1][]{}\newcommand\<[1][]{}\newcommand\\[1][]{}%
   \def\fromto##1##2##3{##3}%
   }{}%

\newcommand{\inlinemathhs}{\sethscode{inlinemathhscode}}

\usepackage{xspace}

\newcommand\labelledinferrule[3]{\inferrule{#1}{#2}~\text{\textsc{#3}}}
\newcommand\ruletype[1]{\center\boxed{#1}}

\newcommand\rulename[2]{\newcommand{#1}{\textsc{#2}\xspace}}
\rulename{\ECSP}{E\_CSP}
\rulename{\EVar}{E\_Var}
\rulename{\ESpliceVar}{E\_Splice\_Var}
\rulename{\ETopSpliceVar}{E\_Top\_Splice\_Var}
\rulename{\EVarTopLevel}{E\_Var\_TopLevel}
\rulename{\ECase}{E\_Case}
\rulename{\EConstr}{E\_Constr}
\rulename{\EAbs}{E\_Abs}
\rulename{\EApp}{E\_App}
\rulename{\ETAbs}{E\_TAbs}
\rulename{\ETApp}{E\_TApp}
\rulename{\EEvApp}{E\_EvApp}
\rulename{\EQuote}{E\_Quote}
\rulename{\ESplice}{E\_Splice}
\rulename{\ESpliceTop}{E\_Splice\_Top}
\rulename{\ERun}{E\_Run}
\rulename{\ELift}{E\_Lift}
\rulename{\ELiftType}{E\_Lift\_Type}
\rulename{\ECAbs}{E\_C\_Abs}
\rulename{\ECApp}{E\_C\_App}

\rulename{\Def}{Def}
\rulename{\Cls}{Cls}
\rulename{\Inst}{Inst}

\rulename{\PExpr}{P\_Expr}
\rulename{\PDef}{P\_Def}
\rulename{\PCls}{P\_Cls}
\rulename{\PInst}{P\_Inst}

\rulename{\CTC}{C\_TC}
\rulename{\CCodeC}{C\_CodeC}

\rulename{\EvDict}{Ev\_Dict}
\rulename{\EvImp}{Ev\_Imp}

\rulename{\TCSP}{T\_CSP}
\rulename{\TVar}{T\_Var}
\rulename{\TConst}{T\_Const}
\rulename{\TArrow}{T\_Arrow}
\rulename{\TCArrow}{T\_CArrow}
\rulename{\TForAll}{T\_For\_All}
\rulename{\TCode}{T\_Code}
\rulename{\TTySplice}{T\_Ty\_Splice}
\rulename{\TCodeT}{T\_CodeT}
\rulename{\TNat}{T\_Nat}
\rulename{\TBool}{T\_Bool}

\rulename{\EVarCSP}{E\_Var\_CSP}
\rulename{\TVarCSP}{T\_Var\_CSP}

\rulename{\CGlobal}{C\_Global}
\rulename{\CLocal}{C\_Local}
\rulename{\CIncr}{C\_Incr}
\rulename{\CDecr}{C\_Decr}

\rulename{\SPEmpty}{SP\_Empty}
\rulename{\SPCons}{SP\_Cons}

\rulename{\CDef}{C\_Def}
\rulename{\CSpDef}{C\_SpDef}

\rulename{\CMain}{CP\_Main}
\rulename{\CPDef}{CP\_Def}
\rulename{\CPSpDef}{CP\_SpDef}

\rulename{\DEQuote}{D\_E\_Quote}
\rulename{\DETBeta}{D\_E\_TBeta}
\rulename{\DETApp}{D\_E\_TApp}
\rulename{\DETLam}{D\_E\_TLam}
\rulename{\DEBeta}{D\_E\_Beta}
\rulename{\DEAppL}{D\_E\_App\_L}
\rulename{\DEAppR}{D\_E\_App\_R}

\rulename{\DSPBody}{D\_SP\_Body}
\rulename{\DSPCons}{D\_SP\_Cons}

\rulename{\DDef}{D\_Def}
\rulename{\DSPDef}{D\_SpDef}
\rulename{\DPDef}{D\_P\_Def}
\rulename{\DPDefBeta}{D\_P\_Def\_Beta}
\rulename{\DPSPDefBeta}{D\_P\_SpDef\_Beta}
\rulename{\DPSPDef}{D\_P\_SpDef}
\rulename{\DPMain}{D\_P\_Main}

\rulename{\CStrip}{C\_Strip}
\rulename{\CEmpty}{C\_Empty}

\theoremstyle{acmdefinition}
\newtheorem{cexample}{Example}

\newtheorem{tsexample}{Example}

\newtheorem{tvexample}{Example}

\newtheorem{iexample}{Example}

\AtBeginEnvironment{figure}{\small}
\AtBeginEnvironment{figure*}{\small}

\begin{document}

\title{A Specification for Typed Template Haskell}
\author{Matthew Pickering}
\affiliation{
  \department{Department of Computer Science}              
  \institution{University of Bristol}            
  \country{United Kingdom}                    
}
\email{matthew.pickering@bristol.ac.uk}          

\author{Andres L\"o{h}}
\affiliation{
\institution{Well-Typed LLP}
}
\email{andres@well-typed.com}

\author{Nicolas Wu}
\affiliation{
  \department{Department of Computing}              
  \institution{Imperial College London}            
  \country{United Kingdom}                    
}
\email{n.wu@imperial.ac.uk}          

\begin{abstract}
Multi-stage programming is a proven technique that provides predictable
performance characteristics by controlling code generation.
We propose a core semantics for Typed Template Haskell, an extension of Haskell
that supports multi staged programming that interacts well with polymorphism
and qualified types.
Our semantics relates a declarative source language with qualified types to a core language
based on the the polymorphic lambda calculus augmented with multi-stage constructs.
\end{abstract}

\ccsdesc[500]{Software and its engineering~General programming languages}
\ccsdesc[300]{Social and professional topics~History of programming languages}

\keywords{staging, polymorphism, constraints }  

\maketitle

\newcommand{\lam}[2]{\lambda #1.\, #2}
\newcommand{\Lam}[2]{\Lambda #1.\, #2}
\newcommand{\tlam}[2]{\forall #1.\, #2}
\newcommand{\app}[2]{#1~#2}
\newcommand{\tapp}[2]{#1~\langle#2\rangle}
\newcommand{\cast}[2]{#1 \triangleright #2}

\newcommand{\ctxext}[1]{\Gamma,#1}
\newcommand{\emptyctx}[0]{\varnothing}
\newcommand{\caseof}[2]{\keyword{case}~#1~\keyword{of}~#2}
\newcommand{\ruleref}[2]{\textsc{#1}\_\textsc{#2}}
\providecommand{\xvdash}[1]{\vdash_{\mkern-5mu\scriptscriptstyle\rule[-.9ex]{0pt}{0pt}#1}}
\providecommand{\sfvdash}[1]{\xvdash{\textsf{#1}}}
\newcommand{\vdashn}[1]{\vdash^{#1}}
\newcommand{\vdasho}[2]{\vdash_{#1}^{#2}}
\newcommand{\vdashct}[1]{\vdasho{ct}{#1}}
\newcommand{\entails}[1]{\vDash^{#1}}
\newcommand{\subst}[2]{\lbrack #1 / #2 \rbrack}
\newcommand{\quot}[1]{\lbrack #1 \rbrack}
\newcommand{\lift}[1]{\textit{Lift}~#1}
\newcommand{\liftop}[1]{\uparrow\!#1}
\newcommand{\liftopt}[1]{\Uparrow\!#1}
\newcommand{\qq}[1]{\llbracket~#1~\rrbracket}

\newcommand{\leadstocsp}{\leadsto}
\newcommand{\leadstocspty}{\leadsto_{ty}}
\newcommand{\leadstocspev}{\leadsto_{ev}}

\newcommand{\colonT}[0]{\colon \hspace{-0.07cm}}
\newcommand{\highlightR}[1]{%
  \colorbox{gray!40}{$\displaystyle#1$}}

\section{Introduction}
\label{sec:introduction}





Producing optimal code is a difficult task that is greatly assisted by \emph{staging
annotations}, a technique which has been extensively studied
and implemented in a variety of languages~%
\cite{%
taha2000metaml,
rompf2010lms,
kiselyov2014design
}. These annotations give programmers fine control
by instructing the compiler to generate code in one stage of compilation that
can be used in another.

The classic example of staging is the \ensuremath{\Varid{power}} function, where the value $n^k$
can be efficiently computed for a fixed $k$ by generating code where the
required multiplications have been unrolled and inlined.  The incarnation of
staged programming in Typed Template Haskell%
\footnote{Typed Template Haskell is a variation of the Template
Haskell~\cite{sheard2002template} that adds types in the style of MetaML~\cite{taha2000metaml}.
The abstract datatype \ensuremath{\Conid{Code}\;\Varid{a}} in this paper is a newtype wrapper around \ensuremath{\Conid{Q}\;(\Conid{TExp}\;\Varid{a})}.
In GHC, typed quotes are implemented by \text{\ttfamily \char91{}\char124{}\char124{}~\char124{}\char124{}\char93{}} and typed splices by \text{\ttfamily \char36{}\char36{}\char40{}~\char41{}}
, rather than \ensuremath{\qq{\cdot }} and \ensuremath{\$(\cdot )}. It was implemented in 2013 by Geoffrey Mainland (who also used \ensuremath{\Varid{power}} as a motivating example) under a proposal by
Simon Peyton Jones.}
benefits from
type classes, one of
the distinguishing features of Haskell~\cite{Hall:1996:TypeClasses,pj1997typeclass}, allowing a
definition that can be reused for any type that is qualified to be numeric:
\begin{hscode}\SaveRestoreHook
\column{B}{@{}>{\hspre}l<{\hspost}@{}}\ColumnHook
\column{10}{@{}>{\hspre}l<{\hspost}@{}}\ColumnHook
\column{14}{@{}>{\hspre}c<{\hspost}@{}}\ColumnHook
\column{14E}{@{}l@{}}\ColumnHook
\column{17}{@{}>{\hspre}l<{\hspost}@{}}\ColumnHook
\column{E}{@{}>{\hspre}l<{\hspost}@{}}\ColumnHook
\>[B]{}\Varid{power}\mathbin{::}\Conid{Num}\;\Varid{a}\Rightarrow \Conid{Int}\to \Conid{Code}\;\Varid{a}\to \Conid{Code}\;\Varid{a}{}\<[E]%
\\
\>[B]{}\Varid{power}\;\mathrm{0}\;{}\<[10]%
\>[10]{}\Varid{cn}{}\<[14]%
\>[14]{}\mathrel{=}{}\<[14E]%
\>[17]{}\qq{\mathrm{1}}{}\<[E]%
\\
\>[B]{}\Varid{power}\;\Varid{k}\;{}\<[10]%
\>[10]{}\Varid{cn}{}\<[14]%
\>[14]{}\mathrel{=}{}\<[14E]%
\>[17]{}\qq{\$(\Varid{cn})\mathbin{*}\$(\Varid{power}\;(\Varid{k}\mathbin{-}\mathrm{1})\;\Varid{cn})}{}\<[E]%
\ColumnHook
\end{hscode}\resethooks
Any value \ensuremath{\Varid{n}\mathbin{::}\Conid{Int}} can
be quoted to create \ensuremath{\qq{\Varid{n}}\mathbin{::}\Conid{Code}\;\Conid{Int}},
then spliced in the
expression \ensuremath{\$(\Varid{power}\;\mathrm{5}\;\qq{\Varid{n}})} to generate
\ensuremath{\Varid{n}\mathbin{*}(\Varid{n}\mathbin{*}(\Varid{n}\mathbin{*}(\Varid{n}\mathbin{*}(\Varid{n}\mathbin{*}\mathrm{1}))))}.
Thanks to type class polymorphism, this works when \ensuremath{\Varid{n}} has any fixed type
that satisfies the \ensuremath{\Conid{Num}} interface, such as \ensuremath{\Conid{Integer}}, \ensuremath{\Conid{Double}} and countless
other types.

It is somewhat surprising, then, that the following function fails to compile
in the latest implementation of Typed Template Haskell in GHC 8.10.1:
\begin{hscode}\SaveRestoreHook
\column{B}{@{}>{\hspre}l<{\hspost}@{}}\ColumnHook
\column{E}{@{}>{\hspre}l<{\hspost}@{}}\ColumnHook
\>[B]{}\Varid{power5}\mathbin{::}\Conid{Num}\;\Varid{a}\Rightarrow \Varid{a}\to \Varid{a}{}\<[E]%
\\
\>[B]{}\Varid{power5}\;\Varid{n}\mathrel{=}\$(\Varid{power}\;\mathrm{5}\;\qq{\Varid{n}}){}\<[E]%
\ColumnHook
\end{hscode}\resethooks
Currently, GHC complains that there is no instance for
\ensuremath{\Conid{Num}\;\Varid{a}} available, which is strange because the type signature explicitly
states that \ensuremath{\Conid{Num}\;\Varid{a}} may be assumed. But this is not the only problem with
this simple example:
in the definition of \ensuremath{\Varid{power}}, the constraint is used
inside a quotation but is bound outside.
As we will see, this is an ad-hoc decision that then leads to subtle inconsistencies.

This paper sets out to formally answer the question of how a language with
polymorphism and qualified types should interact with a multi-stage programming
language.
As well as the fact that all code generation happens at compile time,
these are features that distinguish Typed Template Haskell,
which until now has had no formalism even though it is an extension already
fully integrated into GHC.  We extend previous work that looked at cross-stage
persistence in this setting~\citep{pickering2019multi} by more closely
modelling the reality of the type system implemented in GHC. In particular the
implications of compile-time code generation are considered, as are the
impredicativity restriction and the interaction with instance definitions.

Providing a formalism for the interaction of polymorphism, qualified types, and
multi-stage programming has important consequences for Typed Template Haskell
and other future implementations that combine these features. This
specification resolves the uncertainty about how it should interact with new
language features. It has also been unclear how to fix a large number of bugs
in the implementation because there is no precise semantics it is supposed to
operate under. Presently these deficiencies hampered the adoption of
meta-programming in Haskell as a trusted means of producing code with
predictable performance characteristics. This untapped potential has been
demonstrated to be beneficial in other staged
systems in
MetaML~\citep{taha2000metaml},
Scala's LMS~\citep{rompf2010lms},
and
MetaOCaml~\citep{kiselyov2014design}.
\nw{State that staging allows control of performance without having to look at
core, which is infamously difficult to work with}

At first glance it may appear that constructing programs using only quotations
and splices is restrictive.  However, multi-stage programming is widely
applicable to any domain in which statically known abstractions should be
eliminated. Some particular examples include removing the overhead of using
parser combinators~\cite{jonnalagedda2014staged, krishnaswami2019typed, willis2020parsley}, generic programming~\cite{yallop2017staged} and effect handlers~\cite{schuster2020zero}. Furthermore, there
are certain interesting techniques and tricks which are useful when
constructing multi-stage programs that can vastly improve performance
\citep{taha2004gentle,kiselyov2018reconciling}.

Following a brief introduction to staging with Typed Template Haskell
(Section~\ref{sec:background}), this paper makes the following contributions:
\begin{itemize}
\item{We demonstrate the problems that arise from the subtle interaction between qualified constraints and staging, which highlights the need for a clear specification (Section~\ref{sec:motivation}).}
\item{We introduce a type system for a source language which models Typed
Template Haskell and introduces a new constraint form (Section~\ref{sec:type-system}).}
\item{We introduce an explicitly typed core language which is System F extended
with multi-stage features (Section~\ref{sec:core})}.
\item{We give the elaboration semantics from the source language into the core
language (Section~\ref{sec:elaboration}).}
\end{itemize}
Future directions including type inference, the latest developments in
supporting impredicativity and notes about an implementation are then
discussed~(Section~\ref{sec:future}). This work is put into the context of
related work (Section~\ref{sec:related}), before finally concluding
(Section~\ref{sec:conclusion}).

\section{Background: Multi-Staged Programming}
\label{sec:background}





This section describes the fundamental concepts of Typed Template
Haskell as currently implemented in GHC. In these aspects, the current
implementation coincides with our proposed specification of Typed Template Haskell,
as we are excluding polymorphism and
type classes. This serves as preparation for more interesting
examples that are discussed later (Section~\ref{sec:motivation}).

Typed Template Haskell is an extension to Haskell which implements the two
standard staging annotations of multi-stage programming: \emph{quotes} and
\emph{splices}.  An expression~\ensuremath{\Varid{e}\mathbin{::}\Varid{a}} can be quoted to generate
the expression~\ensuremath{\qq{\Varid{e}}\mathbin{::}\Conid{Code}\;\Varid{a}}. Conversely, an expression~\ensuremath{\Varid{c}\mathbin{::}\Conid{Code}\;\Varid{a}}
can be spliced to extract the expression~\ensuremath{\$(\Varid{c})}. An expression of
type~\ensuremath{\Conid{Code}\;\Varid{a}} is a representation of an expression of type~\ensuremath{\Varid{a}}.
Given these definitions, it may seem that quotes and splices can be
used freely so long as the types align: well-typed problems don't go wrong, as
the old adage says; but things are not so simple for staged programs. As well
as being well-typed, a program must be {well-staged}, whereby the level of variables is considered.

The concept of a level is of fundamental importance in multi-stage programming.
The \emph{level} of an expression is an integer given by the number of quotes that
surround it, minus the number of splices: quotation increases the level and
splicing decreases the level.
Negative levels are
evaluated at compile time, level~\ensuremath{\mathrm{0}} is evaluated at runtime and positive levels
are future unevaluated stages.\footnote{Unlike MetaML~\cite{taha1997multi}, code
generation in Typed Template Haskell is only supported at compile time, and
therefore there is no~\ensuremath{\keyword{run}} operation.}
The goal of designing a staged calculus is to ensure that the levels of the
program can be evaluated in order so that an expression at a particular level
can only be evaluated when all expressions it depends on at previous levels
have been first evaluated.
\al{Would it make sense to define ``stage'' here? ``The goal of designing a staged
calculus is to ensure that the levels of the program can be evaluated in \textbf{stages}
so that \dots ?}

In the simplest setting, a program is \emph{well-staged} if each variable it
mentions is used only at the level in which it is bound. Doing so in any other
stage may simply be impossible, or at least require special attention.
The next three example programs, \ensuremath{\Varid{timely}}, \ensuremath{\Varid{hasty}}, and \ensuremath{\Varid{tardy}}, are all well-typed,
but only the first is well-staged.

Using a variable that was defined in the \emph{same} stage is permitted. A variable
can be introduced locally using a lambda abstraction inside a quotation and then used
freely within that context:
\begin{hscode}\SaveRestoreHook
\column{B}{@{}>{\hspre}l<{\hspost}@{}}\ColumnHook
\column{E}{@{}>{\hspre}l<{\hspost}@{}}\ColumnHook
\>[B]{}\Varid{timely}\mathbin{::}\Conid{Code}\;(\Conid{Int}\to \Conid{Int}){}\<[E]%
\\
\>[B]{}\Varid{timely}\mathrel{=}\qq{\lambda \mathit{x}\to \mathit{x}}{}\<[E]%
\ColumnHook
\end{hscode}\resethooks
Of course, the variable is still subject to the usual scoping rules of abstraction.

Using a variable at a level \emph{before} it is bound is
problematic because at the point we wish to evaluate the prior stage, we will
not yet know the value of the future stage variable and so the evaluation will
get stuck:
\begin{hscode}\SaveRestoreHook
\column{B}{@{}>{\hspre}l<{\hspost}@{}}\ColumnHook
\column{E}{@{}>{\hspre}l<{\hspost}@{}}\ColumnHook
\>[B]{}\Varid{hasty}\mathbin{::}\Conid{Code}\;\Conid{Int}\to \Conid{Int}{}\<[E]%
\\
\>[B]{}\Varid{hasty}\;\Varid{c}\mathrel{=}\$(\Varid{c}){}\<[E]%
\ColumnHook
\end{hscode}\resethooks
Here, we cannot splice~\ensuremath{\Varid{c}} without knowing the concrete representation that~\ensuremath{\Varid{c}}
will be instantiated to.
There is no recovery from this situation without violating the fact that lower
levels must be fully evaluated before higher ones.

Using a variable at a stage \emph{after} it is bound is problematic because the
variable will not generally be in the environment.  It may, for instance, be
bound to a certain known value at compile time but no longer present at runtime.
\begin{hscode}\SaveRestoreHook
\column{B}{@{}>{\hspre}l<{\hspost}@{}}\ColumnHook
\column{E}{@{}>{\hspre}l<{\hspost}@{}}\ColumnHook
\>[B]{}\Varid{tardy}\mathbin{::}\Conid{Int}\to \Conid{Code}\;\Conid{Int}{}\<[E]%
\\
\>[B]{}\Varid{tardy}\;\mathit{x}\mathrel{=}\qq{\mathit{x}}{}\<[E]%
\ColumnHook
\end{hscode}\resethooks
In contrast to the \ensuremath{\Varid{hasty}} example, the situation is not hopeless here: there are ways to
support referencing previous-stage variables, which is called \emph{cross-stage
persistence}~\cite{taha1997multi}.

One option is to interpret a variable from a previous stage using a \emph{lifting}
construct which copies the current value of the variable into a future-stage
representation. As the process of lifting is akin to serialisation, this can be
achieved quite easily for base types such as strings and integers, but more
complex types such as functions are problematic.

Another option is to use \emph{path-based persistence}: for example, a top-level
identifier defined in another module can be persisted, because we can assume that
the other module has been separately compiled, so the top-level identifier is still
available at the same location in future stages.

\nw{I think this commented para is redundant?}

GHC currently implements the restrictions described above. For cross-stage
persistence, it employs both lifting and path-based persistence. Top-level
variables defined in another module can be used at any level. Lifting is
restricted to types that are instances of a type class \ensuremath{\Conid{Lift}} and witnessed
by a method
\begin{hscode}\SaveRestoreHook
\column{B}{@{}>{\hspre}l<{\hspost}@{}}\ColumnHook
\column{E}{@{}>{\hspre}l<{\hspost}@{}}\ColumnHook
\>[B]{}\Varid{lift}\mathbin{::}\Conid{Lift}\;\Varid{a}\Rightarrow \Varid{a}\to \Conid{Code}\;\Varid{a}{}\<[E]%
\ColumnHook
\end{hscode}\resethooks
which notably excludes function types and other abstract types such as~\ensuremath{\Conid{IO}}.
An example such as \ensuremath{\Varid{tardy}} can then be viewed as being implicitly rewritten
to
\begin{hscode}\SaveRestoreHook
\column{B}{@{}>{\hspre}l<{\hspost}@{}}\ColumnHook
\column{E}{@{}>{\hspre}l<{\hspost}@{}}\ColumnHook
\>[B]{}\Varid{tardy'}\mathbin{::}\Conid{Int}\to \Conid{Code}\;\Conid{Int}{}\<[E]%
\\
\>[B]{}\Varid{tardy'}\;\mathit{x}\mathrel{=}\qq{\$(\Varid{lift}\;\mathit{x})}{}\<[E]%
\ColumnHook
\end{hscode}\resethooks
in which the reference of~\ensuremath{\mathit{x}} occurs at the same level as it is bound.
For this reason, our formal languages introduced in Section~\ref{sec:type-system}
and Section~\ref{sec:core}
will consider path-based persistence, but not implicit lifting, as this is
easy to add separately in the same way as GHC currently implements it.

Let us consider the example from the introduction once again:
\begin{hscode}\SaveRestoreHook
\column{B}{@{}>{\hspre}l<{\hspost}@{}}\ColumnHook
\column{10}{@{}>{\hspre}l<{\hspost}@{}}\ColumnHook
\column{14}{@{}>{\hspre}c<{\hspost}@{}}\ColumnHook
\column{14E}{@{}l@{}}\ColumnHook
\column{17}{@{}>{\hspre}l<{\hspost}@{}}\ColumnHook
\column{E}{@{}>{\hspre}l<{\hspost}@{}}\ColumnHook
\>[B]{}\Varid{power}\;\mathrm{0}\;{}\<[10]%
\>[10]{}\Varid{cn}{}\<[14]%
\>[14]{}\mathrel{=}{}\<[14E]%
\>[17]{}\qq{\mathrm{1}}{}\<[E]%
\\
\>[B]{}\Varid{power}\;\Varid{k}\;{}\<[10]%
\>[10]{}\Varid{cn}{}\<[14]%
\>[14]{}\mathrel{=}{}\<[14E]%
\>[17]{}\qq{\$(\Varid{cn})\mathbin{*}\$(\Varid{power}\;(\Varid{k}\mathbin{-}\mathrm{1})\;\Varid{cn})}{}\<[E]%
\ColumnHook
\end{hscode}\resethooks
This example makes use of quotes and splices. The references to the
variables \ensuremath{\Varid{power}}, \ensuremath{\Varid{k}} and \ensuremath{\Varid{cn}} are all at the same level as their
binding occurrence, as they occur within one splice and one quote.
The static information is the exponent~\ensuremath{\Varid{k}} and the run-time information
is the base~\ensuremath{\Varid{cn}}. Therefore by using the staged function the static information
can be eliminated by partially evaluating the function at compile-time by using
a top-level splice. The generated code does not mention the static
information.

When using Typed Template Haskell, it is common to just deal with two stages, a
\al{I've changed ``two levels'' to ``two stages'' here. I think even in Typed Template
Haskell, one is actually using at least three levels frequently: \ensuremath{\mathbin{+}\mathrm{1}}, \ensuremath{\mathrm{0}}, and \ensuremath{\mathbin{-}\mathrm{1}}.}
compile-time stage and a run-time stage. The compile-time stage contains all
the statically known information about the domain, the typing rules ensure that
information from the run-time stage is not needed to evaluate the compile-time
stage and then the program is partially evaluated at compile-time to evaluate
the compile-time fragment to a value.  It is a common misconception that Typed
Template Haskell only supports two stages, but there is no such
restriction.\footnote{There is, however, an artificial restriction about
nesting quotation brackets which makes writing programs with more than two
stages difficult. It is ongoing work to lift this restriction.}

\nw{Maybe kill this paragraph entirely}
\al{I commented it for now, as I also think it does not fit.}





%

\section{Staging with Type Classes and Polymorphism}
\label{sec:motivation}

\al{I've changed the title from ``Motivation: A Problem and an Informal
Solution'' as I totally did not get what ``the'' Problem and what the
Informal Solution is.}
\nw{You made it ``Typed Template Haskell with Type Classes and Polymorphism'', which is better. I think I like ``Problems with Type Classes and Polymorphism'' or ``Understanding Staging with Type Classes and Polymorphism'' more. I want to avoid TTH being the only focus. Besides, our position is that TTH includes type classes and polymorphism by specficiation.}

The examples in the previous section were simple demonstrations of the
importance of considering levels in a well-staged program.
This section discusses more complicated cases which involve class constraints,
top-level splices, instance definitions and type variables. It is here where
our proposed version of Typed Template Haskell deviates from GHC's current
implementation. Therefore, for each example we will report how the program
\emph{should} behave and contrast it with how the current implementation in
GHC behaves.

\nw{I don't think it's important to flag the \ensuremath{\Varid{power}} discussion in the signposting, so I'm commenting it out.}

\subsection{Constraints}
\label{sec:constraints}

Constraints introduced by type classes have the potential to cause staging
errors: type classes are implemented by passing dictionaries as evidence, and the
implicit use of these dictionaries must adhere to the level restrictions discussed in
Section~\ref{sec:background}. This requires a careful
treatment of the interaction between constraints and staging, which GHC
currently does not handle correctly.

\begin{cexample}
\label{ex:c1}

Consider the use of a type class method inside a quotation, similarly
to how~\ensuremath{\Varid{power}} is used in~\ensuremath{\Varid{power5}} in the introduction:
\begin{hscode}\SaveRestoreHook
\column{B}{@{}>{\hspre}l<{\hspost}@{}}\ColumnHook
\column{E}{@{}>{\hspre}l<{\hspost}@{}}\ColumnHook
\>[B]{}\Varid{c}_{1}\mathbin{::}\Conid{Show}\;\Varid{a}\Rightarrow \Conid{Code}\;(\Varid{a}\to \Conid{String}){}\<[E]%
\\
\>[B]{}\Varid{c}_{1}\mathrel{=}\qq{\Varid{show}}{}\<[E]%
\ColumnHook
\end{hscode}\resethooks
Thinking carefully about the levels involved, the signature indicates
that the evidence for \ensuremath{\Conid{Show}\;\Varid{a}}, which is available at stage~\ensuremath{\mathrm{0}}, can
be used to satisfy the evidence needed by \ensuremath{\Conid{Show}\;\Varid{a}} at stage~\ensuremath{\mathrm{1}}.

In the normal dictionary passing implementation of type classes, type class
constraints are elaborated to a function which accepts a dictionary which acts
as evidence for the constraint. Therefore we can assume that the elaborated
version of \ensuremath{\Varid{c}_{1}} looks similar to the following:
\begin{hscode}\SaveRestoreHook
\column{B}{@{}>{\hspre}l<{\hspost}@{}}\ColumnHook
\column{E}{@{}>{\hspre}l<{\hspost}@{}}\ColumnHook
\>[B]{}\Varid{d}_{1}\mathbin{::}\Conid{ShowDict}\;\Varid{a}\to \Conid{Code}\;(\Varid{a}\to \Conid{String}){}\<[E]%
\\
\>[B]{}\Varid{d}_{1}\;\Varid{dShow}\mathrel{=}\qq{\Varid{show}\;\Varid{dShow}}{}\<[E]%
\ColumnHook
\end{hscode}\resethooks
Now this reveals a subtle problem: naively elaborating without considering
the \emph{levels of constraints} has introduced a cross-stage reference where
the dictionary variable~\ensuremath{\Varid{dShow}} is introduced at level~\ensuremath{\mathrm{0}} but used at level~\ensuremath{\mathrm{1}}.
As we have learned in Section~\ref{sec:background}, one remedy of this situation
is to try to make~\ensuremath{\Varid{dShow}} cross-stage persistent by lifting. However, in general,
lifting of dictionaries is not straightforward to implement. Recall that lifting
in GHC is restricted to instances of the \ensuremath{\Conid{Lift}} class which excludes functions --
but type class dictionaries are nearly always a record of functions, so
automatic lifting given the current implementation is not possible.

GHC nevertheless accepts this program, with the underlying problem only being
revealed when subsequently trying to splice the program. We instead argue that
the program~\ensuremath{\Varid{c}_{1}} is ill-typed and should therefore be rejected at compilation time.
Our solution is to introduce a new constraint form, \ensuremath{\Conid{CodeC}\;\Conid{C}}, which indicates
that constraint~\ensuremath{\Conid{C}} is available to be used in the next stage. Using this,
the corrected type signature for example~\ref{ex:c1} is as follows:
\begin{hscode}\SaveRestoreHook
\column{B}{@{}>{\hspre}l<{\hspost}@{}}\ColumnHook
\column{E}{@{}>{\hspre}l<{\hspost}@{}}\ColumnHook
\>[B]{}\Varid{c}_{1}'\mathbin{::}\Conid{CodeC}\;(\Conid{Show}\;\Varid{a})\Rightarrow \Conid{Code}\;(\Varid{a}\to \Conid{String}){}\<[E]%
\\
\>[B]{}\Varid{c}_{1}'\mathrel{=}\qq{\Varid{show}}{}\<[E]%
\ColumnHook
\end{hscode}\resethooks
The \ensuremath{\Conid{CodeC}\;(\Conid{Show}\;\Varid{a})} constraint is introduced at level~\ensuremath{\mathrm{0}} but indicates that the
\ensuremath{\Conid{Show}\;\Varid{a}} constraint will be available to be used at level~\ensuremath{\mathrm{1}}. Therefore the
\ensuremath{\Conid{Show}\;\Varid{a}} constraint can be used to satisfy the \ensuremath{\Varid{show}} method used inside the
quotation.
The corresponding elaborated version is similar to the following:
\begin{hscode}\SaveRestoreHook
\column{B}{@{}>{\hspre}l<{\hspost}@{}}\ColumnHook
\column{E}{@{}>{\hspre}l<{\hspost}@{}}\ColumnHook
\>[B]{}\Varid{d}_{1}'\mathbin{::}\Conid{Code}\;(\Conid{ShowDict}\;\Varid{a})\to \Conid{Code}\;(\Varid{a}\to \Conid{String}){}\<[E]%
\\
\>[B]{}\Varid{d}_{1}'\;\Varid{cdShow}\mathrel{=}\qq{\Varid{show}\;\$(\Varid{cdShow})}{}\<[E]%
\ColumnHook
\end{hscode}\resethooks
As \ensuremath{\Varid{cdShow}} is now the representation of a dictionary, we can splice the representation
inside the quote. The reference to \ensuremath{\Varid{cdShow}} is at the correct level and
the program is well-staged.
\matt{Mention about impredicativity here?}
\nw{Not yet, I think. Let's give more good news first.}
\al{Agreed.}

\end{cexample}

\begin{cexample}
\label{ex:c2}

The example~\ref{ex:c1} uses a locally provided constraint which causes us some
difficulty. We have a different situation if a constraint can be
solved by a concrete global type class instance:
\begin{hscode}\SaveRestoreHook
\column{B}{@{}>{\hspre}l<{\hspost}@{}}\ColumnHook
\column{E}{@{}>{\hspre}l<{\hspost}@{}}\ColumnHook
\>[B]{}\Varid{c}_{2}\mathbin{::}\Conid{Code}\;(\Conid{Int}\to \Conid{String}){}\<[E]%
\\
\>[B]{}\Varid{c}_{2}\mathrel{=}\qq{\Varid{show}}{}\<[E]%
\ColumnHook
\end{hscode}\resethooks
In~\ensuremath{\Varid{c}_{2}}, the global \ensuremath{\Conid{Show}\;\Conid{Int}} instance is used to satisfy the \ensuremath{\Varid{show}} constraint
inside the quotation. Since this elaborates to a reference to a top-level
instance dictionary, the reference can be persisted using path-based persistence.
Therefore it is permitted to use top-level instances to satisfy constraints inside
a quotation, in the same way that it is permitted to refer to top-level variables.

\al{Can the following be phrased better?}
\nw{Maybe improved by stating how GHCs solution could potentially be harmful? Should we say explicitly that our calculus accepts this?}
\matt{It's not guaranteed that the same \ensuremath{\Conid{Show}\;\Conid{Int}} dictionary will be present
in both places using overlapping instances, there is an example if our previous
paper.}
GHC currently correctly accepts this program. However, it does not actually
perform the dictionary translation until the program is spliced, which is harmless
in this case, because the same \ensuremath{\Conid{Show}\;\Conid{Int}} instance is in scope at both points.
\end{cexample}

\subsection{Top-level splices}
\label{sec:top-level}

A splice that appears in a definition at the top-level scope of a module\footnote{%
Do not confuse this use of ``top-level'' with the staging level.}
introduces new scoping challenges
because it can potentially require class constraints to be used at levels prior
to the ones where they are introduced.

Since a top-level splice is evaluated at compile time, it should be clear
that no run-time information must be used in a top-level splice definition.
In particular, the definition should not be permitted to access local variables
or local constraints defined at a higher level.
\al{I changed ``prior'' to ``higher'' because otherwise, it does not make
sense to me.}
This is because local
information which is available only at runtime cannot be used to
influence the execution of the expression during compilation.

\begin{tsexample}
\label{ex:ts1}
In the following example there are two modules.  Module~\ensuremath{\Conid{A}} defines the \ensuremath{\Conid{Lift}} class and
contains the definition of a global instance \ensuremath{\Conid{Lift}\;\Conid{Int}}. The \ensuremath{\Varid{lift}} function
from this instance is used in module~\ensuremath{\Conid{B}} in the definition of \ensuremath{\Varid{ts}_{1}}, which is a
top-level definition. The reference to~\ensuremath{\Varid{lift}} occurs inside a top-level splice,
thus at level~\ensuremath{\mathbin{-}\mathrm{1}}, and so the \ensuremath{\Conid{Lift}\;\Conid{Int}} instance is needed a compile time.

\noindent
\begin{minipage}[t]{0.4\linewidth}
\begin{hscode}\SaveRestoreHook
\column{B}{@{}>{\hspre}l<{\hspost}@{}}\ColumnHook
\column{3}{@{}>{\hspre}l<{\hspost}@{}}\ColumnHook
\column{E}{@{}>{\hspre}l<{\hspost}@{}}\ColumnHook
\>[B]{}\mathbf{module}\;\Conid{A}\;\mathbf{where}{}\<[E]%
\\[\blanklineskip]%
\>[B]{}\mathbf{class}\;\Conid{Lift}\;\Varid{a}\;\mathbf{where}{}\<[E]%
\\
\>[B]{}\hsindent{3}{}\<[3]%
\>[3]{}\Varid{lift}\mathbin{::}\Varid{a}\to \Conid{Code}\;\Varid{a}{}\<[E]%
\\[\blanklineskip]%
\>[B]{}\mathbf{instance}\;\Conid{Lift}\;\Conid{Int}\;\mathbf{where}{}\<[E]%
\\
\>[B]{}\hsindent{3}{}\<[3]%
\>[3]{}\mathbin{...}{}\<[E]%
\ColumnHook
\end{hscode}\resethooks
\end{minipage}
\begin{minipage}[t]{0.4\linewidth}
\begin{hscode}\SaveRestoreHook
\column{B}{@{}>{\hspre}l<{\hspost}@{}}\ColumnHook
\column{E}{@{}>{\hspre}l<{\hspost}@{}}\ColumnHook
\>[B]{}\mathbf{module}\;\Conid{B}\;\mathbf{where}{}\<[E]%
\\
\>[B]{}\mathbf{import}\;\Conid{A}{}\<[E]%
\\[\blanklineskip]%
\>[B]{}\Varid{ts}_{1}\mathbin{::}\Conid{Int}{}\<[E]%
\\
\>[B]{}\Varid{ts}_{1}\mathrel{=}\$(\Varid{lift}\;\mathrm{5}){}\<[E]%
\ColumnHook
\end{hscode}\resethooks
\end{minipage}

\noindent
This program is currently accepted by GHC, and it should be, because the evidence
for a top-level instance is defined in a top-level variable and top-level definitions defined
in other modules are permitted to appear in top-level splices.

\al{I am not terribly happy with the following paragraph. Should this restriction be mentioned
in Section~\ref{sec:background}? Also, the use of ``mirror'' is wrong because the purpose
of this section is to a large extent to describe how GHC is behaving currently.}
The situation is subtly different to top-level quotations as in example~\ref{ex:c2},
because the instance
must be defined in another module which mirrors the restriction
implemented in GHC that top-level definitions can only be used in top-level splices
if they are defined in other modules.
\end{tsexample}

\begin{tsexample}
\label{ex:ts2}

\matt{deleted an incorrect sentence here about instanitating a type}
On the other hand, constraints introduced locally by a type signature for a
top-level definition must not be allowed.  In the following example, the
\ensuremath{\Conid{Lift}\;\Conid{A}} constraint is introduced at level \ensuremath{\mathrm{0}} by the type signature of \ensuremath{\Varid{ts}_{2}}
but used at level \ensuremath{\mathbin{-}\mathrm{1}} in the body:
\begin{hscode}\SaveRestoreHook
\column{B}{@{}>{\hspre}l<{\hspost}@{}}\ColumnHook
\column{E}{@{}>{\hspre}l<{\hspost}@{}}\ColumnHook
\>[B]{}\mathbf{data}\;\Conid{A}\mathrel{=}\Conid{A}{}\<[E]%
\\[\blanklineskip]%
\>[B]{}\Varid{ts}_{2}\mathbin{::}\Conid{Lift}\;\Conid{A}\Rightarrow \Conid{A}{}\<[E]%
\\
\>[B]{}\Varid{ts}_{2}\mathrel{=}\$(\Varid{lift}\;\Conid{A}){}\<[E]%
\ColumnHook
\end{hscode}\resethooks
We assume \ensuremath{\Varid{lift}} is imported from another module as in Example~\ref{ex:ts1}
and therefore cross-stage persistent.
However, the problem is that the dictionary that will be used to implement the~\ensuremath{\Conid{Lift}\;\Conid{A}}
constraint will not be known until runtime.
Therefore the definition of \ensuremath{\Conid{Lift}\;\Conid{A}} is not available to be used
at compile-time in the top-level splice, and GHC correctly rejects this program.
It is evident why this program should be rejected considering the dictionary
translation, in a similar vein to~\ref{ex:c1}. The elaboration would amount
to a future-stage reference inside the splice.
\end{tsexample}

\begin{tsexample}
\label{ex:ts3}
So far we have considered the interaction between constraints and quotations
separately to the interaction between constraints and top-level splices.
The combination of the two reveals further subtle issues.
The following function~\ensuremath{\Varid{ts}_{3}} is at the top-level and
uses a constrained type. This example is both well-typed and well-staged.
\begin{hscode}\SaveRestoreHook
\column{B}{@{}>{\hspre}l<{\hspost}@{}}\ColumnHook
\column{E}{@{}>{\hspre}l<{\hspost}@{}}\ColumnHook
\>[B]{}\Varid{ts}_{3}\mathbin{::}\Conid{Ord}\;\Varid{a}\Rightarrow \Varid{a}\to \Varid{a}\to \Conid{Ordering}{}\<[E]%
\\
\>[B]{}\Varid{ts}_{3}\mathrel{=}\$(\qq{\Varid{compare}}){}\<[E]%
\ColumnHook
\end{hscode}\resethooks
In~\ensuremath{\Varid{ts}_{3}}, the body of the top-level splice is a simple quotation of the \ensuremath{\Varid{compare}}
method. This method requires an \ensuremath{\Conid{Ord}} constraint which is provided by the context
on~\ensuremath{\Varid{ts}_{3}}. The constraint is introduced at level~\ensuremath{\mathrm{0}} and also used at level~\ensuremath{\mathrm{0}}, as
the top-level splice and the quotation cancel each other out. It is therefore
perfectly fine to use the dictionary passed to \ensuremath{\Varid{ts}_{3}} to satisfy the requirements
of~\ensuremath{\Varid{compare}}.

Unfortunately, the current implementation in GHC rejects this program. It generally
excludes local constraints from the scope inside top-level splices, in order to reject
programs like Example~\ref{ex:ts2}. Our
specification accepts the example by tracking the levels of local constraints.
\end{tsexample}

\subsection{The \ensuremath{\Varid{power}} function revisited}
\label{sec:power-revisited}

Having discussed constraints and also their interaction with top-level splices, we
can now fully explain
why \ensuremath{\Varid{power5}} function discussed in Section~\ref{sec:introduction} works in current
GHC when spliced at a concrete type, but fails when spliced with an overloaded type,
whereas it should work for both:
\begin{hscode}\SaveRestoreHook
\column{B}{@{}>{\hspre}l<{\hspost}@{}}\ColumnHook
\column{10}{@{}>{\hspre}l<{\hspost}@{}}\ColumnHook
\column{30}{@{}>{\hspre}l<{\hspost}@{}}\ColumnHook
\column{E}{@{}>{\hspre}l<{\hspost}@{}}\ColumnHook
\>[B]{}\Varid{power5}{}\<[10]%
\>[10]{}\mathbin{::}\Conid{Int}\to \Conid{Int}{}\<[30]%
\>[30]{}\mbox{\onelinecomment  Option M}{}\<[E]%
\\
\>[B]{}\Varid{power5}{}\<[10]%
\>[10]{}\mathbin{::}\Conid{Num}\;\Varid{a}\Rightarrow \Varid{a}\to \Varid{a}{}\<[30]%
\>[30]{}\mbox{\onelinecomment  Option P}{}\<[E]%
\\
\>[B]{}\Varid{power5}\;\Varid{n}\mathrel{=}\$(\Varid{power}\;\mathrm{5}\;\qq{\Varid{n}}){}\<[E]%
\ColumnHook
\end{hscode}\resethooks
The two problems are that GHC accepts \ensuremath{\Varid{power}} at the incorrect type
\begin{hscode}\SaveRestoreHook
\column{B}{@{}>{\hspre}l<{\hspost}@{}}\ColumnHook
\column{E}{@{}>{\hspre}l<{\hspost}@{}}\ColumnHook
\>[B]{}\Varid{power}\mathbin{::}\Conid{Num}\;\Varid{a}\Rightarrow \Conid{Int}\to \Conid{Code}\;\Varid{a}\to \Conid{Code}\;\Varid{a}{}\<[E]%
\ColumnHook
\end{hscode}\resethooks
(as in Example~\ref{ex:c1})
and that it does not actually perform a dictionary translation for this until
this is spliced. Option~M works in GHC, because it finds the \ensuremath{\Conid{Num}\;\Conid{Int}} global
instance when splicing, similarly to Example~\ref{ex:ts1}, and everything is
(accidentally) fine. However, Option~P fails because the local constraint is
not made available in the splice as in Example~\ref{ex:ts2}, and even if it was,
it would be a reference at the wrong level.

As with Example~\ref{ex:c1}, we argue that the function \ensuremath{\Varid{power}}
should instead only be accepted at type
\begin{hscode}\SaveRestoreHook
\column{B}{@{}>{\hspre}l<{\hspost}@{}}\ColumnHook
\column{E}{@{}>{\hspre}l<{\hspost}@{}}\ColumnHook
\>[B]{}\Varid{power}\mathbin{::}\Conid{CodeC}\;(\Conid{Num}\;\Varid{a})\Rightarrow \Conid{Int}\to \Conid{Code}\;\Varid{a}\to \Conid{Code}\;\Varid{a}{}\<[E]%
\ColumnHook
\end{hscode}\resethooks
Then, Option~M is fine due to the cross-stage persistence of the global
\ensuremath{\Conid{Num}\;\Conid{Int}} instance declaration; and Option~P works as well, as the program will
elaborate to code that is similar to:
\begin{hscode}\SaveRestoreHook
\column{B}{@{}>{\hspre}l<{\hspost}@{}}\ColumnHook
\column{E}{@{}>{\hspre}l<{\hspost}@{}}\ColumnHook
\>[B]{}\Varid{power5'}\mathbin{::}\Conid{NumDict}\;\Varid{a}\to \Varid{a}\to \Varid{a}{}\<[E]%
\\
\>[B]{}\Varid{power5'}\;\Varid{dNum}\;\Varid{n}\mathrel{=}\$(\Varid{power}\;\qq{\Varid{dNum}}\;\mathrm{5}\;\qq{\Varid{n}}){}\<[E]%
\ColumnHook
\end{hscode}\resethooks
By quoting \ensuremath{\Varid{dNum}}, the argument to \ensuremath{\Varid{power}} is a representation
of a dictionary as required, and reference is at the correct level.

\subsection{Instance Definitions}
\label{sec:instances}

The final challenge to do with constraints is dealing with instance definitions
which use top-level splices. This situation is of particular interest as there
are already special typing rules in GHC for instance methods which bring into
scope the instance currently being defined in the body of the instance
definition. This commonly happens in recursive datatypes where the instance
declaration must be recursively defined.

\begin{iexample}
\label{ex:i1}

In the same manner as a top-level splice, the body of an instance method
cannot use a local constraint, in particular the instance currently
being defined or any of the instance head in order to influence the code generation
in a top-level splice. On the other hand, the generated code should certainly
be permitted to use the instances from the instance context and the currently
defined instance. In fact, this is necessary in order to generate the majority
of instances for recursive datatypes!
Here is a instance definition that is defined recursively.
\begin{hscode}\SaveRestoreHook
\column{B}{@{}>{\hspre}l<{\hspost}@{}}\ColumnHook
\column{3}{@{}>{\hspre}l<{\hspost}@{}}\ColumnHook
\column{13}{@{}>{\hspre}c<{\hspost}@{}}\ColumnHook
\column{13E}{@{}l@{}}\ColumnHook
\column{16}{@{}>{\hspre}l<{\hspost}@{}}\ColumnHook
\column{E}{@{}>{\hspre}l<{\hspost}@{}}\ColumnHook
\>[B]{}\mathbf{data}\;\Conid{Stream}\mathrel{=}\Conid{Cons}\;\{\mskip1.5mu \Varid{hd}\mathbin{::}\Conid{Int},\Varid{tl}\mathbin{::}\Conid{Stream}\mskip1.5mu\}{}\<[E]%
\\[\blanklineskip]%
\>[B]{}\mathbf{instance}\;\Conid{Eq}\;\Conid{Stream}\;\mathbf{where}{}\<[E]%
\\
\>[B]{}\hsindent{3}{}\<[3]%
\>[3]{}\Varid{s}_{1}\doubleequals \Varid{s}_{2}{}\<[13]%
\>[13]{}\mathrel{=}{}\<[13E]%
\>[16]{}\Varid{hd}\;\Varid{s}_{1}\doubleequals \Varid{hd}\;\Varid{s}_{2}\mathrel{\&\&}\Varid{tl}\;\Varid{s}_{1}\doubleequals \Varid{tl}\;\Varid{s}_{2}{}\<[E]%
\ColumnHook
\end{hscode}\resethooks
The instance for \ensuremath{\Conid{Eq}\;\Conid{Stream}} compares the tails of the streams using the
equality instance for \ensuremath{\Conid{Eq}\;\Conid{Stream}} that is currently being defined. This code works
well, as it should, so there is reason to believe that a staged version should
also be definable.
\end{iexample}

\begin{iexample}
\label{ex:i2}
The following is just a small variation of Example~\ref{ex:i1} where the body
of the method is wrapped in a splice and a quote:
\begin{hscode}\SaveRestoreHook
\column{B}{@{}>{\hspre}l<{\hspost}@{}}\ColumnHook
\column{3}{@{}>{\hspre}l<{\hspost}@{}}\ColumnHook
\column{E}{@{}>{\hspre}l<{\hspost}@{}}\ColumnHook
\>[B]{}\mathbf{instance}\;\Conid{Eq}\;\Conid{Stream}\;\mathbf{where}{}\<[E]%
\\
\>[B]{}\hsindent{3}{}\<[3]%
\>[3]{}(\doubleequals )\mathrel{=}\$(\qq{\lambda \Varid{s}_{1}\;\Varid{s}_{2}\to \Varid{hd}\;\Varid{s}_{1}\doubleequals \Varid{hd}\;\Varid{s}_{2}\mathrel{\&\&}\Varid{tl}\;\Varid{s}_{1}\doubleequals \Varid{tl}\;\Varid{s}_{2}}){}\<[E]%
\ColumnHook
\end{hscode}\resethooks
This program should be accepted and equivalent to~\ref{ex:i1},
because in general, \ensuremath{\$(\qq{\Varid{e}})\mathrel{=}\Varid{e}}.
However, it is presently rejected by GHC, with the claim that the recursive use
of \ensuremath{(\doubleequals )} for comparing the tails is not stage-correct, when in principle
it could and should be made cross-stage persistent.
\al{Is this explanation sufficient?}
\end{iexample}

\begin{iexample}
\label{ex:i3}
Yet another variant of Example~\ref{ex:i2} is to try to move the recursion
out of the instance declaration, as follows:
\begin{hscode}\SaveRestoreHook
\column{B}{@{}>{\hspre}l<{\hspost}@{}}\ColumnHook
\column{E}{@{}>{\hspre}l<{\hspost}@{}}\ColumnHook
\>[B]{}\Varid{eqStream}\mathbin{::}\Conid{CodeC}\;(\Conid{Eq}\;\Conid{Stream})\Rightarrow \Conid{Code}\;(\Conid{Stream}\to \Conid{Stream}\to \Conid{Bool}){}\<[E]%
\\
\>[B]{}\Varid{eqStream}\mathrel{=}\qq{\lambda \Varid{s}_{1}\;\Varid{s}_{2}\to \Varid{hd}\;\Varid{s}_{1}\doubleequals \Varid{hd}\;\Varid{s}_{2}\mathrel{\&\&}\Varid{tl}\;\Varid{s}_{1}\doubleequals \Varid{tl}\;\Varid{s}_{2}}{}\<[E]%
\ColumnHook
\end{hscode}\resethooks
Then, in a different module, we should be able to say:
\begin{hscode}\SaveRestoreHook
\column{B}{@{}>{\hspre}l<{\hspost}@{}}\ColumnHook
\column{3}{@{}>{\hspre}l<{\hspost}@{}}\ColumnHook
\column{E}{@{}>{\hspre}l<{\hspost}@{}}\ColumnHook
\>[B]{}\mathbf{instance}\;\Conid{Eq}\;\Conid{Stream}\;\mathbf{where}{}\<[E]%
\\
\>[B]{}\hsindent{3}{}\<[3]%
\>[3]{}(\doubleequals )\mathrel{=}\$(\Varid{eqStream}){}\<[E]%
\ColumnHook
\end{hscode}\resethooks
It is important that the definition of \ensuremath{\Varid{eqStream}} uses the new constraint
form \ensuremath{\Conid{CodeC}\;(\Conid{Eq}\;\Conid{Stream})} so that the definition of \ensuremath{\Varid{eqStream}}
is well-staged. As \ensuremath{\Conid{Eq}\;\Conid{Stream}} is the instance which is
currently being defined, the \ensuremath{\Conid{Eq}\;\Conid{Stream}} constraint is introduced at level 0 when typing the
instance definition. Therefore, in a top-level splice the local constraint can only be used
by functions which require a \ensuremath{\Conid{CodeC}\;(\Conid{Eq}\;\Conid{Stream})} constraint.
This example is similar to \ensuremath{\Varid{power}} in
Section~\ref{sec:power-revisited} but the local constraint is introduced
implicitly by the instance definition.

It is somewhat ironic that while one of the major use cases of untyped
Template Haskell is generating instance definitions such as this one,
it seems impossible to use the current implementation of Typed Template
Haskell for the same purpose.
\end{iexample}

\subsection{Type Variables}
\label{sec:type-variables}

\al{This subsection is quite weak. The example does not make sense
(\ensuremath{\Varid{get}} defined in terms of \ensuremath{\Varid{get}}?). Furthermore, it could easily be
``fixed'' anyway because the type variable is used at a single level.
The second half of this section is very technical and abstract, and
I think it distracts from the flow of the paper. Also, the main
argument of not having case distinctions on types prompts the question
is we have difficulties handling equality constraints and/or type
families. I'll defer this for now, but will try to look at this
again.}

The previous examples showed that type constraints require special attention in
order to ensure correct staging. It is therefore natural to consider type
variables as well, and indeed previous work by~\citet{pickering2019multi} ensured that type
variables are were level-aware in order to ensure a sound translation.
However, our calculus will show that this is a conservative position in a
language based on the polymorphic lambda calculus which enforces a phase distinction~\cite{cardelli1988phase} where
typechecking happens entirely prior to running any stages of a program.

\begin{tvexample}
\label{ex:tv1}
Consider the following example that turns a list of quoted
values into a quoted list of those values:
\begin{hscode}\SaveRestoreHook
\column{B}{@{}>{\hspre}l<{\hspost}@{}}\ColumnHook
\column{16}{@{}>{\hspre}l<{\hspost}@{}}\ColumnHook
\column{E}{@{}>{\hspre}l<{\hspost}@{}}\ColumnHook
\>[B]{}\Varid{list}\mathbin{::}\forall \Varid{a}\hsforall .[\mskip1.5mu \Conid{Code}\;\Varid{a}\mskip1.5mu]\to \Conid{Code}\;[\mskip1.5mu \Varid{a}\mskip1.5mu]{}\<[E]%
\\
\>[B]{}\Varid{list}\;[\mskip1.5mu \mskip1.5mu]{}\<[16]%
\>[16]{}\mathrel{=}\qq{[\mskip1.5mu \mskip1.5mu]\;\mathord{@}\Varid{a}}{}\<[E]%
\\
\>[B]{}\Varid{list}\;(\mathit{x}\mathbin{:}\Varid{xs}){}\<[16]%
\>[16]{}\mathrel{=}\qq{(\mathbin{:})\;\mathord{@}\Varid{a}\;\$(\mathit{x})\;\$(\Varid{list}\;\mathord{@}\Varid{a}\;\Varid{xs})}{}\<[E]%
\ColumnHook
\end{hscode}\resethooks
The type variable~\ensuremath{\Varid{a}} is bound at level $0$ and used both at at level $1$ (in
its application to the type constructors), and at level $0$ (in its recursive
application to the \ensuremath{\Varid{list}} function).
If we took the
lead from values then this program would be rejected by the typechecker.
However, because of the in-built phase distinction, evaluation
cannot get stuck on an unknown type variable as all the types will be
fixed in the program before execution starts.  We just need to make sure that
the substitution operation can substitute a type inside a quotation.

As it happens, this definition is correctly accepted by GHC, which is
promising. However, it cannot actually be used since GHC produces compile error
when given an expression such as \ensuremath{\$(\Varid{list}\;[\mskip1.5mu \qq{\Conid{True}},\qq{\Conid{False}}\mskip1.5mu])}, since it complains that the type variable \ensuremath{\Varid{a}} is not in scope during type
checking, and eventually generates an internal error in GHC itself.
\end{tvexample}

\matt{Axed this as it amounts to saying "there is a phase-distinction"}
%

\nw{I've axed the next para which talks about the solution, since we don't do that in this section and I want this part to be brief.}

\subsection{Intermediate Summary}

This section has uncovered the fact that the treatment of
constraints is rather arbitrary in the current implementation of GHC.
The examples we discussed have been motivated by three different ad-hoc
restrictions that the current implementation exhibits.
Firstly, prior stage constraint references (Section~\ref{sec:constraints}) are
permitted without any checks.
Secondly, top-level splices
(Section~\ref{sec:top-level})
are typechecked in an environment isolated from the
local constraint environment which is an overly conservative restriction.
Thirdly, in an instance definition (Section~\ref{sec:instances}), any reference
to the instance currently being defined inside a top-level splice is rejected
even if it is level-correct.
Additionally
we showed that although
constraints have a role to play in understanding staging, type variables need
no special treatment beyond the correct use of type application
(Section~\ref{sec:type-variables}).

Do the problems uncovered threaten the soundness of the language? In the case
of prior-stage references, they do. The current representation form of
quotations in GHC is untyped and therefore any evidence is erased from the internal
representation of a quotation.
The instance is therefore ``persisted by inference'', which operates under
the assumption that enough information is present in an untyped representation
to re-infer all the contextual type information erased after typechecking.
This assumption leads to unsoundness as generated programs will fail to
typecheck, as discussed already by~\citet{pickering2019multi}.
Future-stage references are forbidden by conservative checks,
which we will aim to make more precise in our calculus in order to accept more programs.

The problems observed so far are the result of interactions between class
constraints and staging.  Our goal now will be to develop a formal calculus to
model and resolve these problems. The approach will be to
introduce a source language  that includes type constraints in addition to
quotes and splices (Section~\ref{sec:type-system}), and
a core language that is a variant of the explicit polymorphic lambda calculus with multi-stage constructs
(Section~\ref{sec:core}). We then show how the source language can be
translated into a core language where constraints have been elaborated away
into a representation form for a quotation that is typed and where all
contextual type information is saved (Section~\ref{sec:elaboration}).





\section{Source Language}
\label{sec:type-system}

The source language we introduce has been designed to incorporate the essential
features of a language with metaprogramming and qualified types that is able to
correctly handle the situations discussed in Section~\ref{sec:motivation}.
We try to stay faithful to GHC's current implementation of Typed Template Haskell
where possible, with the addition of the quoted constraint form \ensuremath{\Conid{CodeC}} for the reasons
discussed in Section~\ref{sec:motivation}.
The key features of this language are:
\al{I'm not sure what the status of this list is. Points 1--3 are really about
our contributions. Point 4 is more general, which is why I added point 5. Still
feels a bit arbitrary though.}
\begin{enumerate}
\item{Values \emph{and constraints} are always indexed by the level at which
they are introduced to ensure that they are well-staged.}
\item{The constraint form~\ensuremath{\Conid{CodeC}} indicates that a constraint is
available at the next stage.}
\item{Types (including type variables) are not level-indexed and can be used at any level.}
\item{There is no \ensuremath{\keyword{run}} operation in the language. All evaluation of code is performed
at compile-time by mean of top-level splices which imply the existence of negative levels.}
\item{Path-based cross-stage persistence is supported.}
\nw{I removed the discussion of impredicativity here because it is advanced material that hasn't yet been defined. It is now incorporated into the description of syntax.}
\end{enumerate}

\nw{My gut feeling is that we've compared to GHC's TTH enough already. I want
to remove this paragraph and instead say something pithy: ``These features have
been carefully chosen to allow a specification of Typed Template Haskell
that does not suffer from the shortcomings of the current implementation.''}

\noindent
These features have been carefully chosen to allow a specification of Typed
Template Haskell that addresses all the shortcomings identified in Section~\ref{sec:motivation}
while staying close to the general flavour of meta-programming that
is enabled in its current implementation in GHC.

\subsection{Syntax}

\begingroup
\def\ghl#1{}





\begin{figure*}

\invisiblecomments
\aligncolumn{130}{r}%
\savecolumns
\begin{hscode}\SaveRestoreHook
\column{B}{@{}>{\hspre}l<{\hspost}@{}}\ColumnHook
\column{18}{@{}>{\hspre}c<{\hspost}@{}}\ColumnHook
\column{18E}{@{}l@{}}\ColumnHook
\column{23}{@{}>{\hspre}l<{\hspost}@{}}\ColumnHook
\column{125}{@{}>{\hspre}l<{\hspost}@{}}\ColumnHook
\column{130}{@{}>{\hspre}l<{\hspost}@{}}\ColumnHook
\column{E}{@{}>{\hspre}l<{\hspost}@{}}\ColumnHook
\>[B]{}\Varid{pgm}{}\<[18]%
\>[18]{}\mathbin{::=}{}\<[18E]%
\>[23]{}\Varid{e}\mid \Varid{def};\Varid{pgm}\mid \Varid{cls};\Varid{pgm}\mid \Varid{inst};\Varid{pgm}\;{}\<[E]%
\\
\>[B]{}\Varid{def}{}\<[18]%
\>[18]{}\mathbin{::=}{}\<[18E]%
\>[23]{}\mathbf{def}~\Varid{k}\mathrel{=}\Varid{e}\;{}\<[E]%
\\
\>[B]{}\Varid{cls}{}\<[18]%
\>[18]{}\mathbin{::=}{}\<[18E]%
\>[23]{}\mathbf{class}\;~\Conid{TC}~\Varid{a}\;\mathbf{where}\;\{\mskip1.5mu \Varid{k}\mathbin{::}\sigma \mskip1.5mu\}\;{}\<[E]%
\\
\>[B]{}\Varid{inst}{}\<[18]%
\>[18]{}\mathbin{::=}{}\<[18E]%
\>[23]{}\mathbf{instance}\;~\overline{\Conid{C}}\Rightarrow \Conid{TC}~\tau \;\mathbf{where}\;\{\mskip1.5mu \Varid{k}\mathrel{=}\Varid{e}\mskip1.5mu\}\;{}\<[E]%
\\[\blanklineskip]%
\>[B]{}\Varid{e}{}\<[18]%
\>[18]{}\mathbin{::=}{}\<[18E]%
\>[125]{}\quad {}\<[130]%
\>[130]{}\mbox{\onelinecomment  expressions}{}\<[E]%
\\
\>[18]{} {}\<[18E]%
\>[23]{}\mathit{x}\mid \Varid{k}{}\<[125]%
\>[125]{}\quad {}\<[130]%
\>[130]{}\mbox{\onelinecomment  variables / globals}{}\<[E]%
\\
\>[18]{}\mid {}\<[18E]%
\>[23]{}\lambda \mathit{x}\colonT \tau .\Varid{e}\mid \Varid{e}~\Varid{e}{}\<[125]%
\>[125]{}\quad {}\<[130]%
\>[130]{}\mbox{\onelinecomment  abstraction / application}{}\<[E]%
\\
\>[18]{}\mid {}\<[18E]%
\>[23]{}\qq{\Varid{e}}\mid \$(\Varid{e}){}\<[125]%
\>[125]{}\quad \;{}\<[130]%
\>[130]{}\mbox{\onelinecomment  quotation / splice}{}\<[E]%
\\[\blanklineskip]%
\>[B]{}\tau {}\<[18]%
\>[18]{}\mathbin{::=}{}\<[18E]%
\>[125]{}\quad {}\<[130]%
\>[130]{}\mbox{\onelinecomment  types}{}\<[E]%
\\
\>[18]{} {}\<[18E]%
\>[23]{}\Varid{a}\mid \Conid{H}{}\<[125]%
\>[125]{}\quad {}\<[130]%
\>[130]{}\mbox{\onelinecomment  variables / constants}{}\<[E]%
\\
\>[18]{}\mid {}\<[18E]%
\>[23]{}\tau \to \tau {}\<[125]%
\>[125]{}\quad {}\<[130]%
\>[130]{}\mbox{\onelinecomment  functions}{}\<[E]%
\\
\>[18]{}\mid {}\<[18E]%
\>[23]{}\Conid{Code}\;\tau {}\<[125]%
\>[125]{}\quad \;{}\<[130]%
\>[130]{}\mbox{\onelinecomment  representation}{}\<[E]%
\\
\>[B]{}\rho {}\<[18]%
\>[18]{}\mathbin{::=}{}\<[18E]%
\>[23]{}\tau \mid \Conid{C}\Rightarrow \rho {}\<[125]%
\>[125]{}\quad {}\<[130]%
\>[130]{}\mbox{\onelinecomment  qualified types}{}\<[E]%
\ColumnHook
\end{hscode}\resethooks
\spacecorrectnosep
\aligncolumn{130}{r}%
\restorecolumns
\begin{hscode}\SaveRestoreHook
\column{B}{@{}>{\hspre}l<{\hspost}@{}}\ColumnHook
\column{18}{@{}>{\hspre}c<{\hspost}@{}}\ColumnHook
\column{18E}{@{}l@{}}\ColumnHook
\column{23}{@{}>{\hspre}l<{\hspost}@{}}\ColumnHook
\column{125}{@{}>{\hspre}c<{\hspost}@{}}\ColumnHook
\column{125E}{@{}l@{}}\ColumnHook
\column{130}{@{}>{\hspre}l<{\hspost}@{}}\ColumnHook
\column{E}{@{}>{\hspre}l<{\hspost}@{}}\ColumnHook
\>[B]{}\sigma {}\<[18]%
\>[18]{}\mathbin{::=}{}\<[18E]%
\>[23]{}\rho \mid \forall \Varid{a}.\sigma {}\<[125]%
\>[125]{}\quad {}\<[125E]%
\>[130]{}\mbox{\onelinecomment  quantification}{}\<[E]%
\ColumnHook
\end{hscode}\resethooks
\spacecorrectsep
\aligncolumn{130}{r}%
\restorecolumns
\begin{hscode}\SaveRestoreHook
\column{B}{@{}>{\hspre}l<{\hspost}@{}}\ColumnHook
\column{18}{@{}>{\hspre}c<{\hspost}@{}}\ColumnHook
\column{18E}{@{}l@{}}\ColumnHook
\column{23}{@{}>{\hspre}l<{\hspost}@{}}\ColumnHook
\column{125}{@{}>{\hspre}c<{\hspost}@{}}\ColumnHook
\column{125E}{@{}l@{}}\ColumnHook
\column{130}{@{}>{\hspre}l<{\hspost}@{}}\ColumnHook
\column{E}{@{}>{\hspre}l<{\hspost}@{}}\ColumnHook
\>[B]{}\Conid{C}{}\<[18]%
\>[18]{}\mathbin{::=}{}\<[18E]%
\>[23]{}\Conid{TC}~\tau \mid \Conid{CodeC}\;\Conid{C}{}\<[125]%
\>[125]{}\quad {}\<[125E]%
\>[130]{}\mbox{\onelinecomment  constraint / representation}{}\<[E]%
\ColumnHook
\end{hscode}\resethooks
\spacecorrectnosep
\aligncolumn{132}{r}%
\restorecolumns
\begin{hscode}\SaveRestoreHook
\column{B}{@{}>{\hspre}l<{\hspost}@{}}\ColumnHook
\column{18}{@{}>{\hspre}c<{\hspost}@{}}\ColumnHook
\column{18E}{@{}l@{}}\ColumnHook
\column{23}{@{}>{\hspre}l<{\hspost}@{}}\ColumnHook
\column{127}{@{}>{\hspre}c<{\hspost}@{}}\ColumnHook
\column{127E}{@{}l@{}}\ColumnHook
\column{132}{@{}>{\hspre}l<{\hspost}@{}}\ColumnHook
\column{E}{@{}>{\hspre}l<{\hspost}@{}}\ColumnHook
\>[B]{}\Gamma {}\<[18]%
\>[18]{}\mathbin{::=}{}\<[18E]%
\>[23]{}\bullet \mid \Gamma ,\mathit{x}\colonT (\tau ,\Varid{n})\mid \Gamma ,\Varid{a}\mid \Gamma , \ghl{\Varid{ev}\colonT\!\!}(\Conid{C},\Varid{n}){}\<[127]%
\>[127]{}\quad {}\<[127E]%
\>[132]{}\mbox{\onelinecomment  type environment}{}\<[E]%
\ColumnHook
\end{hscode}\resethooks
\spacecorrectnosep
\aligncolumn{132}{r}%
\restorecolumns
\begin{hscode}\SaveRestoreHook
\column{B}{@{}>{\hspre}l<{\hspost}@{}}\ColumnHook
\column{18}{@{}>{\hspre}c<{\hspost}@{}}\ColumnHook
\column{18E}{@{}l@{}}\ColumnHook
\column{23}{@{}>{\hspre}l<{\hspost}@{}}\ColumnHook
\column{127}{@{}>{\hspre}c<{\hspost}@{}}\ColumnHook
\column{127E}{@{}l@{}}\ColumnHook
\column{132}{@{}>{\hspre}l<{\hspost}@{}}\ColumnHook
\column{E}{@{}>{\hspre}l<{\hspost}@{}}\ColumnHook
\>[B]{}\Conid{P}{}\<[18]%
\>[18]{}\mathbin{::=}{}\<[18E]%
\>[23]{}\bullet \mid \Conid{P}, \ghl{\Varid{ev}\colonT\!\!}(\forall a.\overline{\Conid{C}}\Rightarrow \Conid{C})\mid \Conid{P},\Varid{k}\colonT\sigma {}\<[127]%
\>[127]{}\quad {}\<[127E]%
\>[132]{}\mbox{\onelinecomment  program environment}{}\<[E]%
\ColumnHook
\end{hscode}\resethooks
\caption{Source Syntax}
\label{fig:syntax}
\end{figure*}
\endgroup

The syntax of the source language (Figure~\ref{fig:syntax}) models programs
with type classes and metaprogramming features.  This syntax
closely follows the models of languages with type classes of
\citet{bottu2017quantified} and \citet{chakravarty2005associated},
but with the addition of multi-stage features and without equality or
quantified constraints.

A program \ensuremath{\Varid{pgm}} is a sequence of value definitions \ensuremath{\Varid{def}}, class definitions
\ensuremath{\Varid{cls}}, and instance definitions \ensuremath{\Varid{inst}} followed by a top-level expression \ensuremath{\Varid{e}}.

Top-level definitions \ensuremath{\Varid{def}} are added to the calculus to model separate compilation
and path-based cross-stage persistence in a way similar to what GHC currently
implements: only variables previously defined in a
top-level definition can be referenced at arbitrary levels.

Both type classes and instances are in the language, but simplified as far as
possible: type class definitions \ensuremath{\Varid{cls}} have precisely one method and no superclasses;
instance definitions \ensuremath{\Varid{inst}} are permitted to have an instance context.

The expression language \ensuremath{\Varid{e}} is a standard $\lambda$-calculus
with the addition of the two multi-stage constructs, quotation \ensuremath{\qq{\Varid{e}}} and splicing
\ensuremath{\$(\Varid{e})} which can be used to separate a program into stages.

Just as in Haskell, our language is \emph{predicative}: type variables that appear as
class parameters and in quantifiers range only over monotypes. This is modelled
by the use of a Damas-Milner style \nw{Cite Damas-Milner?}
type system which distinguishes between monotypes \ensuremath{\tau}, qualified types \ensuremath{\rho} and
polytypes \ensuremath{\sigma}. The type argument to the representation constructor \ensuremath{\Conid{Code}\;\tau}
must be a monotype.

A constraint \ensuremath{\Conid{C}} is either a type class constraint \ensuremath{\Conid{TC}\;\tau}, or a representation of a constraint \ensuremath{\Conid{CodeC}\;\Conid{C}}.


The environment \ensuremath{\Gamma } is used for locally introduced information, including
value variables \ensuremath{\mathit{x}\mathbin{:}(\tau,\Varid{n})}, type variables \ensuremath{\Varid{a}}, and local type class axioms
\ensuremath{(\Conid{C},\Varid{n})}. The environment keeps track of the (integer) level \ensuremath{\Varid{n}} that value and
constraint variables are introduced at; the typing rules ensure that the
variables are only used at the current level.

The program theory~\ensuremath{\Conid{P}} is an environment of the type class axioms introduced by
instance declarations and of type information for names introduced by top-level
definitions. The axioms are used to dictate whether the usage of a
type class method is allowed or not. The names indicate which variables can
be used in a cross-stage persistent fashion.

\subsection{Typing Rules}

The typing rules proceed in a predicable way for anyone familiar with
qualified type systems. The difference is that the source expression typing
judgement (Figure~\ref{fig:expr-typing}) and constraint entailment judgement
(Figure~\ref{fig:constraint-entailment}) are now indexed by a level.
Furthermore, since these rules are the basis for elaboration into the core
language, they have
been combined with the elaboration rules, as highlighted by a
\colorbox{gray!25}{grey}
box whose contents
can be safely ignored until the discussion of elaboration~(Section~\ref{sec:elaboration}).

The typing rules also refer to type and constraint formation rules
(Figures~\ref{fig:ty-kinding-elaboration} and~\ref{fig:constraint-formation}),
which check just that type variables are well-scoped (our language does not
include an advanced kind system).






\begin{figure*}
\inlinemathhs
\ruletype{%
\begin{hscode}\SaveRestoreHook
\column{B}{@{}>{\hspre}l<{\hspost}@{}}\ColumnHook
\column{E}{@{}>{\hspre}l<{\hspost}@{}}\ColumnHook
\>[B]{}\Conid{P};\Gamma  \vdashn{\Varid{n}} \Varid{e} : \sigma\ghl{\leadsto\Varid{t}~\vert~\Conid{TSP}}{}\<[E]%
\ColumnHook
\end{hscode}\resethooks
}
\begin{mathpar}
\labelledinferrule{%
\begin{hscode}\SaveRestoreHook
\column{B}{@{}>{\hspre}l<{\hspost}@{}}\ColumnHook
\column{E}{@{}>{\hspre}l<{\hspost}@{}}\ColumnHook
\>[B]{}\mathit{x}\colonT (\tau,\Varid{n}) \in \Gamma {}\<[E]%
\ColumnHook
\end{hscode}\resethooks
}{%
\begin{hscode}\SaveRestoreHook
\column{B}{@{}>{\hspre}l<{\hspost}@{}}\ColumnHook
\column{E}{@{}>{\hspre}l<{\hspost}@{}}\ColumnHook
\>[B]{}\Conid{P};\Gamma  \vdashn{\Varid{n}} \mathit{x} : \tau\ghl{\leadsto\mathit{x}~\vert~\bullet }{}\<[E]%
\ColumnHook
\end{hscode}\resethooks
}%
{\EVar}
\and
\labelledinferrule{%
\begin{hscode}\SaveRestoreHook
\column{B}{@{}>{\hspre}l<{\hspost}@{}}\ColumnHook
\column{E}{@{}>{\hspre}l<{\hspost}@{}}\ColumnHook
\>[B]{}\Varid{k}\colonT \sigma \in \Conid{P}{}\<[E]%
\ColumnHook
\end{hscode}\resethooks
}{%
\begin{hscode}\SaveRestoreHook
\column{B}{@{}>{\hspre}l<{\hspost}@{}}\ColumnHook
\column{E}{@{}>{\hspre}l<{\hspost}@{}}\ColumnHook
\>[B]{}\Conid{P};\Gamma  \vdashn{\Varid{n}} \Varid{k} : \sigma\ghl{\leadsto\Varid{k}~\vert~\bullet }{}\<[E]%
\ColumnHook
\end{hscode}\resethooks
}{\EVarTopLevel}
\and
\labelledinferrule{%
\begin{hscode}\SaveRestoreHook
\column{B}{@{}>{\hspre}l<{\hspost}@{}}\ColumnHook
\column{E}{@{}>{\hspre}l<{\hspost}@{}}\ColumnHook
\>[B]{}\Conid{P};\Gamma ,\mathit{x}\colonT (\tau_{1},\Varid{n}) \vdashn{\Varid{n}} \Varid{e} : \tau_{2}\ghl{\leadsto\Varid{t}~\vert~\Conid{TSP}}{}\<[E]%
\ColumnHook
\end{hscode}\resethooks
\and
\begin{hscode}\SaveRestoreHook
\column{B}{@{}>{\hspre}l<{\hspost}@{}}\ColumnHook
\column{E}{@{}>{\hspre}l<{\hspost}@{}}\ColumnHook
\>[B]{}\Gamma \vdash_{\mathsf{ty}}\tau_{1}\ghl{\leadsto\tau_{1}'}{}\<[E]%
\ColumnHook
\end{hscode}\resethooks
}{%
\begin{hscode}\SaveRestoreHook
\column{B}{@{}>{\hspre}l<{\hspost}@{}}\ColumnHook
\column{E}{@{}>{\hspre}l<{\hspost}@{}}\ColumnHook
\>[B]{}\Conid{P};\Gamma  \vdashn{\Varid{n}} \lambda \mathit{x}\colonT \tau_{1}.\Varid{e} : \tau_{1}\to \tau_{2}\ghl{\leadsto\lambda \mathit{x}\colonT \tau_{1}'.\Varid{t}~\vert~\Conid{TSP}}{}\<[E]%
\ColumnHook
\end{hscode}\resethooks
}{\EAbs}
\and
\labelledinferrule{%
\begin{hscode}\SaveRestoreHook
\column{B}{@{}>{\hspre}l<{\hspost}@{}}\ColumnHook
\column{E}{@{}>{\hspre}l<{\hspost}@{}}\ColumnHook
\>[B]{}\Conid{P};\Gamma  \vdashn{\Varid{n}} \Varid{e}_{1} : \tau_{1}\to \tau_{2}\ghl{\leadsto\Varid{t}_{1}~\vert~\Conid{TSP}_1}{}\<[E]%
\ColumnHook
\end{hscode}\resethooks
\and
\begin{hscode}\SaveRestoreHook
\column{B}{@{}>{\hspre}l<{\hspost}@{}}\ColumnHook
\column{E}{@{}>{\hspre}l<{\hspost}@{}}\ColumnHook
\>[B]{}\Conid{P};\Gamma  \vdashn{\Varid{n}} \Varid{e}_{2} : \tau_{1}\ghl{\leadsto\Varid{t}_{2}~\vert~\Conid{TSP}_2}{}\<[E]%
\ColumnHook
\end{hscode}\resethooks
}{%
\begin{hscode}\SaveRestoreHook
\column{B}{@{}>{\hspre}l<{\hspost}@{}}\ColumnHook
\column{E}{@{}>{\hspre}l<{\hspost}@{}}\ColumnHook
\>[B]{}\Conid{P};\Gamma  \vdashn{\Varid{n}} \Varid{e}_{1}~\Varid{e}_{2} : \tau_{2}\ghl{\leadsto\Varid{t}_{1}~\Varid{t}_{2}~\vert~\Conid{TSP}_1 \cup \Conid{TSP}_2}{}\<[E]%
\ColumnHook
\end{hscode}\resethooks
}{\EApp}
\and
\labelledinferrule{%
\begin{hscode}\SaveRestoreHook
\column{B}{@{}>{\hspre}l<{\hspost}@{}}\ColumnHook
\column{E}{@{}>{\hspre}l<{\hspost}@{}}\ColumnHook
\>[B]{}\Varid{a} \notin \Gamma {}\<[E]%
\ColumnHook
\end{hscode}\resethooks
\and
\begin{hscode}\SaveRestoreHook
\column{B}{@{}>{\hspre}l<{\hspost}@{}}\ColumnHook
\column{E}{@{}>{\hspre}l<{\hspost}@{}}\ColumnHook
\>[B]{}\Conid{P};\Gamma ,\Varid{a} \vdashn{\Varid{n}} \Varid{e} : \sigma\ghl{\leadsto\Varid{t}~\vert~\Conid{TSP}}{}\<[E]%
\ColumnHook
\end{hscode}\resethooks
}{%
\begin{hscode}\SaveRestoreHook
\column{B}{@{}>{\hspre}l<{\hspost}@{}}\ColumnHook
\column{E}{@{}>{\hspre}l<{\hspost}@{}}\ColumnHook
\>[B]{}\Conid{P};\Gamma  \vdashn{\Varid{n}} \Varid{e} : \forall \Varid{a}.\sigma\ghl{\leadsto\Lambda \Varid{a}.\Varid{t}~\vert~\Conid{TSP}}{}\<[E]%
\ColumnHook
\end{hscode}\resethooks
}{\ETAbs}
\and
\labelledinferrule{%
\begin{hscode}\SaveRestoreHook
\column{B}{@{}>{\hspre}l<{\hspost}@{}}\ColumnHook
\column{E}{@{}>{\hspre}l<{\hspost}@{}}\ColumnHook
\>[B]{}\Conid{P};\Gamma  \vdashn{\Varid{n}} \Varid{e} : \forall \Varid{a}.\sigma\ghl{\leadsto\Varid{t}~\vert~\Conid{TSP}}{}\<[E]%
\ColumnHook
\end{hscode}\resethooks
\and
\begin{hscode}\SaveRestoreHook
\column{B}{@{}>{\hspre}l<{\hspost}@{}}\ColumnHook
\column{E}{@{}>{\hspre}l<{\hspost}@{}}\ColumnHook
\>[B]{}\Gamma \vdash_{\mathsf{ty}}\tau\ghl{\leadsto\tau^\prime}{}\<[E]%
\ColumnHook
\end{hscode}\resethooks
}{%
\begin{hscode}\SaveRestoreHook
\column{B}{@{}>{\hspre}l<{\hspost}@{}}\ColumnHook
\column{E}{@{}>{\hspre}l<{\hspost}@{}}\ColumnHook
\>[B]{}\Conid{P};\Gamma  \vdashn{\Varid{n}} \Varid{e} : \sigma[\tau/\Varid{a}]\ghl{\leadsto\Varid{e}~\langle \tau^\prime\rangle ~\vert~\Conid{TSP}}{}\<[E]%
\ColumnHook
\end{hscode}\resethooks
}{\ETApp}
\and
\labelledinferrule{%
\begin{hscode}\SaveRestoreHook
\column{B}{@{}>{\hspre}l<{\hspost}@{}}\ColumnHook
\column{E}{@{}>{\hspre}l<{\hspost}@{}}\ColumnHook
\>[B]{}\Conid{P};\Gamma  \vdashn{\Varid{n}\mathbin{+}\mathrm{1}} \Varid{e} : \tau\ghl{\leadsto\Varid{t}~\vert~\Conid{TSP}}{}\<[E]%
\ColumnHook
\end{hscode}\resethooks
}{%
\begin{hscode}\SaveRestoreHook
\column{B}{@{}>{\hspre}l<{\hspost}@{}}\ColumnHook
\column{54}{@{}>{\hspre}l<{\hspost}@{}}\ColumnHook
\column{E}{@{}>{\hspre}l<{\hspost}@{}}\ColumnHook
\>[B]{}\Conid{P};\Gamma  \vdashn{\Varid{n}} \qq{\Varid{e}} : \Conid{Code}\;\tau\ghl{\leadsto{}\<[54]%
\>[54]{}\qq{\Varid{t}}_{\Conid{TSP}_{\Varid{n}}}~\vert~\lfloor\Conid{TSP}\rfloor^{\Varid{n}}}{}\<[E]%
\ColumnHook
\end{hscode}\resethooks
}{\EQuote}
\and
\labelledinferrule{%
\begin{hscode}\SaveRestoreHook
\column{B}{@{}>{\hspre}l<{\hspost}@{}}\ColumnHook
\column{E}{@{}>{\hspre}l<{\hspost}@{}}\ColumnHook
\>[B]{}\Conid{P};\Gamma  \vdashn{\Varid{n}\mathbin{-}\mathrm{1}} \Varid{e} : \Conid{Code}\;\tau\ghl{\leadsto\Varid{t}~\vert~\Conid{TSP}}{}\<[E]%
\ColumnHook
\end{hscode}\resethooks
\and
\begin{hscode}\SaveRestoreHook
\column{B}{@{}>{\hspre}l<{\hspost}@{}}\ColumnHook
\column{E}{@{}>{\hspre}l<{\hspost}@{}}\ColumnHook
\>[B]{}\ghl{\mathsf{fresh}~\Varid{sp}}{}\<[E]%
\ColumnHook
\end{hscode}\resethooks
\and
\begin{hscode}\SaveRestoreHook
\column{B}{@{}>{\hspre}l<{\hspost}@{}}\ColumnHook
\column{E}{@{}>{\hspre}l<{\hspost}@{}}\ColumnHook
\>[B]{}\Gamma \vdash_{\mathsf{ty}}\tau\ghl{\leadsto\tau^\prime}{}\<[E]%
\ColumnHook
\end{hscode}\resethooks
\and
\begin{hscode}\SaveRestoreHook
\column{B}{@{}>{\hspre}l<{\hspost}@{}}\ColumnHook
\column{E}{@{}>{\hspre}l<{\hspost}@{}}\ColumnHook
\>[B]{}\ghl{\Gamma \leadsto\Delta}{}\<[E]%
\ColumnHook
\end{hscode}\resethooks
}{%
\begin{hscode}\SaveRestoreHook
\column{B}{@{}>{\hspre}l<{\hspost}@{}}\ColumnHook
\column{E}{@{}>{\hspre}l<{\hspost}@{}}\ColumnHook
\>[B]{}\Conid{P};\Gamma  \vdashn{\Varid{n}} \$(\Varid{e}) : \tau\ghl{\leadsto\Varid{sp}~\vert~\{\Delta\vdash\Varid{sp}\colonT\tau^\prime=\Varid{t}\} \cup_{\Varid{n}\mathbin{-}\mathrm{1}}\Conid{TSP}}{}\<[E]%
\ColumnHook
\end{hscode}\resethooks
}{\ESplice}
\and
\labelledinferrule{%
\begin{hscode}\SaveRestoreHook
\column{B}{@{}>{\hspre}l<{\hspost}@{}}\ColumnHook
\column{E}{@{}>{\hspre}l<{\hspost}@{}}\ColumnHook
\>[B]{}\Gamma \vdash_{\mathsf{ct}}\Conid{C}\ghl{\leadsto\tau}{}\<[E]%
\ColumnHook
\end{hscode}\resethooks
\and
\begin{hscode}\SaveRestoreHook
\column{B}{@{}>{\hspre}l<{\hspost}@{}}\ColumnHook
\column{E}{@{}>{\hspre}l<{\hspost}@{}}\ColumnHook
\>[B]{}\Conid{P};\Gamma , \ghl{\Varid{ev}\colonT\!\!}(\Conid{C},\Varid{n}) \vdashn{\Varid{n}} \Varid{e} : \rho\ghl{\leadsto\Varid{t}~\vert~\Conid{TSP}}{}\<[E]%
\ColumnHook
\end{hscode}\resethooks
\and
\begin{hscode}\SaveRestoreHook
\column{B}{@{}>{\hspre}l<{\hspost}@{}}\ColumnHook
\column{E}{@{}>{\hspre}l<{\hspost}@{}}\ColumnHook
\>[B]{}\ghl{\mathsf{fresh}~\Varid{ev}}{}\<[E]%
\ColumnHook
\end{hscode}\resethooks
}{%
\begin{hscode}\SaveRestoreHook
\column{B}{@{}>{\hspre}l<{\hspost}@{}}\ColumnHook
\column{E}{@{}>{\hspre}l<{\hspost}@{}}\ColumnHook
\>[B]{}\Conid{P};\Gamma  \vdashn{\Varid{n}} \Varid{e} : \Conid{C}\Rightarrow \rho\ghl{\leadsto\lambda \Varid{ev}\colonT \tau.\Varid{t}~\vert~\Conid{TSP}}{}\<[E]%
\ColumnHook
\end{hscode}\resethooks
}{\ECAbs}
\and
\labelledinferrule{%
\begin{hscode}\SaveRestoreHook
\column{B}{@{}>{\hspre}l<{\hspost}@{}}\ColumnHook
\column{E}{@{}>{\hspre}l<{\hspost}@{}}\ColumnHook
\>[B]{}\Conid{P};\Gamma  \vdashn{\Varid{n}} \Varid{e} : \Conid{C}\Rightarrow \rho\ghl{\leadsto\Varid{t}_{1}~\vert~\Conid{TSP}_1}{}\<[E]%
\ColumnHook
\end{hscode}\resethooks
\and
\begin{hscode}\SaveRestoreHook
\column{B}{@{}>{\hspre}l<{\hspost}@{}}\ColumnHook
\column{E}{@{}>{\hspre}l<{\hspost}@{}}\ColumnHook
\>[B]{}\Conid{P};\Gamma \entails{\Varid{n}}\Conid{C}\ghl{\leadsto\Varid{t}_{2}~\vert~\Conid{TSP}_2}{}\<[E]%
\ColumnHook
\end{hscode}\resethooks
}{%
\begin{hscode}\SaveRestoreHook
\column{B}{@{}>{\hspre}l<{\hspost}@{}}\ColumnHook
\column{E}{@{}>{\hspre}l<{\hspost}@{}}\ColumnHook
\>[B]{}\Conid{P};\Gamma  \vdashn{\Varid{n}} \Varid{e} : \rho\ghl{\leadsto\Varid{t}_{1}~\Varid{t}_{2}~\vert~\Conid{TSP}_1 \cup \Conid{TSP}_2}{}\<[E]%
\ColumnHook
\end{hscode}\resethooks
}{\ECApp}
\and

\end{mathpar}
\caption{Source Expression Typing \ghl{\text{with Elaboration Semantics}}}
\label{fig:expr-elaboration}
\label{fig:expr-typing}
\end{figure*}

\subsubsection{Levels}

A large part of the rules are indexed by their (integer) level.
It is standard to index the expression judgement at a specific level but less
standard to index
the constraint entailment judgement. The purpose of
the level index is to ensure that expression variables and constraints
can only be used at the level they are introduced. The result is a program which
is separated into stages in such a way that when imbued with operational semantics,
the fragments at stage~\ensuremath{\Varid{n}} will be evaluated before the fragments at stage~\ensuremath{\Varid{n}\mathbin{+}\mathrm{1}}.

The full program is checked at level \ensuremath{\mathrm{0}}.
Quotation ($\EQuote$) increases the level by one and splicing ($\ESplice$)
decreases the level.
It is permitted to reach negative levels in this source language.
The negative levels indicate a stage which should evaluate at compile-time.

In more detail, the consequence of the level-indexing is that:
\begin{itemize}
\item A variable can be introduced at any specific level \ensuremath{\Varid{n}} by a
  $\lambda$-abstraction (\EAbs) and used at precisely that level (\EVar).

\item Variables introduced by top-level definitions
  can be used at any level (\EVarTopLevel).

\item Type variables can be introduced (\ETAbs) and used at any level (\ETApp).

\item Constraints can be introduced at any level (\ECAbs), but in practice, due
  to the system being predicative, this level will always be \ensuremath{\mathrm{0}} (see
  Section~\ref{sec:level-of-constraints}).
  Constraints are also introduced into the program logic by instance declarations.
  Constraints are appropriately eliminated by application~(\ECApp).
  The entailment relation~(Figure~\ref{fig:constraint-entailment}) ensures
  that a constraint is available at the current~stage.
\end{itemize}

Note that we do not allow implicit lifting as discussed in Section~\ref{sec:background}
and implemented in GHC. Explicit lifting via a \ensuremath{\Conid{Lift}} class can easily be expressed in
this language, and implicit lifting, if desired, can be orthogonally added as an additional
elaboration step.

The rules furthermore implement the top-level splice restriction present in GHC.
A top-level splice will require its body to be at level \ensuremath{\mathbin{-}\mathrm{1}} and thereby rule out
the use of any variables defined outside of the splice, with the exception of
top-level variables.

\subsubsection{Level of constraints}
\label{sec:level-of-constraints}

Inspecting the \ECAbs rule independently it would appear that a local
constraint can be introduced at any level and then be used at that level.
In fact, the \ECAbs rule can only ever be applied at level~\ensuremath{\mathrm{0}} because the \EQuote
rule requires that the enclosed expression has a \ensuremath{\tau}~type. This forbids the
\ensuremath{\rho}~type as produced by \ECAbs and more closely models the restriction
to predicative types as present in GHC.

The new constraint form \ensuremath{\Conid{CodeC}\;\Conid{C}} can be used to allow local constraints
to be used at positive levels in \ECApp. The rules \CIncr and \CDecr in constraint
entailment can convert from \ensuremath{\Conid{CodeC}\;\Conid{C}} and \ensuremath{\Conid{C}} and vice versa. During elaboration,
these rules will insert splices and quotes around the dictionary arguments as needed.

Constraints satisfied by instance declarations can be used at any level by means of~\CGlobal.

\begingroup
\def\ghl#1{}%
Let us illustrate this by revisiting Example~\ref{ex:c1} and considering the
expression \ensuremath{\qq{\Varid{show}}}. In this case, we assume the program environment to contain
the type of \ensuremath{\Varid{show}}:
\begin{hscode}\SaveRestoreHook
\column{B}{@{}>{\hspre}l<{\hspost}@{}}\ColumnHook
\column{E}{@{}>{\hspre}l<{\hspost}@{}}\ColumnHook
\>[B]{}\Conid{P}\mathrel{=}\bullet ,\Varid{show}\colonT\forall \Varid{a}.\Conid{Show}~\Varid{a}\Rightarrow \Varid{a}\to \Conid{String}{}\<[E]%
\ColumnHook
\end{hscode}\resethooks
From this we can use \EVarTopLevel to conclude that
\begin{hscode}\SaveRestoreHook
\column{B}{@{}>{\hspre}l<{\hspost}@{}}\ColumnHook
\column{E}{@{}>{\hspre}l<{\hspost}@{}}\ColumnHook
\>[B]{}\Conid{P};\Gamma  \vdashn{\mathrm{1}} \Varid{show} : \forall \Varid{a}.\Conid{Show}~\Varid{a}\Rightarrow \Varid{a}\to \Conid{String}{}\<[E]%
\ColumnHook
\end{hscode}\resethooks
Because \ensuremath{\qq{\Varid{show}}} must have type~\ensuremath{\Conid{Code}\;\tau} for some monotype~\ensuremath{\tau}, we cannot apply
\EQuote yet, but must first apply \ETApp and \ECApp. We can conclude
\begin{hscode}\SaveRestoreHook
\column{B}{@{}>{\hspre}l<{\hspost}@{}}\ColumnHook
\column{E}{@{}>{\hspre}l<{\hspost}@{}}\ColumnHook
\>[B]{}\Conid{P};\Gamma  \vdashn{\mathrm{1}} \Varid{show} : \Varid{a}\to \Conid{String}{}\<[E]%
\ColumnHook
\end{hscode}\resethooks
if the entailment
\begin{hscode}\SaveRestoreHook
\column{B}{@{}>{\hspre}l<{\hspost}@{}}\ColumnHook
\column{E}{@{}>{\hspre}l<{\hspost}@{}}\ColumnHook
\>[B]{}\Conid{P};\Gamma \entails{\mathrm{1}}\Conid{Show}~\Varid{a}{}\<[E]%
\ColumnHook
\end{hscode}\resethooks
holds. As we do not assume any instances for \ensuremath{\Conid{Show}} in this example, the constraint
can only be justified via the local environment \ensuremath{\Gamma }. Because, as discussed above,
any constraint can only have been introduced at level~\ensuremath{\mathrm{0}}, the only way to move a local
constraint from level~\ensuremath{\mathrm{0}} to level~\ensuremath{\mathrm{1}} is rule \CIncr, which requires us to show
\begin{hscode}\SaveRestoreHook
\column{B}{@{}>{\hspre}l<{\hspost}@{}}\ColumnHook
\column{E}{@{}>{\hspre}l<{\hspost}@{}}\ColumnHook
\>[B]{}\Conid{P};\Gamma \entails{\mathrm{0}}\Conid{CodeC}\;(\Conid{Show}~\Varid{a}){}\<[E]%
\ColumnHook
\end{hscode}\resethooks
This in turn follows from \CLocal if we assume that \ensuremath{\Gamma } contains \ensuremath{\Conid{CodeC}\;(\Conid{Show}\;\Varid{a})}:
\begin{hscode}\SaveRestoreHook
\column{B}{@{}>{\hspre}l<{\hspost}@{}}\ColumnHook
\column{E}{@{}>{\hspre}l<{\hspost}@{}}\ColumnHook
\>[B]{}\Gamma \mathrel{=}\bullet , \ghl{\mathit{x}\colonT\!\!}(\Conid{CodeC}\;(\Conid{Show}~\Varid{a}),\mathrm{0}){}\<[E]%
\ColumnHook
\end{hscode}\resethooks
At this point, we know that
\begin{hscode}\SaveRestoreHook
\column{B}{@{}>{\hspre}l<{\hspost}@{}}\ColumnHook
\column{E}{@{}>{\hspre}l<{\hspost}@{}}\ColumnHook
\>[B]{}\Conid{P};\Gamma  \vdashn{\mathrm{0}} \qq{\Varid{show}} : \Conid{Code}\;(\Varid{a}\to \Conid{String}){}\<[E]%
\ColumnHook
\end{hscode}\resethooks
and by applying \ECAbs and \ETAbs, we obtain
\begin{hscode}\SaveRestoreHook
\column{B}{@{}>{\hspre}l<{\hspost}@{}}\ColumnHook
\column{E}{@{}>{\hspre}l<{\hspost}@{}}\ColumnHook
\>[B]{}\Conid{P};\bullet  \vdashn{\mathrm{0}} \qq{\Varid{show}} : \forall \Varid{a}.\Conid{CodeC}\;(\Conid{Show}~\Varid{a})\Rightarrow \Conid{Code}\;(\Varid{a}\to \Conid{String}){}\<[E]%
\ColumnHook
\end{hscode}\resethooks
as desired.
\endgroup






\begin{figure*}
\inlinemathhs
\ruletype{%
\begin{hscode}\SaveRestoreHook
\column{B}{@{}>{\hspre}l<{\hspost}@{}}\ColumnHook
\column{E}{@{}>{\hspre}l<{\hspost}@{}}\ColumnHook
\>[B]{}\Conid{P};\Gamma \entails{\Varid{n}}\Conid{C}\ghl{\leadsto\Varid{t}~\vert~\Conid{TSP}}{}\<[E]%
\ColumnHook
\end{hscode}\resethooks
}%
\begin{mathpar}
\labelledinferrule{%
\begin{hscode}\SaveRestoreHook
\column{B}{@{}>{\hspre}l<{\hspost}@{}}\ColumnHook
\column{E}{@{}>{\hspre}l<{\hspost}@{}}\ColumnHook
\>[B]{}\ghl{\Varid{ev}\colonT\!\!}(\forall \Varid{a}.\overline{\Conid{C}_{\Varid{i}}}\Rightarrow \Conid{C}) \in \Conid{P}{}\<[E]%
\ColumnHook
\end{hscode}\resethooks
\and
\begin{hscode}\SaveRestoreHook
\column{B}{@{}>{\hspre}l<{\hspost}@{}}\ColumnHook
\column{E}{@{}>{\hspre}l<{\hspost}@{}}\ColumnHook
\>[B]{}\Gamma \vdash_{\mathsf{ty}}\tau\ghl{\leadsto\tau^\prime}{}\<[E]%
\ColumnHook
\end{hscode}\resethooks
\and
\begin{hscode}\SaveRestoreHook
\column{B}{@{}>{\hspre}l<{\hspost}@{}}\ColumnHook
\column{E}{@{}>{\hspre}l<{\hspost}@{}}\ColumnHook
\>[B]{}\overline{\Conid{P};\Gamma \entails{\Varid{n}}\Conid{C}_{\Varid{i}}[\tau/\Varid{a}]\ghl{\leadsto\Varid{t}_{\Varid{i}}~\vert~\Conid{TSP}_i}}{}\<[E]%
\ColumnHook
\end{hscode}\resethooks
}{%
\begin{hscode}\SaveRestoreHook
\column{B}{@{}>{\hspre}l<{\hspost}@{}}\ColumnHook
\column{94}{@{}>{\hspre}l<{\hspost}@{}}\ColumnHook
\column{E}{@{}>{\hspre}l<{\hspost}@{}}\ColumnHook
\>[B]{}\Conid{P};\Gamma \entails{\Varid{n}}\Conid{C}[\tau/\Varid{a}]\ghl{\leadsto\Varid{ev}~\langle \tau^\prime\rangle ~\overline{\Varid{t}_{\Varid{i}}}~\vert~{}\<[94]%
\>[94]{}\bigcup_i\Conid{TSP}_i}{}\<[E]%
\ColumnHook
\end{hscode}\resethooks
}{\CGlobal}
\and
\labelledinferrule{%
\begin{hscode}\SaveRestoreHook
\column{B}{@{}>{\hspre}l<{\hspost}@{}}\ColumnHook
\column{E}{@{}>{\hspre}l<{\hspost}@{}}\ColumnHook
\>[B]{}\ghl{\Varid{ev}\colonT\!\!} (\Conid{C},\Varid{n}) \in \Gamma {}\<[E]%
\ColumnHook
\end{hscode}\resethooks
}{%
\begin{hscode}\SaveRestoreHook
\column{B}{@{}>{\hspre}l<{\hspost}@{}}\ColumnHook
\column{E}{@{}>{\hspre}l<{\hspost}@{}}\ColumnHook
\>[B]{}\Conid{P};\Gamma \entails{\Varid{n}}\Conid{C}\ghl{\leadsto\Varid{ev}~\vert~\bullet }{}\<[E]%
\ColumnHook
\end{hscode}\resethooks
}{\CLocal}
\and
\labelledinferrule{%
\begin{hscode}\SaveRestoreHook
\column{B}{@{}>{\hspre}l<{\hspost}@{}}\ColumnHook
\column{E}{@{}>{\hspre}l<{\hspost}@{}}\ColumnHook
\>[B]{}\Conid{P};\Gamma \entails{\Varid{n}\mathbin{+}\mathrm{1}}\Conid{C}\ghl{\leadsto\Varid{t}~\vert~\Conid{TSP}}{}\<[E]%
\ColumnHook
\end{hscode}\resethooks
}{%
\begin{hscode}\SaveRestoreHook
\column{B}{@{}>{\hspre}l<{\hspost}@{}}\ColumnHook
\column{E}{@{}>{\hspre}l<{\hspost}@{}}\ColumnHook
\>[B]{}\Conid{P};\Gamma \entails{\Varid{n}}\Conid{CodeC}\;\Conid{C}\ghl{\leadsto\qq{\Varid{t}}_{\Conid{TSP}_{\Varid{n}}}~\vert~\lfloor\Conid{TSP}\rfloor^{\Varid{n}}}{}\<[E]%
\ColumnHook
\end{hscode}\resethooks
}{\CDecr}
\and
\labelledinferrule{%
\begin{hscode}\SaveRestoreHook
\column{B}{@{}>{\hspre}l<{\hspost}@{}}\ColumnHook
\column{E}{@{}>{\hspre}l<{\hspost}@{}}\ColumnHook
\>[B]{}\Conid{P};\Gamma \entails{\Varid{n}\mathbin{-}\mathrm{1}}\Conid{CodeC}\;\Conid{C}\ghl{\leadsto\Varid{t}~\vert~\Conid{TSP}}{}\<[E]%
\ColumnHook
\end{hscode}\resethooks
 \and
\begin{hscode}\SaveRestoreHook
\column{B}{@{}>{\hspre}l<{\hspost}@{}}\ColumnHook
\column{E}{@{}>{\hspre}l<{\hspost}@{}}\ColumnHook
\>[B]{}\Gamma \vdash_{\mathsf{ct}}\Conid{C}\ghl{\leadsto\tau}{}\<[E]%
\ColumnHook
\end{hscode}\resethooks
\and
\begin{hscode}\SaveRestoreHook
\column{B}{@{}>{\hspre}l<{\hspost}@{}}\ColumnHook
\column{E}{@{}>{\hspre}l<{\hspost}@{}}\ColumnHook
\>[B]{}\ghl{\mathsf{fresh}~\Varid{sp}}{}\<[E]%
\ColumnHook
\end{hscode}\resethooks
\and
\begin{hscode}\SaveRestoreHook
\column{B}{@{}>{\hspre}l<{\hspost}@{}}\ColumnHook
\column{E}{@{}>{\hspre}l<{\hspost}@{}}\ColumnHook
\>[B]{}\ghl{\Gamma \leadsto\Delta}{}\<[E]%
\ColumnHook
\end{hscode}\resethooks
}{%
\begin{hscode}\SaveRestoreHook
\column{B}{@{}>{\hspre}l<{\hspost}@{}}\ColumnHook
\column{E}{@{}>{\hspre}l<{\hspost}@{}}\ColumnHook
\>[B]{}\Conid{P};\Gamma \entails{\Varid{n}}\Conid{C}\ghl{\leadsto\Varid{sp}~\vert~\{\Delta\vdash\Varid{sp}\colonT\tau=\Varid{t}\} \cup_{\Varid{n}\mathbin{-}\mathrm{1}}\Conid{TSP}}{}\<[E]%
\ColumnHook
\end{hscode}\resethooks
}{\CIncr}
\end{mathpar}

%
\caption{Source Constraints \ghl{\text{with Elaboration}}}
\label{fig:constraint-elaboration}
\label{fig:constraint-entailment}
\end{figure*}

\subsubsection{Program Typing}

A program (Figure~\ref{fig:program-typing}) is a sequence of value, class and instances declarations followed by
an expression. The declaration forms extend the program theory which is used
to typecheck subsequent definitions. Value definitions extend the list of
top-level definitions available at all stages. The judgement makes it clear that
the top-level of the program is level~\ensuremath{\mathrm{0}} and that each expression is checked in
an empty local environment.

Class definitions extend the program theory with the qualified class method.
Instance definitions extend the environment with an axiom for the specific instance
which is being defined. The rule checks that the class method is of the type specified in the
class definition.

Notice that before the type class method is checked, the local environment~\ensuremath{\Gamma } is
extended with an axiom for the instance we are currently checking. This is
important to allow recursive definitions~(Example~\ref{ex:i1}) to be defined in a natural fashion.
The constraint is introduced at level~\ensuremath{\mathrm{0}}, the top-level of the program. It is important
to introduce the constraint to the local environment rather than program theory
because this constraint should not be available at negative levels.
If the constraint was introduced to the program theory, then it could be used
at negative levels, which would amount to attempting to use the instance
in order to affect the definition of said instance. This nuance in the typing
rules explains why Example~\ref{ex:i2} is accepted and the necessity of the \ensuremath{\Conid{CodeC}}
constraint in Example~\ref{ex:i3}.






\begin{figure*}
\inlinemathhs
\ruletype{%
\begin{hscode}\SaveRestoreHook
\column{B}{@{}>{\hspre}l<{\hspost}@{}}\ColumnHook
\column{E}{@{}>{\hspre}l<{\hspost}@{}}\ColumnHook
\>[B]{}\ghl{\Varid{pgm}~\vert~}\Conid{P}\vdash_{\mathsf{def}}\Varid{def} : \Conid{P}\ghl{\leadsto\Varid{pgm}}{}\<[E]%
\ColumnHook
\end{hscode}\resethooks
}
\begin{mathpar}
\labelledinferrule{%
\begin{hscode}\SaveRestoreHook
\column{B}{@{}>{\hspre}l<{\hspost}@{}}\ColumnHook
\column{E}{@{}>{\hspre}l<{\hspost}@{}}\ColumnHook
\>[B]{}\Conid{P};\bullet  \vdashn{\mathrm{0}} \Varid{e} : \sigma\ghl{\leadsto\Varid{t}~\vert~\Conid{TSP}}{}\<[E]%
\ColumnHook
\end{hscode}\resethooks
\and
\begin{hscode}\SaveRestoreHook
\column{B}{@{}>{\hspre}l<{\hspost}@{}}\ColumnHook
\column{E}{@{}>{\hspre}l<{\hspost}@{}}\ColumnHook
\>[B]{}\bullet \vdash_{\mathsf{ty}}\sigma\ghl{\leadsto\tau}{}\<[E]%
\ColumnHook
\end{hscode}\resethooks
\and
\ghl{
\begin{hscode}\SaveRestoreHook
\column{B}{@{}>{\hspre}l<{\hspost}@{}}\ColumnHook
\column{E}{@{}>{\hspre}l<{\hspost}@{}}\ColumnHook
\>[B]{}\Varid{d}\mathrel{=}\mathbf{def}~\Varid{k} \colonT \tau\mathrel{=}\Varid{t}{}\<[E]%
\ColumnHook
\end{hscode}\resethooks
}
  \and
\ghl{
\begin{hscode}\SaveRestoreHook
\column{B}{@{}>{\hspre}l<{\hspost}@{}}\ColumnHook
\column{E}{@{}>{\hspre}l<{\hspost}@{}}\ColumnHook
\>[B]{}\Varid{ds}\mathrel{=}\Varid{collapse}(\mathbin{-}\mathrm{1},\Conid{TSP},\Varid{d};\Varid{pgm}){}\<[E]%
\ColumnHook
\end{hscode}\resethooks
}
}{%
\begin{hscode}\SaveRestoreHook
\column{B}{@{}>{\hspre}l<{\hspost}@{}}\ColumnHook
\column{E}{@{}>{\hspre}l<{\hspost}@{}}\ColumnHook
\>[B]{}\ghl{\Varid{pgm}~\vert~}\Conid{P}\vdash_{\mathsf{def}}\mathbf{def}~\Varid{k}\mathrel{=}\Varid{e} : \Conid{P},\Varid{k}\colonT\sigma\ghl{\leadsto\Varid{ds}}{}\<[E]%
\ColumnHook
\end{hscode}\resethooks
}{\Def}
\end{mathpar}
\caption{Source Definition Typing \ghl{\text{with Elaboration}}}
\label{fig:definition-typing}
\end{figure*}






\begin{figure*}
\inlinemathhs
\ruletype{%
\begin{hscode}\SaveRestoreHook
\column{B}{@{}>{\hspre}l<{\hspost}@{}}\ColumnHook
\column{E}{@{}>{\hspre}l<{\hspost}@{}}\ColumnHook
\>[B]{}\ghl{\Varid{pgm}~\vert~}\Conid{P}\vdash_{\mathsf{cls}}\Varid{cls} : \Conid{P}\ghl{\leadsto\Varid{pgm}}{}\<[E]%
\ColumnHook
\end{hscode}\resethooks
}
\begin{mathpar}
\labelledinferrule{%
\begin{hscode}\SaveRestoreHook
\column{B}{@{}>{\hspre}l<{\hspost}@{}}\ColumnHook
\column{E}{@{}>{\hspre}l<{\hspost}@{}}\ColumnHook
\>[B]{}\bullet ,\Varid{a}\vdash_{\mathsf{ty}}\sigma{}\<[E]%
\ColumnHook
\end{hscode}\resethooks
}{%
\begin{hscode}\SaveRestoreHook
\column{B}{@{}>{\hspre}l<{\hspost}@{}}\ColumnHook
\column{E}{@{}>{\hspre}l<{\hspost}@{}}\ColumnHook
\>[B]{}\ghl{\Varid{pgm}~\vert~}\Conid{P}\vdash_{\mathsf{cls}}\;\mathbf{class}\;~\Conid{TC}~\Varid{a}\;\mathbf{where}\;\{\mskip1.5mu \Varid{k}\mathbin{::}\sigma\mskip1.5mu\} : \Conid{P},\Varid{k}\colonT\forall \Varid{a}.\Conid{TC}~\Varid{a}\Rightarrow \sigma\ghl{\leadsto\Varid{pgm}}{}\<[E]%
\ColumnHook
\end{hscode}\resethooks
}{\Cls}
\end{mathpar}
\caption{Source Class Typing \ghl{\text{with Elaboration}}}
\label{fig:class-typing}
\end{figure*}






\begin{figure*}
\inlinemathhs
\ruletype{%
\begin{hscode}\SaveRestoreHook
\column{B}{@{}>{\hspre}l<{\hspost}@{}}\ColumnHook
\column{E}{@{}>{\hspre}l<{\hspost}@{}}\ColumnHook
\>[B]{}\ghl{\Varid{pgm}~\vert~}\Conid{P}\vdash_{\mathsf{inst}}\Varid{inst} : \Conid{P}\ghl{\leadsto\Varid{pgm}}{}\<[E]%
\ColumnHook
\end{hscode}\resethooks
}
\begin{mathpar}
\labelledinferrule{%
\begin{hscode}\SaveRestoreHook
\column{B}{@{}>{\hspre}l<{\hspost}@{}}\ColumnHook
\column{E}{@{}>{\hspre}l<{\hspost}@{}}\ColumnHook
\>[B]{}\mathbf{class}\;~\Conid{TC}~\Varid{a}\;\mathbf{where}\;\{\mskip1.5mu \Varid{k}\mathbin{::}\sigma\mskip1.5mu\}{}\<[E]%
\ColumnHook
\end{hscode}\resethooks
\and
\begin{hscode}\SaveRestoreHook
\column{B}{@{}>{\hspre}l<{\hspost}@{}}\ColumnHook
\column{E}{@{}>{\hspre}l<{\hspost}@{}}\ColumnHook
\>[B]{}\overline{\Varid{b\char95 j}}\mathrel{=}\mathsf{fv}(\tau){}\<[E]%
\ColumnHook
\end{hscode}\resethooks
\and
\begin{hscode}\SaveRestoreHook
\column{B}{@{}>{\hspre}l<{\hspost}@{}}\ColumnHook
\column{E}{@{}>{\hspre}l<{\hspost}@{}}\ColumnHook
\>[B]{}\bullet \vdash_{\mathsf{ty}}\sigma[\tau/\Varid{a}]\ghl{\leadsto\tau^\prime}{}\<[E]%
\ColumnHook
\end{hscode}\resethooks
\and
\begin{hscode}\SaveRestoreHook
\column{B}{@{}>{\hspre}l<{\hspost}@{}}\ColumnHook
\column{14}{@{}>{\hspre}l<{\hspost}@{}}\ColumnHook
\column{E}{@{}>{\hspre}l<{\hspost}@{}}\ColumnHook
\>[B]{}\overline{{}\<[14]%
\>[14]{}\bullet ,\overline{\Varid{b\char95 j}}\vdash_{\mathsf{ct}}\Conid{C}_{\Varid{i}}\ghl{\leadsto\tau_{\Varid{i}}}}{}\<[E]%
\ColumnHook
\end{hscode}\resethooks
\\\and
\begin{hscode}\SaveRestoreHook
\column{B}{@{}>{\hspre}l<{\hspost}@{}}\ColumnHook
\column{E}{@{}>{\hspre}l<{\hspost}@{}}\ColumnHook
\>[B]{}\Conid{P};\bullet ,\overline{\Varid{b\char95 j}}, \ghl{\Varid{ev}\colonT\!\!}(\Conid{TC}~\tau,\mathrm{0}),\overline{(\Conid{C}_{\Varid{i}},\mathrm{0})} \vdashn{\mathrm{0}} \Varid{e} : \sigma[\tau/\Varid{a}]\ghl{\leadsto\Varid{t}~\vert~\Conid{TSP}}{}\<[E]%
\ColumnHook
\end{hscode}\resethooks
\and \\
\ghl{
\begin{hscode}\SaveRestoreHook
\column{B}{@{}>{\hspre}l<{\hspost}@{}}\ColumnHook
\column{68}{@{}>{\hspre}l<{\hspost}@{}}\ColumnHook
\column{E}{@{}>{\hspre}l<{\hspost}@{}}\ColumnHook
\>[B]{}\Varid{dI}\mathrel{=}\mathbf{def}~\Varid{ev} \colonT (\forall \overline{\Varid{b\char95 j}}.\overline{\tau_{\Varid{i}}}\to \tau^\prime)\mathrel{=}{}\<[68]%
\>[68]{}\Varid{t}{}\<[E]%
\ColumnHook
\end{hscode}\resethooks
}
\and
\ghl{
\begin{hscode}\SaveRestoreHook
\column{B}{@{}>{\hspre}l<{\hspost}@{}}\ColumnHook
\column{E}{@{}>{\hspre}l<{\hspost}@{}}\ColumnHook
\>[B]{}\Varid{ds}\mathrel{=}\Varid{collapse}(\mathbin{-}\mathrm{1},\Conid{TSP},\Varid{dI};\Varid{pgm}){}\<[E]%
\ColumnHook
\end{hscode}\resethooks
}
\and
\begin{hscode}\SaveRestoreHook
\column{B}{@{}>{\hspre}l<{\hspost}@{}}\ColumnHook
\column{E}{@{}>{\hspre}l<{\hspost}@{}}\ColumnHook
\>[B]{}\ghl{\mathsf{fresh}~\Varid{ev}}{}\<[E]%
\ColumnHook
\end{hscode}\resethooks
}{%
\begin{hscode}\SaveRestoreHook
\column{B}{@{}>{\hspre}l<{\hspost}@{}}\ColumnHook
\column{E}{@{}>{\hspre}l<{\hspost}@{}}\ColumnHook
\>[B]{}\ghl{\Varid{pgm}~\vert~}\Conid{P}\vdash_{\mathsf{inst}}\;\mathbf{instance}\;~\overline{\Conid{C}_{\Varid{i}}}\Rightarrow \Conid{TC}~\tau\;\mathbf{where}\;\{\mskip1.5mu \Varid{k}\mathrel{=}\Varid{e}\mskip1.5mu\} : \Conid{P}, \ghl{\Varid{ev}\colonT\!\!}(\forall \overline{\Varid{b\char95 j}}.\overline{\Conid{C}_{\Varid{i}}}\Rightarrow \Conid{TC}~\tau)\ghl{\leadsto\Varid{ds}}{}\<[E]%
\ColumnHook
\end{hscode}\resethooks
}{\Inst}
\end{mathpar}
\caption{Source Instance Typing \ghl{\text{with Elaboration}}}
\label{fig:instance-typing}
\end{figure*}






\begin{figure*}
\inlinemathhs
\ruletype{%
\begin{hscode}\SaveRestoreHook
\column{B}{@{}>{\hspre}l<{\hspost}@{}}\ColumnHook
\column{E}{@{}>{\hspre}l<{\hspost}@{}}\ColumnHook
\>[B]{}\Conid{P}\vdash_{\mathsf{pgm}}\Varid{pgm} : \sigma\ghl{\leadsto\Varid{pgm}}{}\<[E]%
\ColumnHook
\end{hscode}\resethooks
}
\begin{mathpar}
\labelledinferrule{%
\begin{hscode}\SaveRestoreHook
\column{B}{@{}>{\hspre}l<{\hspost}@{}}\ColumnHook
\column{E}{@{}>{\hspre}l<{\hspost}@{}}\ColumnHook
\>[B]{}\Conid{P};\bullet  \vdashn{\mathrm{0}} \Varid{e} : \sigma\ghl{\leadsto\Varid{t}~\vert~\Conid{TSP}}{}\<[E]%
\ColumnHook
\end{hscode}\resethooks
\and
\begin{hscode}\SaveRestoreHook
\column{B}{@{}>{\hspre}l<{\hspost}@{}}\ColumnHook
\column{E}{@{}>{\hspre}l<{\hspost}@{}}\ColumnHook
\>[B]{}\bullet \vdash_{\mathsf{ty}}\sigma\ghl{\leadsto\tau}{}\<[E]%
\ColumnHook
\end{hscode}\resethooks
\and
\ghl{
\begin{hscode}\SaveRestoreHook
\column{B}{@{}>{\hspre}l<{\hspost}@{}}\ColumnHook
\column{E}{@{}>{\hspre}l<{\hspost}@{}}\ColumnHook
\>[B]{}\Varid{p}\mathrel{=}\Varid{t}\colonT \tau{}\<[E]%
\ColumnHook
\end{hscode}\resethooks
}
\and
\ghl{
\begin{hscode}\SaveRestoreHook
\column{B}{@{}>{\hspre}l<{\hspost}@{}}\ColumnHook
\column{E}{@{}>{\hspre}l<{\hspost}@{}}\ColumnHook
\>[B]{}\Varid{ds}\mathrel{=}\Varid{collapse}(\mathbin{-}\mathrm{1},\Conid{TSP},\Varid{p}){}\<[E]%
\ColumnHook
\end{hscode}\resethooks
}
}{%
\begin{hscode}\SaveRestoreHook
\column{B}{@{}>{\hspre}l<{\hspost}@{}}\ColumnHook
\column{E}{@{}>{\hspre}l<{\hspost}@{}}\ColumnHook
\>[B]{}\Conid{P}\vdash_{\mathsf{pgm}}\Varid{e} : \sigma\ghl{\leadsto\Varid{ds}}{}\<[E]%
\ColumnHook
\end{hscode}\resethooks
}{\PExpr}
\and
\labelledinferrule{%
\begin{hscode}\SaveRestoreHook
\column{B}{@{}>{\hspre}l<{\hspost}@{}}\ColumnHook
\column{E}{@{}>{\hspre}l<{\hspost}@{}}\ColumnHook
\>[B]{}\Conid{P}_{2}\vdash_{\mathsf{pgm}}\Varid{pgm} : \sigma\ghl{\leadsto\Varid{pgm}}{}\<[E]%
\ColumnHook
\end{hscode}\resethooks
\and
\begin{hscode}\SaveRestoreHook
\column{B}{@{}>{\hspre}l<{\hspost}@{}}\ColumnHook
\column{E}{@{}>{\hspre}l<{\hspost}@{}}\ColumnHook
\>[B]{}\ghl{\Varid{pgm}~\vert~}\Conid{P}_{1}\vdash_{\mathsf{def}}\Varid{def} : \Conid{P}_{2}\ghl{\leadsto\Varid{pgm}_{\mathrm{1}}}{}\<[E]%
\ColumnHook
\end{hscode}\resethooks
}{%
\begin{hscode}\SaveRestoreHook
\column{B}{@{}>{\hspre}l<{\hspost}@{}}\ColumnHook
\column{E}{@{}>{\hspre}l<{\hspost}@{}}\ColumnHook
\>[B]{}\Conid{P}_{1}\vdash_{\mathsf{pgm}}\Varid{def};\Varid{pgm} : \sigma\ghl{\leadsto\Varid{pgm}_{\mathrm{1}}}{}\<[E]%
\ColumnHook
\end{hscode}\resethooks
}{\PDef}
\and
\labelledinferrule{%
\begin{hscode}\SaveRestoreHook
\column{B}{@{}>{\hspre}l<{\hspost}@{}}\ColumnHook
\column{E}{@{}>{\hspre}l<{\hspost}@{}}\ColumnHook
\>[B]{}\Conid{P}_{2}\vdash_{\mathsf{pgm}}\Varid{pgm} : \sigma\ghl{\leadsto\Varid{pgm}}{}\<[E]%
\ColumnHook
\end{hscode}\resethooks
\and
\begin{hscode}\SaveRestoreHook
\column{B}{@{}>{\hspre}l<{\hspost}@{}}\ColumnHook
\column{E}{@{}>{\hspre}l<{\hspost}@{}}\ColumnHook
\>[B]{}\ghl{\Varid{pgm}~\vert~}\Conid{P}_{1}\vdash_{\mathsf{cls}}\Varid{cls} : \Conid{P}_{2}\ghl{\leadsto\Varid{pgm}_{\mathrm{2}}}{}\<[E]%
\ColumnHook
\end{hscode}\resethooks
}{%
\begin{hscode}\SaveRestoreHook
\column{B}{@{}>{\hspre}l<{\hspost}@{}}\ColumnHook
\column{E}{@{}>{\hspre}l<{\hspost}@{}}\ColumnHook
\>[B]{}\Conid{P}_{1}\vdash_{\mathsf{pgm}}\Varid{cls};\Varid{pgm} : \sigma\ghl{\leadsto\Varid{pgm}_{\mathrm{1}}}{}\<[E]%
\ColumnHook
\end{hscode}\resethooks
}{\PCls}
\and
\labelledinferrule{%
\begin{hscode}\SaveRestoreHook
\column{B}{@{}>{\hspre}l<{\hspost}@{}}\ColumnHook
\column{E}{@{}>{\hspre}l<{\hspost}@{}}\ColumnHook
\>[B]{}\Conid{P}_{2}\vdash_{\mathsf{pgm}}\Varid{pgm} : \sigma\ghl{\leadsto\Varid{pgm}}{}\<[E]%
\ColumnHook
\end{hscode}\resethooks
\and
\begin{hscode}\SaveRestoreHook
\column{B}{@{}>{\hspre}l<{\hspost}@{}}\ColumnHook
\column{E}{@{}>{\hspre}l<{\hspost}@{}}\ColumnHook
\>[B]{}\ghl{\Varid{pgm}~\vert~}\Conid{P}_{1}\vdash_{\mathsf{inst}}\Varid{inst} : \Conid{P}_{2}\ghl{\leadsto\Varid{pgm}_{\mathrm{1}}}{}\<[E]%
\ColumnHook
\end{hscode}\resethooks
}{%
\begin{hscode}\SaveRestoreHook
\column{B}{@{}>{\hspre}l<{\hspost}@{}}\ColumnHook
\column{E}{@{}>{\hspre}l<{\hspost}@{}}\ColumnHook
\>[B]{}\Conid{P}_{1}\vdash_{\mathsf{pgm}}\Varid{inst};\Varid{pgm} : \sigma\ghl{\leadsto\Varid{pgm}_{\mathrm{1}}}{}\<[E]%
\ColumnHook
\end{hscode}\resethooks
}{\PInst}
\end{mathpar}
\caption{Source Program Typing \ghl{\text{with Elaboration}}}
\label{fig:program-typing}
\end{figure*}






\begin{figure*}
\inlinemathhs
\ruletype{%
\begin{hscode}\SaveRestoreHook
\column{B}{@{}>{\hspre}l<{\hspost}@{}}\ColumnHook
\column{E}{@{}>{\hspre}l<{\hspost}@{}}\ColumnHook
\>[B]{}\Gamma \vdash_{\mathsf{ty}}\sigma\ghl{\leadsto\tau}{}\<[E]%
\ColumnHook
\end{hscode}\resethooks
}
\begin{mathpar}
\labelledinferrule{%
\begin{hscode}\SaveRestoreHook
\column{B}{@{}>{\hspre}l<{\hspost}@{}}\ColumnHook
\column{E}{@{}>{\hspre}l<{\hspost}@{}}\ColumnHook
\>[B]{}\Varid{a} \in \Gamma {}\<[E]%
\ColumnHook
\end{hscode}\resethooks
}{%
\begin{hscode}\SaveRestoreHook
\column{B}{@{}>{\hspre}l<{\hspost}@{}}\ColumnHook
\column{E}{@{}>{\hspre}l<{\hspost}@{}}\ColumnHook
\>[B]{}\Gamma \vdash_{\mathsf{ty}}\Varid{a}\ghl{\leadsto\Varid{a}}{}\<[E]%
\ColumnHook
\end{hscode}\resethooks
}{\TVar}
\and
\labelledinferrule{%
\begin{hscode}\SaveRestoreHook
\ColumnHook
\end{hscode}\resethooks
}{%
\begin{hscode}\SaveRestoreHook
\column{B}{@{}>{\hspre}l<{\hspost}@{}}\ColumnHook
\column{E}{@{}>{\hspre}l<{\hspost}@{}}\ColumnHook
\>[B]{}\Gamma \vdash_{\mathsf{ty}}\Conid{H}\ghl{\leadsto\Conid{H}}{}\<[E]%
\ColumnHook
\end{hscode}\resethooks
}{\TConst}
\and
\labelledinferrule{%
\begin{hscode}\SaveRestoreHook
\column{B}{@{}>{\hspre}l<{\hspost}@{}}\ColumnHook
\column{E}{@{}>{\hspre}l<{\hspost}@{}}\ColumnHook
\>[B]{}\Gamma \vdash_{\mathsf{ty}}\tau_{1}\ghl{\leadsto\tau_{1}'}{}\<[E]%
\ColumnHook
\end{hscode}\resethooks
\and
\begin{hscode}\SaveRestoreHook
\column{B}{@{}>{\hspre}l<{\hspost}@{}}\ColumnHook
\column{E}{@{}>{\hspre}l<{\hspost}@{}}\ColumnHook
\>[B]{}\Gamma \vdash_{\mathsf{ty}}\tau_{2}\ghl{\leadsto\tau_{2}'}{}\<[E]%
\ColumnHook
\end{hscode}\resethooks
}{%
\begin{hscode}\SaveRestoreHook
\column{B}{@{}>{\hspre}l<{\hspost}@{}}\ColumnHook
\column{E}{@{}>{\hspre}l<{\hspost}@{}}\ColumnHook
\>[B]{}\Gamma \vdash_{\mathsf{ty}}\tau_{1}\to \tau_{2}\ghl{\leadsto\tau_{1}'\to \tau_{2}'}{}\<[E]%
\ColumnHook
\end{hscode}\resethooks
}{\TArrow}
\and
\labelledinferrule{%
\begin{hscode}\SaveRestoreHook
\column{B}{@{}>{\hspre}l<{\hspost}@{}}\ColumnHook
\column{E}{@{}>{\hspre}l<{\hspost}@{}}\ColumnHook
\>[B]{}\Gamma \vdash_{\mathsf{ct}}\Conid{C}\ghl{\leadsto\tau_{1}}{}\<[E]%
\ColumnHook
\end{hscode}\resethooks
\and
\begin{hscode}\SaveRestoreHook
\column{B}{@{}>{\hspre}l<{\hspost}@{}}\ColumnHook
\column{E}{@{}>{\hspre}l<{\hspost}@{}}\ColumnHook
\>[B]{}\Gamma \vdash_{\mathsf{ty}}\rho\ghl{\leadsto\tau_{2}}{}\<[E]%
\ColumnHook
\end{hscode}\resethooks
}{%
\begin{hscode}\SaveRestoreHook
\column{B}{@{}>{\hspre}l<{\hspost}@{}}\ColumnHook
\column{E}{@{}>{\hspre}l<{\hspost}@{}}\ColumnHook
\>[B]{}\Gamma \vdash_{\mathsf{ty}}\Conid{C}\Rightarrow \rho\ghl{\leadsto\tau_{1}\to \tau_{2}}{}\<[E]%
\ColumnHook
\end{hscode}\resethooks
}{\TCArrow}
\and
\labelledinferrule{%
\begin{hscode}\SaveRestoreHook
\column{B}{@{}>{\hspre}l<{\hspost}@{}}\ColumnHook
\column{E}{@{}>{\hspre}l<{\hspost}@{}}\ColumnHook
\>[B]{}\Gamma ,\Varid{a}\vdash_{\mathsf{ty}}\sigma\ghl{\leadsto\tau}{}\<[E]%
\ColumnHook
\end{hscode}\resethooks
}{%
\begin{hscode}\SaveRestoreHook
\column{B}{@{}>{\hspre}l<{\hspost}@{}}\ColumnHook
\column{E}{@{}>{\hspre}l<{\hspost}@{}}\ColumnHook
\>[B]{}\Gamma \vdash_{\mathsf{ty}}\forall \Varid{a}.\sigma\ghl{\leadsto\forall \Varid{a}.\tau}{}\<[E]%
\ColumnHook
\end{hscode}\resethooks
}{\TForAll}
\and
\labelledinferrule{%
\begin{hscode}\SaveRestoreHook
\column{B}{@{}>{\hspre}l<{\hspost}@{}}\ColumnHook
\column{E}{@{}>{\hspre}l<{\hspost}@{}}\ColumnHook
\>[B]{}\Gamma \vdash_{\mathsf{ty}}\tau\ghl{\leadsto\tau^\prime}{}\<[E]%
\ColumnHook
\end{hscode}\resethooks
}{%
\begin{hscode}\SaveRestoreHook
\column{B}{@{}>{\hspre}l<{\hspost}@{}}\ColumnHook
\column{E}{@{}>{\hspre}l<{\hspost}@{}}\ColumnHook
\>[B]{}\Gamma \vdash_{\mathsf{ty}}\Conid{Code}\;\tau\ghl{\leadsto\Conid{Code}\;\tau^\prime}{}\<[E]%
\ColumnHook
\end{hscode}\resethooks
}{\TCode}
\end{mathpar}
\caption{Source Type Formation \ghl{\text{with Elaboration}}}
\label{fig:ty-kinding-elaboration}
\end{figure*}

\nw{TODO: Constraint Formation with Elaboration is slightly different}
\al{I think this is ok.}





\begin{figure*}
\inlinemathhs
\ruletype{%
\begin{hscode}\SaveRestoreHook
\column{B}{@{}>{\hspre}l<{\hspost}@{}}\ColumnHook
\column{E}{@{}>{\hspre}l<{\hspost}@{}}\ColumnHook
\>[B]{}\Gamma \vdash_{\mathsf{ct}}\Conid{C}\ghl{\leadsto\tau}{}\<[E]%
\ColumnHook
\end{hscode}\resethooks
}
\begin{mathpar}
\labelledinferrule{%
\begin{hscode}\SaveRestoreHook
\column{B}{@{}>{\hspre}l<{\hspost}@{}}\ColumnHook
\column{E}{@{}>{\hspre}l<{\hspost}@{}}\ColumnHook
\>[B]{}\mathbf{class}\;~\Conid{TC}~\Varid{a}\;\mathbf{where}\;\{\mskip1.5mu \Varid{k}\mathbin{::}\sigma\mskip1.5mu\}{}\<[E]%
\ColumnHook
\end{hscode}\resethooks
\and
\begin{hscode}\SaveRestoreHook
\column{B}{@{}>{\hspre}l<{\hspost}@{}}\ColumnHook
\column{E}{@{}>{\hspre}l<{\hspost}@{}}\ColumnHook
\>[B]{}\Gamma \vdash_{\mathsf{ty}}\sigma[\tau/\Varid{a}]\ghl{\leadsto\tau^\prime}{}\<[E]%
\ColumnHook
\end{hscode}\resethooks
}{%
\begin{hscode}\SaveRestoreHook
\column{B}{@{}>{\hspre}l<{\hspost}@{}}\ColumnHook
\column{E}{@{}>{\hspre}l<{\hspost}@{}}\ColumnHook
\>[B]{}\Gamma \vdash_{\mathsf{ct}}\Conid{TC}~\tau\ghl{\leadsto\tau^\prime}{}\<[E]%
\ColumnHook
\end{hscode}\resethooks
}{\CTC}
\and
\labelledinferrule{%
\begin{hscode}\SaveRestoreHook
\column{B}{@{}>{\hspre}l<{\hspost}@{}}\ColumnHook
\column{E}{@{}>{\hspre}l<{\hspost}@{}}\ColumnHook
\>[B]{}\Gamma \vdash_{\mathsf{ct}}\Conid{C}\ghl{\leadsto\tau}{}\<[E]%
\ColumnHook
\end{hscode}\resethooks
}{%
\begin{hscode}\SaveRestoreHook
\column{B}{@{}>{\hspre}l<{\hspost}@{}}\ColumnHook
\column{E}{@{}>{\hspre}l<{\hspost}@{}}\ColumnHook
\>[B]{}\Gamma \vdash_{\mathsf{ct}}\Conid{CodeC}\;\Conid{C}\ghl{\leadsto\Conid{Code}\;\tau}{}\<[E]%
\ColumnHook
\end{hscode}\resethooks
}{\CCodeC}
\end{mathpar}
\caption{Source Constraint Formation \ghl{\text{with Elaboration}}}
\label{fig:constraint-formation}
\end{figure*}

\al{This is going to be controversial, but I'm dropping the separate
examples section tentatively for the integrated E1 reprisal which can
probably be shortened further. I don't see the second one as sufficiently
different.}





\section{The Core Language}
\label{sec:core}

In this section we describe an explicitly typed core language which is suitable
as a compilation target for the declarative source language we described in
the Section~\ref{sec:type-system}.

\newcommand{\qqe}[2]{\llbracket~#1~\rrbracket_{#2}}
\newcommand{\ecase}[3]{\mathbf{case}~#1~\mathbf{of}~#2 \to #3 }

\subsection{Syntax}

The syntax for the core language is presented in Figure~\ref{fig:core-syntax}.
It is a variant of the explicit polymorphic lambda calculus with
multi-stage constructs, top-level definitions and
top-level splice definitions.

Quotes and splices are represented by the syntactic form \ensuremath{\qq{\Varid{e}}_{\Conid{SP}}}, which
is a quotation with an associated splice environment which binds \emph{splice variables}
for each splice point within the quoted expression. A splice point is
where the result of evaluating a splice will be inserted. Splice variables to represent
the splice points are bound
by splice environments and top-level splice definitions. The expression syntax contains
no splices, splices are modelled using the splice environments which are attached
to quotations and top-level splice definitions.

The splice environment maps a splice point~\ensuremath{\Varid{sp}} to a local type
environment~\ensuremath{\Delta}, a type~\ensuremath{\tau} and an expression~\ensuremath{\Varid{e}} which we write as
$\Delta \vdash sp \colonT \tau = e$. The typing rules will ensure that the
expression~\ensuremath{\Varid{e}} has type~\ensuremath{\Conid{Code}\;\tau}. The purpose of the environment \ensuremath{\Delta} is
to support open code representations which lose their lexical scoping when lifted
from the quotation.

A top-level splice definition is used to support elaborating from negative levels
in the source language. These top level declarations are of
the form \ensuremath{\mathbf{spdef}~\Delta \vdashn{\Varid{n}}\Varid{sp} \colonT \tau\mathrel{=}\Varid{e}} which indicates that the expression~\ensuremath{\Varid{e}}
will have type~\ensuremath{\Conid{Code}\;\tau} at level~\ensuremath{\Varid{n}} in environment~\ensuremath{\Delta}. Top-level splice
definitions also explicitly record the level of the original splice so that
the declaration can be typechecked at the correct level. The level index is not
necessary for the quotation splice environments because the level of the whole environment
is fixed by the level at which the quotation appears.

\subsection{Typing Rules}

The expression typing rules for the core language are for the most part the
same as those in the source language. The rules that differ are shown in Figure~\ref{fig:core-expr-typing}.

The splice environment typing rules are given in Figure~\ref{fig:splice-env-typing}.
A splice environment is well-typed if each of its definitions is well-typed.
We check that the body of each definition has type \ensuremath{\Conid{Code}\;\tau} in an environment extended
by \ensuremath{\Delta}. In the \EQuote rule, the body of the quotation is checked in an environment
\ensuremath{\Gamma }
extended with the contents of the splice environment. A splice variable \ensuremath{\Varid{sp}} is associated
with an environment \ensuremath{\Delta}, a type \ensuremath{\tau} and a level \ensuremath{\Varid{n}} in
\ensuremath{\Gamma }.
When a splice variable is used, the \ESpliceVar rule ensures that the claimed local
environment aligns with the actual environment, the type of the variable is correct
and the level matches the surrounding context.

Top-level splice definitions are typed in a similar manner in Figure~\ref{fig:core-defn-typing}.
The body of the definition is checked to have type \ensuremath{\Conid{Code}\;\tau}
at level \ensuremath{\Varid{n}} in environment \ensuremath{\Gamma }, the
program theory is then extended with a splice variable \ensuremath{\Varid{sp}\mathbin{:}(\Gamma ,\tau)}
which can be used in the remainder of the program.

\subsection{Dynamic Semantics}

\al{It would be good to actually refer to the Figures from here.}

The introduction of splice environments makes the evaluation order of
the core calculus evident and hence a suitable target for compilation.

Usually in order to ensure a well-staged evaluation order, the reduction relation
must be level-indexed to evaluate splices inside quotations. In our calculus,
there is no need to do this because the splices have already been lifted outside
of the quotation during the elaboration process. This style is less convenient
to program with, but easy to reason about and implement.

In realistic implementations, the quotations are compiled to a representation form
for which implementing substitution can be difficult. By lifting the splices outside
of the representation, the representation does not need to be inspected or
traversed before it is spliced back into the program. The representation can
be treated in an opaque manner which gives us more implementation freedom about
its form.
In our calculus this is evidenced by the fact there is no reduction rule which
reduces inside a quotation.

The program evaluation semantics evaluate each declaration in turn from top to
bottom. Top-level definitions are evaluated to values and substituted into the
rest of the program. Top-level splice definitions are evaluated to a value of
type \ensuremath{\Conid{Code}\;\tau} which has the form \ensuremath{\qq{\Varid{e}}_{\Conid{SP}}}.
The splice variable is bound to the value \ensuremath{\Varid{e}} with the substitution \ensuremath{\Conid{SP}} applied
and then substituted into the remainder of the program.

Using splice environments and top-level splice definitions is reminiscent of the approach taken in logically inspired
languages by \citet{nanevski2002meta} and \citet{davies2001modal}.

\subsection{Module Restriction}

In GHC, the module restriction dictates that only identifiers bound in other
modules can be used inside top-level splices. This restriction is modelled in
our calculus by the restriction that only identifiers previously bound in
top-level definitions can be used inside a top-level splice. The intention
is therefore to consider each top-level definition to be defined in its
own ``module'' which is completely evaluated before moving onto the next definition.

This reflects the situation in a compiler such as GHC which supports separate
compilation. When compiling a program which uses multiple modules,
one module may contain a top-level splice which is then used inside another
top-level splice in a different module. Although syntactically both top-level
splices occur at level~\ensuremath{\mathbin{-}\mathrm{1}}, they effectively occur at different stages.
In the same way as different modules occur at different stages due to separate
compilation, in our calculus, each top-level definition can be considered to
be evaluated at a new stage.

At this point it is worthwhile to consider what exactly we mean by compile-time
and run-time. So far we have stated that the intention is for splices at negative
levels to represent compile-time evaluation so we should state how we intend
this statement to be understood in our formalism. The elaboration procedure
will elaborate each top-level definition to zero or more splice definitions
(one for each top-level splice it contains) followed by a normal value definition.
Then, during the evaluation of the core program the splice definitions will be evaluated
prior to the top-level definition that originally contained them.

The meaning of ``compile-time'' is therefore that the evaluation of the top-level
splice happens before the top-level definition is evaluated. It is also possible
to imagine a semantics which partially evaluates a program to a residual by
computing and removing as many splice definitions as possible.

If we were to more precisely model a module as a collection of definitions then
the typing rules could be modified to only allow definitions from previously
defined modules to be used at the top-level and all splice definitions could be
grouped together at the start of a module definition before any of the value definitions.
Then it would be clearly possible to evaluate all of the splice definitions for
a module before commencing to evaluate the module definitions.





\begin{figure*}
\invisiblecomments
\aligncolumn{162}{r}%
\savecolumns
\begin{hscode}\SaveRestoreHook
\column{B}{@{}>{\hspre}l<{\hspost}@{}}\ColumnHook
\column{19}{@{}>{\hspre}c<{\hspost}@{}}\ColumnHook
\column{19E}{@{}l@{}}\ColumnHook
\column{24}{@{}>{\hspre}l<{\hspost}@{}}\ColumnHook
\column{E}{@{}>{\hspre}l<{\hspost}@{}}\ColumnHook
\>[B]{}\Varid{pgm}{}\<[19]%
\>[19]{}\mathbin{::=}{}\<[19E]%
\>[24]{}\Varid{e}\colonT \tau\mid \Varid{def};\Varid{pgm}\mid \Varid{spdef};\Varid{pgm}{}\<[E]%
\ColumnHook
\end{hscode}\resethooks
\spacecorrectsep
\aligncolumn{162}{r}%
\restorecolumns
\begin{hscode}\SaveRestoreHook
\column{B}{@{}>{\hspre}l<{\hspost}@{}}\ColumnHook
\column{19}{@{}>{\hspre}c<{\hspost}@{}}\ColumnHook
\column{19E}{@{}l@{}}\ColumnHook
\column{24}{@{}>{\hspre}l<{\hspost}@{}}\ColumnHook
\column{157}{@{}>{\hspre}c<{\hspost}@{}}\ColumnHook
\column{157E}{@{}l@{}}\ColumnHook
\column{162}{@{}>{\hspre}l<{\hspost}@{}}\ColumnHook
\column{E}{@{}>{\hspre}l<{\hspost}@{}}\ColumnHook
\>[B]{}\Varid{def}{}\<[19]%
\>[19]{}\mathbin{::=}{}\<[19E]%
\>[24]{}\mathbf{def}~\Varid{k} \colonT \tau\mathrel{=}\Varid{e}\;{}\<[E]%
\\
\>[B]{}\Varid{spdef}{}\<[19]%
\>[19]{}\mathbin{::=}{}\<[19E]%
\>[24]{}\mathbf{spdef}~\Delta \vdashn{\Varid{n}}\Varid{sp} \colonT \tau\mathrel{=}\Varid{e}\;{}\<[E]%
\\[\blanklineskip]%
\>[B]{}\Varid{e},\Varid{t}{}\<[19]%
\>[19]{}\mathbin{::=}{}\<[19E]%
\>[157]{}\quad {}\<[157E]%
\>[162]{}\mbox{\onelinecomment  elaborated expressions}{}\<[E]%
\\
\>[19]{} {}\<[19E]%
\>[24]{}\mathit{x}\mid \Varid{sp}\mid \Varid{k}{}\<[157]%
\>[157]{}\quad {}\<[157E]%
\>[162]{}\mbox{\onelinecomment  variables / splice variables / globals}{}\<[E]%
\\
\>[19]{}\mid {}\<[19E]%
\>[24]{}\lambda \mathit{x}\colonT \tau.\Varid{e}\mid \Varid{e}~\Varid{e}{}\<[157]%
\>[157]{}\quad {}\<[157E]%
\>[162]{}\mbox{\onelinecomment  abstraction / application}{}\<[E]%
\\
\>[19]{}\mid {}\<[19E]%
\>[24]{}\Lambda a.\Varid{e}\mid \Varid{e}~\langle \tau\rangle {}\<[157]%
\>[157]{}\quad {}\<[157E]%
\>[162]{}\mbox{\onelinecomment  type abstraction / application}{}\<[E]%
\\
\>[19]{}\mid {}\<[19E]%
\>[24]{}\qq{\Varid{e}}_{\Conid{SP}}{}\<[157]%
\>[157]{}\quad {}\<[157E]%
\>[162]{}\mbox{\onelinecomment  quotation}{}\<[E]%
\ColumnHook
\end{hscode}\resethooks
\spacecorrectsep
\aligncolumn{162}{r}%
\restorecolumns
\begin{hscode}\SaveRestoreHook
\column{B}{@{}>{\hspre}l<{\hspost}@{}}\ColumnHook
\column{19}{@{}>{\hspre}c<{\hspost}@{}}\ColumnHook
\column{19E}{@{}l@{}}\ColumnHook
\column{24}{@{}>{\hspre}l<{\hspost}@{}}\ColumnHook
\column{157}{@{}>{\hspre}c<{\hspost}@{}}\ColumnHook
\column{157E}{@{}l@{}}\ColumnHook
\column{162}{@{}>{\hspre}l<{\hspost}@{}}\ColumnHook
\column{E}{@{}>{\hspre}l<{\hspost}@{}}\ColumnHook
\>[B]{}\Conid{SP}{}\<[19]%
\>[19]{}\mathbin{::=}{}\<[19E]%
\>[24]{}\bullet \mid \Conid{SP},\Delta\vdash \Varid{sp}\colonT\tau\mathrel{=}\Varid{e}{}\<[157]%
\>[157]{}\quad {}\<[157E]%
\>[162]{}\mbox{\onelinecomment  splice environment}{}\<[E]%
\ColumnHook
\end{hscode}\resethooks
\spacecorrectsep
\aligncolumn{162}{r}%
\restorecolumns
\begin{hscode}\SaveRestoreHook
\column{B}{@{}>{\hspre}l<{\hspost}@{}}\ColumnHook
\column{19}{@{}>{\hspre}c<{\hspost}@{}}\ColumnHook
\column{19E}{@{}l@{}}\ColumnHook
\column{24}{@{}>{\hspre}l<{\hspost}@{}}\ColumnHook
\column{157}{@{}>{\hspre}c<{\hspost}@{}}\ColumnHook
\column{157E}{@{}l@{}}\ColumnHook
\column{162}{@{}>{\hspre}l<{\hspost}@{}}\ColumnHook
\column{E}{@{}>{\hspre}l<{\hspost}@{}}\ColumnHook
\>[B]{}\tau{}\<[19]%
\>[19]{}\mathbin{::=}{}\<[19E]%
\>[157]{}\quad {}\<[157E]%
\>[162]{}\mbox{\onelinecomment  core types}{}\<[E]%
\\
\>[19]{} {}\<[19E]%
\>[24]{}\Varid{a}\mid \Conid{H}{}\<[157]%
\>[157]{}\quad {}\<[157E]%
\>[162]{}\mbox{\onelinecomment  variables / constants}{}\<[E]%
\\
\>[19]{}\mid {}\<[19E]%
\>[24]{}\tau\to \tau{}\<[157]%
\>[157]{}\quad {}\<[157E]%
\>[162]{}\mbox{\onelinecomment  functions}{}\<[E]%
\\
\>[19]{}\mid {}\<[19E]%
\>[24]{}\Conid{Code}\;\tau{}\<[157]%
\>[157]{}\quad {}\<[157E]%
\>[162]{}\mbox{\onelinecomment  representation}{}\<[E]%
\\
\>[19]{}\mid {}\<[19E]%
\>[24]{}\forall a.\tau{}\<[157]%
\>[157]{}\quad {}\<[157E]%
\>[162]{}\mbox{\onelinecomment  quantification}{}\<[E]%
\ColumnHook
\end{hscode}\resethooks
\spacecorrectsep
\aligncolumn{167}{r}%
\restorecolumns
\begin{hscode}\SaveRestoreHook
\column{B}{@{}>{\hspre}l<{\hspost}@{}}\ColumnHook
\column{19}{@{}>{\hspre}c<{\hspost}@{}}\ColumnHook
\column{19E}{@{}l@{}}\ColumnHook
\column{24}{@{}>{\hspre}l<{\hspost}@{}}\ColumnHook
\column{162}{@{}>{\hspre}c<{\hspost}@{}}\ColumnHook
\column{162E}{@{}l@{}}\ColumnHook
\column{167}{@{}>{\hspre}l<{\hspost}@{}}\ColumnHook
\column{E}{@{}>{\hspre}l<{\hspost}@{}}\ColumnHook
\>[B]{}\Gamma ,\Delta{}\<[19]%
\>[19]{}\mathbin{::=}{}\<[19E]%
\>[24]{}\bullet \mid \Gamma ,\mathit{x}\colonT (\tau,\Varid{n})\mid \Gamma ,\Varid{sp}\colonT (\Delta,\tau,\Varid{n})\mid \Gamma ,\Varid{a}{}\<[162]%
\>[162]{}\quad {}\<[162E]%
\>[167]{}\mbox{\onelinecomment  type environment}{}\<[E]%
\ColumnHook
\end{hscode}\resethooks
\spacecorrectnosep
\aligncolumn{167}{r}%
\restorecolumns
\begin{hscode}\SaveRestoreHook
\column{B}{@{}>{\hspre}l<{\hspost}@{}}\ColumnHook
\column{19}{@{}>{\hspre}c<{\hspost}@{}}\ColumnHook
\column{19E}{@{}l@{}}\ColumnHook
\column{24}{@{}>{\hspre}l<{\hspost}@{}}\ColumnHook
\column{162}{@{}>{\hspre}c<{\hspost}@{}}\ColumnHook
\column{162E}{@{}l@{}}\ColumnHook
\column{167}{@{}>{\hspre}l<{\hspost}@{}}\ColumnHook
\column{E}{@{}>{\hspre}l<{\hspost}@{}}\ColumnHook
\>[B]{}\Conid{P}{}\<[19]%
\>[19]{}\mathbin{::=}{}\<[19E]%
\>[24]{}\bullet \mid \Conid{P},\Varid{k}\colonT \tau\mid \Conid{P},\Varid{sp}\colonT (\Delta,\tau){}\<[162]%
\>[162]{}\quad {}\<[162E]%
\>[167]{}\mbox{\onelinecomment  program environment}{}\<[E]%
\ColumnHook
\end{hscode}\resethooks
\caption{Core Language Syntax}
\label{fig:core-syntax}
\end{figure*}

\begin{figure*}
\inlinemathhs
\center\boxed{%
\begin{hscode}\SaveRestoreHook
\column{B}{@{}>{\hspre}l<{\hspost}@{}}\ColumnHook
\column{E}{@{}>{\hspre}l<{\hspost}@{}}\ColumnHook
\>[B]{}\Conid{P};\Gamma \vdashn{\Varid{n}} \Varid{e} : \tau{}\<[E]%
\ColumnHook
\end{hscode}\resethooks
}
\begin{mathpar}
\labelledinferrule{%
\begin{hscode}\SaveRestoreHook
\column{B}{@{}>{\hspre}l<{\hspost}@{}}\ColumnHook
\column{E}{@{}>{\hspre}l<{\hspost}@{}}\ColumnHook
\>[B]{}\Delta\vdash\Varid{sp}\colonT (\tau,\Varid{n}) \in \Gamma{}\<[E]%
\ColumnHook
\end{hscode}\resethooks
\and
\begin{hscode}\SaveRestoreHook
\column{B}{@{}>{\hspre}l<{\hspost}@{}}\ColumnHook
\column{E}{@{}>{\hspre}l<{\hspost}@{}}\ColumnHook
\>[B]{}\Delta\subseteq \Gamma{}\<[E]%
\ColumnHook
\end{hscode}\resethooks
}{%
\begin{hscode}\SaveRestoreHook
\column{B}{@{}>{\hspre}l<{\hspost}@{}}\ColumnHook
\column{E}{@{}>{\hspre}l<{\hspost}@{}}\ColumnHook
\>[B]{}\Conid{P};\Gamma \vdashn{\Varid{n}} \Varid{sp} : \tau{}\<[E]%
\ColumnHook
\end{hscode}\resethooks
}{\text{\ESpliceVar}}
\and
\labelledinferrule{%
\begin{hscode}\SaveRestoreHook
\column{B}{@{}>{\hspre}l<{\hspost}@{}}\ColumnHook
\column{E}{@{}>{\hspre}l<{\hspost}@{}}\ColumnHook
\>[B]{}\Varid{sp}\colonT (\Delta,\tau) \in \Conid{P}{}\<[E]%
\ColumnHook
\end{hscode}\resethooks
\and
\begin{hscode}\SaveRestoreHook
\column{B}{@{}>{\hspre}l<{\hspost}@{}}\ColumnHook
\column{E}{@{}>{\hspre}l<{\hspost}@{}}\ColumnHook
\>[B]{}\Delta\subseteq \Gamma{}\<[E]%
\ColumnHook
\end{hscode}\resethooks
}{%
\begin{hscode}\SaveRestoreHook
\column{B}{@{}>{\hspre}l<{\hspost}@{}}\ColumnHook
\column{E}{@{}>{\hspre}l<{\hspost}@{}}\ColumnHook
\>[B]{}\Conid{P};\Gamma \vdashn{\Varid{n}} \Varid{sp} : \tau{}\<[E]%
\ColumnHook
\end{hscode}\resethooks
}{\text{\ETopSpliceVar}}
\and
\labelledinferrule{%
\begin{hscode}\SaveRestoreHook
\column{B}{@{}>{\hspre}l<{\hspost}@{}}\ColumnHook
\column{E}{@{}>{\hspre}l<{\hspost}@{}}\ColumnHook
\>[B]{}\Conid{P};\Gamma,\Varid{a} \vdashn{\Varid{n}} \Varid{e} : \tau{}\<[E]%
\ColumnHook
\end{hscode}\resethooks
}{%
\begin{hscode}\SaveRestoreHook
\column{B}{@{}>{\hspre}l<{\hspost}@{}}\ColumnHook
\column{E}{@{}>{\hspre}l<{\hspost}@{}}\ColumnHook
\>[B]{}\Conid{P};\Gamma \vdashn{\Varid{n}} \Lambda \Varid{a}.\Varid{e} : \forall \Varid{a}.\tau{}\<[E]%
\ColumnHook
\end{hscode}\resethooks
}{\text{\ETAbs}}
\and
\labelledinferrule{%
\begin{hscode}\SaveRestoreHook
\column{B}{@{}>{\hspre}l<{\hspost}@{}}\ColumnHook
\column{E}{@{}>{\hspre}l<{\hspost}@{}}\ColumnHook
\>[B]{}\Conid{P};\Gamma \vdashn{\Varid{n}} \Varid{e} : \forall \Varid{a}.\tau_{2}{}\<[E]%
\ColumnHook
\end{hscode}\resethooks
}{%
\begin{hscode}\SaveRestoreHook
\column{B}{@{}>{\hspre}l<{\hspost}@{}}\ColumnHook
\column{E}{@{}>{\hspre}l<{\hspost}@{}}\ColumnHook
\>[B]{}\Conid{P};\Gamma \vdashn{\Varid{n}} \Varid{e}~\langle \tau_{1}\rangle  : \tau_{2}[\tau_{1}/\Varid{a}]{}\<[E]%
\ColumnHook
\end{hscode}\resethooks
}{\text{\ETApp}}
\and
\labelledinferrule{%
\begin{hscode}\SaveRestoreHook
\column{B}{@{}>{\hspre}l<{\hspost}@{}}\ColumnHook
\column{E}{@{}>{\hspre}l<{\hspost}@{}}\ColumnHook
\>[B]{}\Conid{P};\Gamma, \overline{\Varid{sp}_{\Varid{i}} \colonT (\Delta_{\Varid{i}}, \tau_{\Varid{i}}, \Varid{n}\mathbin{+}\mathrm{1})} \vdashn{\Varid{n}\mathbin{+}\mathrm{1}} \Varid{e} : \tau{}\<[E]%
\ColumnHook
\end{hscode}\resethooks
\and
\begin{hscode}\SaveRestoreHook
\column{B}{@{}>{\hspre}l<{\hspost}@{}}\ColumnHook
\column{E}{@{}>{\hspre}l<{\hspost}@{}}\ColumnHook
\>[B]{}\Conid{P};\Gamma \vdashn{\Varid{n}}_{\mathsf{SP}}\Conid{SP}{}\<[E]%
\ColumnHook
\end{hscode}\resethooks
\and
\begin{hscode}\SaveRestoreHook
\column{B}{@{}>{\hspre}l<{\hspost}@{}}\ColumnHook
\column{E}{@{}>{\hspre}l<{\hspost}@{}}\ColumnHook
\>[B]{}\Conid{SP}\mathrel{=}\overline{\Delta_{\Varid{i}} \vdash \Varid{sp}_{\Varid{i}} \colonT \tau_{\Varid{i}}\mathrel{=}\Varid{e}_{\Varid{i}}}{}\<[E]%
\ColumnHook
\end{hscode}\resethooks
}{%
\begin{hscode}\SaveRestoreHook
\column{B}{@{}>{\hspre}l<{\hspost}@{}}\ColumnHook
\column{E}{@{}>{\hspre}l<{\hspost}@{}}\ColumnHook
\>[B]{}\Conid{P};\Gamma \vdashn{\Varid{n}} \qq{\Varid{e}}_{\Conid{SP}} : \Conid{Code}\;\tau{}\<[E]%
\ColumnHook
\end{hscode}\resethooks
}{\text{\EQuote}}
\end{mathpar}
\caption{Core Expression Typing}
\label{fig:core-expr-typing}
\end{figure*}
\begin{figure*}
\inlinemathhs
\center\boxed{%
\begin{hscode}\SaveRestoreHook
\column{B}{@{}>{\hspre}l<{\hspost}@{}}\ColumnHook
\column{E}{@{}>{\hspre}l<{\hspost}@{}}\ColumnHook
\>[B]{}\Conid{P};\Gamma \vdashn{\Varid{n}}_{\mathsf{SP}}\Conid{SP}{}\<[E]%
\ColumnHook
\end{hscode}\resethooks
}
\begin{mathpar}
\labelledinferrule{ }{%
\begin{hscode}\SaveRestoreHook
\column{B}{@{}>{\hspre}l<{\hspost}@{}}\ColumnHook
\column{E}{@{}>{\hspre}l<{\hspost}@{}}\ColumnHook
\>[B]{}\Conid{P};\Gamma \vdashn{\Varid{n}}_{\mathsf{SP}}\bullet {}\<[E]%
\ColumnHook
\end{hscode}\resethooks
}{\SPEmpty}
\and
\labelledinferrule{%
\begin{hscode}\SaveRestoreHook
\column{B}{@{}>{\hspre}l<{\hspost}@{}}\ColumnHook
\column{E}{@{}>{\hspre}l<{\hspost}@{}}\ColumnHook
\>[B]{}\Conid{P};\Gamma \vdashn{\Varid{n}}_{\mathsf{SP}}\Conid{SP}{}\<[E]%
\ColumnHook
\end{hscode}\resethooks
\and
\begin{hscode}\SaveRestoreHook
\column{B}{@{}>{\hspre}l<{\hspost}@{}}\ColumnHook
\column{E}{@{}>{\hspre}l<{\hspost}@{}}\ColumnHook
\>[B]{}\Conid{P};\Gamma,\Delta \vdashn{\Varid{n}} \Varid{e} : \Conid{Code}\;\tau{}\<[E]%
\ColumnHook
\end{hscode}\resethooks
}{%
\begin{hscode}\SaveRestoreHook
\column{B}{@{}>{\hspre}l<{\hspost}@{}}\ColumnHook
\column{E}{@{}>{\hspre}l<{\hspost}@{}}\ColumnHook
\>[B]{}\Conid{P};\Gamma \vdashn{\Varid{n}}_{\mathsf{SP}}\Conid{SP},\Delta\vdash \Varid{sp}\colonT\tau\mathrel{=}\Varid{e}{}\<[E]%
\ColumnHook
\end{hscode}\resethooks
}{\SPCons}
\end{mathpar}
\caption{Splice Environment Typing}
\label{fig:splice-env-typing}
\end{figure*}
\begin{figure*}
\begin{minipage}[t]{.5\linewidth}
\inlinemathhs
\center\boxed{%
\begin{hscode}\SaveRestoreHook
\column{B}{@{}>{\hspre}l<{\hspost}@{}}\ColumnHook
\column{E}{@{}>{\hspre}l<{\hspost}@{}}\ColumnHook
\>[B]{}\Conid{P}\vdash_{\mathsf{def}}\Varid{def}\mathbin{:}\Conid{P}{}\<[E]%
\ColumnHook
\end{hscode}\resethooks
}
\begin{mathpar}
\labelledinferrule{%
\begin{hscode}\SaveRestoreHook
\column{B}{@{}>{\hspre}l<{\hspost}@{}}\ColumnHook
\column{E}{@{}>{\hspre}l<{\hspost}@{}}\ColumnHook
\>[B]{}\Conid{P};\bullet  \vdashn{\mathrm{0}} \Varid{e} : \tau{}\<[E]%
\ColumnHook
\end{hscode}\resethooks
}{%
\begin{hscode}\SaveRestoreHook
\column{B}{@{}>{\hspre}l<{\hspost}@{}}\ColumnHook
\column{E}{@{}>{\hspre}l<{\hspost}@{}}\ColumnHook
\>[B]{}\Conid{P}\vdash_{\mathsf{def}}\mathbf{def}~\Varid{k} \colonT \tau\mathrel{=}\Varid{e}\mathbin{:}\Conid{P},\Varid{k}\colonT \tau{}\<[E]%
\ColumnHook
\end{hscode}\resethooks
}{\CDef}
\end{mathpar}
\end{minipage}%
\begin{minipage}[t]{.5\linewidth}
\inlinemathhs
\center\boxed{%
\begin{hscode}\SaveRestoreHook
\column{B}{@{}>{\hspre}l<{\hspost}@{}}\ColumnHook
\column{E}{@{}>{\hspre}l<{\hspost}@{}}\ColumnHook
\>[B]{}\Conid{P}\vdash_{\mathsf{spdef}}\Varid{spdef}\mathbin{:}\Conid{P}{}\<[E]%
\ColumnHook
\end{hscode}\resethooks
}
\begin{mathpar}
\labelledinferrule{%
\begin{hscode}\SaveRestoreHook
\column{B}{@{}>{\hspre}l<{\hspost}@{}}\ColumnHook
\column{E}{@{}>{\hspre}l<{\hspost}@{}}\ColumnHook
\>[B]{}\Conid{P};\Gamma \vdashn{\Varid{n}} \Varid{e} : \Conid{Code}\;\tau{}\<[E]%
\ColumnHook
\end{hscode}\resethooks
}{%
\begin{hscode}\SaveRestoreHook
\column{B}{@{}>{\hspre}l<{\hspost}@{}}\ColumnHook
\column{E}{@{}>{\hspre}l<{\hspost}@{}}\ColumnHook
\>[B]{}\Conid{P}\vdash_{\mathsf{spdef}}\mathbf{spdef}~\Gamma \vdashn{\Varid{n}}\Varid{sp} \colonT \tau\mathrel{=}\Varid{e}\mathbin{:}\Conid{P},\Varid{sp}\colonT (\Gamma,\tau){}\<[E]%
\ColumnHook
\end{hscode}\resethooks
}{C\_SpDef}
\end{mathpar}
\label{fig:splice-defn-typing}
\end{minipage}
\caption{Core Definition Typing}
\label{fig:core-defn-typing}
\end{figure*}
\begin{figure*}
\inlinemathhs
\center\boxed{%
\begin{hscode}\SaveRestoreHook
\column{B}{@{}>{\hspre}l<{\hspost}@{}}\ColumnHook
\column{E}{@{}>{\hspre}l<{\hspost}@{}}\ColumnHook
\>[B]{}\Conid{P}\vdash_{\mathsf{pgm}}\Varid{pgm}{}\<[E]%
\ColumnHook
\end{hscode}\resethooks
}
\begin{mathpar}
\labelledinferrule{%
\begin{hscode}\SaveRestoreHook
\column{B}{@{}>{\hspre}l<{\hspost}@{}}\ColumnHook
\column{E}{@{}>{\hspre}l<{\hspost}@{}}\ColumnHook
\>[B]{}\Conid{P};\bullet  \vdashn{\mathrm{0}} \Varid{e} : \tau{}\<[E]%
\ColumnHook
\end{hscode}\resethooks
}{%
\begin{hscode}\SaveRestoreHook
\column{B}{@{}>{\hspre}l<{\hspost}@{}}\ColumnHook
\column{E}{@{}>{\hspre}l<{\hspost}@{}}\ColumnHook
\>[B]{}\Conid{P}\vdash_{\mathsf{pgm}}\Varid{e}\colonT \tau{}\<[E]%
\ColumnHook
\end{hscode}\resethooks
}{\CMain}
\and
\labelledinferrule{%
\begin{hscode}\SaveRestoreHook
\column{B}{@{}>{\hspre}l<{\hspost}@{}}\ColumnHook
\column{E}{@{}>{\hspre}l<{\hspost}@{}}\ColumnHook
\>[B]{}\Conid{P}_{1}\vdash_{\mathsf{def}}\Varid{def}\mathbin{:}\Conid{P}_{2}{}\<[E]%
\ColumnHook
\end{hscode}\resethooks
\and
\begin{hscode}\SaveRestoreHook
\column{B}{@{}>{\hspre}l<{\hspost}@{}}\ColumnHook
\column{E}{@{}>{\hspre}l<{\hspost}@{}}\ColumnHook
\>[B]{}\Conid{P}_{2}\vdash_{\mathsf{pgm}}\Varid{pgm}{}\<[E]%
\ColumnHook
\end{hscode}\resethooks
}{%
\begin{hscode}\SaveRestoreHook
\column{B}{@{}>{\hspre}l<{\hspost}@{}}\ColumnHook
\column{E}{@{}>{\hspre}l<{\hspost}@{}}\ColumnHook
\>[B]{}\Conid{P}_{1}\vdash_{\mathsf{pgm}}\Varid{def};\Varid{pgm}{}\<[E]%
\ColumnHook
\end{hscode}\resethooks
}{\CPDef}
\and
\labelledinferrule{%
\begin{hscode}\SaveRestoreHook
\column{B}{@{}>{\hspre}l<{\hspost}@{}}\ColumnHook
\column{E}{@{}>{\hspre}l<{\hspost}@{}}\ColumnHook
\>[B]{}\Conid{P}_{1}\vdash_{\mathsf{spdef}}\Varid{spdef}\mathbin{:}\Conid{P}_{2}{}\<[E]%
\ColumnHook
\end{hscode}\resethooks
\and
\begin{hscode}\SaveRestoreHook
\column{B}{@{}>{\hspre}l<{\hspost}@{}}\ColumnHook
\column{E}{@{}>{\hspre}l<{\hspost}@{}}\ColumnHook
\>[B]{}\Conid{P}_{2}\vdash_{\mathsf{pgm}}\Varid{pgm}{}\<[E]%
\ColumnHook
\end{hscode}\resethooks
}{%
\begin{hscode}\SaveRestoreHook
\column{B}{@{}>{\hspre}l<{\hspost}@{}}\ColumnHook
\column{E}{@{}>{\hspre}l<{\hspost}@{}}\ColumnHook
\>[B]{}\Conid{P}_{1}\vdash_{\mathsf{pgm}}\Varid{spdef};\Varid{pgm}{}\<[E]%
\ColumnHook
\end{hscode}\resethooks
}{\CPSpDef}
\end{mathpar}{}
\caption{Core Program Typing}
\label{fig:core-program-typing}
\end{figure*}

\begin{figure*}
\inlinemathhs
\ruletype{%
\begin{hscode}\SaveRestoreHook
\column{B}{@{}>{\hspre}l<{\hspost}@{}}\ColumnHook
\column{E}{@{}>{\hspre}l<{\hspost}@{}}\ColumnHook
\>[B]{}\Varid{e}\leadsto\Varid{e}{}\<[E]%
\ColumnHook
\end{hscode}\resethooks
}
\begin{mathpar}
\labelledinferrule{%
\begin{hscode}\SaveRestoreHook
\column{B}{@{}>{\hspre}l<{\hspost}@{}}\ColumnHook
\column{E}{@{}>{\hspre}l<{\hspost}@{}}\ColumnHook
\>[B]{}\Varid{e}_{1}\leadsto\Varid{e}_{1}'{}\<[E]%
\ColumnHook
\end{hscode}\resethooks
}{%
\begin{hscode}\SaveRestoreHook
\column{B}{@{}>{\hspre}l<{\hspost}@{}}\ColumnHook
\column{E}{@{}>{\hspre}l<{\hspost}@{}}\ColumnHook
\>[B]{}\Varid{e}_{1}~\Varid{e}_{2}\leadsto\Varid{e}_{1}'~\Varid{e}_{2}{}\<[E]%
\ColumnHook
\end{hscode}\resethooks
}{\DEAppL}
\and
\labelledinferrule{%
\begin{hscode}\SaveRestoreHook
\column{B}{@{}>{\hspre}l<{\hspost}@{}}\ColumnHook
\column{E}{@{}>{\hspre}l<{\hspost}@{}}\ColumnHook
\>[B]{}\Varid{e}_{2}\leadsto\Varid{e}_{2}'{}\<[E]%
\ColumnHook
\end{hscode}\resethooks
}{%
\begin{hscode}\SaveRestoreHook
\column{B}{@{}>{\hspre}l<{\hspost}@{}}\ColumnHook
\column{E}{@{}>{\hspre}l<{\hspost}@{}}\ColumnHook
\>[B]{}\Varid{e}_{1}~\Varid{e}_{2}\leadsto\Varid{e}_{1}~\Varid{e}_{2}'{}\<[E]%
\ColumnHook
\end{hscode}\resethooks
}{\DEAppR}
\and
\labelledinferrule{ }
{%
\begin{hscode}\SaveRestoreHook
\column{B}{@{}>{\hspre}l<{\hspost}@{}}\ColumnHook
\column{E}{@{}>{\hspre}l<{\hspost}@{}}\ColumnHook
\>[B]{}(\lambda \mathit{x}\colonT \tau.\Varid{e})~\Varid{v}\leadsto\Varid{e}[\Varid{v}/\mathit{x}]{}\<[E]%
\ColumnHook
\end{hscode}\resethooks
}{\DEBeta}
\and
\labelledinferrule{%
\begin{hscode}\SaveRestoreHook
\column{B}{@{}>{\hspre}l<{\hspost}@{}}\ColumnHook
\column{E}{@{}>{\hspre}l<{\hspost}@{}}\ColumnHook
\>[B]{}\Varid{e}\leadsto\Varid{e'}{}\<[E]%
\ColumnHook
\end{hscode}\resethooks
}{%
\begin{hscode}\SaveRestoreHook
\column{B}{@{}>{\hspre}l<{\hspost}@{}}\ColumnHook
\column{E}{@{}>{\hspre}l<{\hspost}@{}}\ColumnHook
\>[B]{}\Varid{e}~\langle \tau\rangle \leadsto\Varid{e'}~\langle \tau\rangle {}\<[E]%
\ColumnHook
\end{hscode}\resethooks
}{\DETApp}
\and
\labelledinferrule{ }
{%
\begin{hscode}\SaveRestoreHook
\column{B}{@{}>{\hspre}l<{\hspost}@{}}\ColumnHook
\column{E}{@{}>{\hspre}l<{\hspost}@{}}\ColumnHook
\>[B]{}(\Lambda \Varid{a}.\Varid{e})~\langle \tau\rangle \leadsto\Varid{e}[\tau/\Varid{a}]{}\<[E]%
\ColumnHook
\end{hscode}\resethooks
}{\DETBeta}
\and
\labelledinferrule{%
\begin{hscode}\SaveRestoreHook
\column{B}{@{}>{\hspre}l<{\hspost}@{}}\ColumnHook
\column{E}{@{}>{\hspre}l<{\hspost}@{}}\ColumnHook
\>[B]{}\Conid{SP}\leadsto\Conid{SP'}{}\<[E]%
\ColumnHook
\end{hscode}\resethooks
}{%
\begin{hscode}\SaveRestoreHook
\column{B}{@{}>{\hspre}l<{\hspost}@{}}\ColumnHook
\column{E}{@{}>{\hspre}l<{\hspost}@{}}\ColumnHook
\>[B]{}\qq{\Varid{e}}_{\Conid{SP}}\leadsto\qq{\Varid{e}}_{\Conid{SP'}}{}\<[E]%
\ColumnHook
\end{hscode}\resethooks
}{\DEQuote}
\end{mathpar}
\caption{Dynamic Semantics}
\label{fig:dyn-expr}
\end{figure*}

\begin{figure*}
\inlinemathhs
\ruletype{%
\begin{hscode}\SaveRestoreHook
\column{B}{@{}>{\hspre}l<{\hspost}@{}}\ColumnHook
\column{E}{@{}>{\hspre}l<{\hspost}@{}}\ColumnHook
\>[B]{}\Conid{SP}\leadsto\Conid{SP}{}\<[E]%
\ColumnHook
\end{hscode}\resethooks
}
\begin{mathpar}
\labelledinferrule{%
\begin{hscode}\SaveRestoreHook
\column{B}{@{}>{\hspre}l<{\hspost}@{}}\ColumnHook
\column{E}{@{}>{\hspre}l<{\hspost}@{}}\ColumnHook
\>[B]{}\Conid{SP}\leadsto\Conid{SP'}{}\<[E]%
\ColumnHook
\end{hscode}\resethooks
}{%
\begin{hscode}\SaveRestoreHook
\column{B}{@{}>{\hspre}l<{\hspost}@{}}\ColumnHook
\column{E}{@{}>{\hspre}l<{\hspost}@{}}\ColumnHook
\>[B]{}\Conid{SP},\Delta\vdash \Varid{sp}\colonT\tau\mathrel{=}\Varid{e}\leadsto\Conid{SP'},\Delta\vdash \Varid{sp}\colonT\tau\mathrel{=}\Varid{e}{}\<[E]%
\ColumnHook
\end{hscode}\resethooks
}{\DSPBody}
\and
\labelledinferrule{%
\begin{hscode}\SaveRestoreHook
\column{B}{@{}>{\hspre}l<{\hspost}@{}}\ColumnHook
\column{E}{@{}>{\hspre}l<{\hspost}@{}}\ColumnHook
\>[B]{}\Varid{e}\leadsto\Varid{e'}{}\<[E]%
\ColumnHook
\end{hscode}\resethooks
}{%
\begin{hscode}\SaveRestoreHook
\column{B}{@{}>{\hspre}l<{\hspost}@{}}\ColumnHook
\column{E}{@{}>{\hspre}l<{\hspost}@{}}\ColumnHook
\>[B]{}\Conid{SP},\Delta\vdash \Varid{sp}\colonT\tau\mathrel{=}\Varid{e}\leadsto\Conid{SP},\Delta\vdash \Varid{sp}\colonT\tau\mathrel{=}\Varid{e'}{}\<[E]%
\ColumnHook
\end{hscode}\resethooks
}{\DSPCons}
\end{mathpar}
\caption{Dynamic Semantics for Splice Environment}
\label{fig:dyn-splice-env}
\end{figure*}

\begin{figure*}
\begin{minipage}[t]{.5\linewidth}
\inlinemathhs
\ruletype{%
\begin{hscode}\SaveRestoreHook
\column{B}{@{}>{\hspre}l<{\hspost}@{}}\ColumnHook
\column{E}{@{}>{\hspre}l<{\hspost}@{}}\ColumnHook
\>[B]{}\Varid{def}\leadsto\Varid{def}{}\<[E]%
\ColumnHook
\end{hscode}\resethooks
}
\begin{mathpar}
\labelledinferrule{%
\begin{hscode}\SaveRestoreHook
\column{B}{@{}>{\hspre}l<{\hspost}@{}}\ColumnHook
\column{E}{@{}>{\hspre}l<{\hspost}@{}}\ColumnHook
\>[B]{}\Varid{e}\leadsto\Varid{e'}{}\<[E]%
\ColumnHook
\end{hscode}\resethooks
}{%
\begin{hscode}\SaveRestoreHook
\column{B}{@{}>{\hspre}l<{\hspost}@{}}\ColumnHook
\column{E}{@{}>{\hspre}l<{\hspost}@{}}\ColumnHook
\>[B]{}\mathbf{def}~\Varid{k} \colonT \tau\mathrel{=}\Varid{e}\leadsto\mathbf{def}~\Varid{k} \colonT \tau\mathrel{=}\Varid{e'}{}\<[E]%
\ColumnHook
\end{hscode}\resethooks
}{\DDef}
\end{mathpar}
\end{minipage}%
\begin{minipage}[t]{.5\linewidth}
\inlinemathhs
\ruletype{%
\begin{hscode}\SaveRestoreHook
\column{B}{@{}>{\hspre}l<{\hspost}@{}}\ColumnHook
\column{E}{@{}>{\hspre}l<{\hspost}@{}}\ColumnHook
\>[B]{}\Varid{spdef}\leadsto\Varid{spdef}{}\<[E]%
\ColumnHook
\end{hscode}\resethooks
}
\begin{mathpar}
\labelledinferrule{%
\begin{hscode}\SaveRestoreHook
\column{B}{@{}>{\hspre}l<{\hspost}@{}}\ColumnHook
\column{E}{@{}>{\hspre}l<{\hspost}@{}}\ColumnHook
\>[B]{}\Varid{e}\leadsto\Varid{e'}{}\<[E]%
\ColumnHook
\end{hscode}\resethooks
}{%
\begin{hscode}\SaveRestoreHook
\column{B}{@{}>{\hspre}l<{\hspost}@{}}\ColumnHook
\column{E}{@{}>{\hspre}l<{\hspost}@{}}\ColumnHook
\>[B]{}\mathbf{spdef}~\Gamma \vdashn{\Varid{n}}\Varid{sp} \colonT \tau\mathrel{=}\Varid{e}\leadsto\mathbf{spdef}~\Gamma \vdashn{\Varid{n}}\Varid{sp} \colonT \tau\mathrel{=}\Varid{e'}{}\<[E]%
\ColumnHook
\end{hscode}\resethooks
}
{\DSPDef}
\end{mathpar}
\end{minipage}
\caption{Dynamic Semantics for Definitions}
\label{fig:dyn-defn}
\end{figure*}
\begin{figure*}
\inlinemathhs
\ruletype{%
\begin{hscode}\SaveRestoreHook
\column{B}{@{}>{\hspre}l<{\hspost}@{}}\ColumnHook
\column{E}{@{}>{\hspre}l<{\hspost}@{}}\ColumnHook
\>[B]{}\Varid{pgm}\leadsto\Varid{pgm}{}\<[E]%
\ColumnHook
\end{hscode}\resethooks
}
\begin{mathpar}
\labelledinferrule{%
\begin{hscode}\SaveRestoreHook
\column{B}{@{}>{\hspre}l<{\hspost}@{}}\ColumnHook
\column{E}{@{}>{\hspre}l<{\hspost}@{}}\ColumnHook
\>[B]{}\Varid{def}\leadsto\Varid{def'}{}\<[E]%
\ColumnHook
\end{hscode}\resethooks
}{%
\begin{hscode}\SaveRestoreHook
\column{B}{@{}>{\hspre}l<{\hspost}@{}}\ColumnHook
\column{E}{@{}>{\hspre}l<{\hspost}@{}}\ColumnHook
\>[B]{}\Varid{def};\Varid{pgm}\leadsto\Varid{def'};\Varid{pgm}{}\<[E]%
\ColumnHook
\end{hscode}\resethooks
}{\DPDef}
\and
\labelledinferrule{
\begin{hscode}\SaveRestoreHook
\column{B}{@{}>{\hspre}l<{\hspost}@{}}\ColumnHook
\column{E}{@{}>{\hspre}l<{\hspost}@{}}\ColumnHook
\>[B]{}\Varid{spdef}\leadsto\Varid{spdef'}{}\<[E]%
\ColumnHook
\end{hscode}\resethooks
}{%
\begin{hscode}\SaveRestoreHook
\column{B}{@{}>{\hspre}l<{\hspost}@{}}\ColumnHook
\column{E}{@{}>{\hspre}l<{\hspost}@{}}\ColumnHook
\>[B]{}\Varid{spdef};\Varid{pgm}\leadsto\Varid{spdef'};\Varid{pgm}{}\<[E]%
\ColumnHook
\end{hscode}\resethooks
}{\DPSPDef }
          \and
\labelledinferrule{%
\begin{hscode}\SaveRestoreHook
\column{B}{@{}>{\hspre}l<{\hspost}@{}}\ColumnHook
\column{E}{@{}>{\hspre}l<{\hspost}@{}}\ColumnHook
\>[B]{}\Varid{e}\leadsto\Varid{e'}{}\<[E]%
\ColumnHook
\end{hscode}\resethooks
}{%
\begin{hscode}\SaveRestoreHook
\column{B}{@{}>{\hspre}l<{\hspost}@{}}\ColumnHook
\column{E}{@{}>{\hspre}l<{\hspost}@{}}\ColumnHook
\>[B]{}\Varid{e}\colonT \tau\leadsto\Varid{e'}\colonT \tau{}\<[E]%
\ColumnHook
\end{hscode}\resethooks
}{\DPMain }
\and
\labelledinferrule{ }
{%
\begin{hscode}\SaveRestoreHook
\column{B}{@{}>{\hspre}l<{\hspost}@{}}\ColumnHook
\column{E}{@{}>{\hspre}l<{\hspost}@{}}\ColumnHook
\>[B]{}\mathbf{def}~\Varid{k} \colonT \tau\mathrel{=}\Varid{v};\Varid{pgm}\leadsto\Varid{pgm}[\Varid{v}/\Varid{k}]{}\<[E]%
\ColumnHook
\end{hscode}\resethooks
}{\DPDefBeta }
\and
\labelledinferrule{ }
{%
\begin{hscode}\SaveRestoreHook
\column{B}{@{}>{\hspre}l<{\hspost}@{}}\ColumnHook
\column{E}{@{}>{\hspre}l<{\hspost}@{}}\ColumnHook
\>[B]{}\mathbf{spdef}~\Delta \vdashn{\Varid{n}}\Varid{sp} \colonT \tau\mathrel{=}\qq{\Varid{e}}_{\Conid{SP}};\Varid{pgm}\leadsto\Varid{pgm}[\Varid{e}[\Conid{SP}]/\Varid{sp}]{}\<[E]%
\ColumnHook
\end{hscode}\resethooks
}{\DPSPDefBeta}
\end{mathpar}
\caption{Dynamic Semantics for Program}
\label{fig:dyn-pgm}
\end{figure*}






\section{Elaboration}
\label{sec:elaboration}

In this section we describe the process of elaboration from the source language
to the core language. Elaboration starts from a well-typed term
in the source language and hence is defined by induction
over the typing derivation tree (Figure~\ref{fig:expr-elaboration}).

\subsection{Elaboration procedure}

There are three key aspects of the elaboration procedure:
\matt{Add figure references}
\begin{enumerate}
\item{Splices at positive levels are removed in favour of a splice environment. The elaboration process returns
a splice environment which is attached to the quotation form.}
\item{Splices at non-positive levels are elaborated to top-level splice definitions, which are bound
prior to the expression within which they were originally contained.}
\item{Type class constraints are converted to explicit dictionary passing. We describe how to
understand the new constraint form \ensuremath{\Conid{CodeC}\;\Conid{C}} in terms of quotation.}
\end{enumerate}

In order to support elaboration, all implicit evidence from
the program theory and local environment are annotated with variables. The idea
is that in the elaborated program, the evidence for each particular construct will
be bound to a variable of that name. Top-level evidence by top-level variables and
local evidence by $\lambda$-bound variables. The modifications to the
typing rules for the source language syntax necessary in order to explain
elaboration are highlighted by a \colorbox{gray!25}{grey} box.
\begingroup
\invisiblecomments
\aligncolumn{132}{r}
\begin{hscode}\SaveRestoreHook
\column{B}{@{}>{\hspre}l<{\hspost}@{}}\ColumnHook
\column{18}{@{}>{\hspre}c<{\hspost}@{}}\ColumnHook
\column{18E}{@{}l@{}}\ColumnHook
\column{23}{@{}>{\hspre}l<{\hspost}@{}}\ColumnHook
\column{127}{@{}>{\hspre}l<{\hspost}@{}}\ColumnHook
\column{132}{@{}>{\hspre}l<{\hspost}@{}}\ColumnHook
\column{E}{@{}>{\hspre}l<{\hspost}@{}}\ColumnHook
\>[B]{}\Gamma {}\<[18]%
\>[18]{}\mathbin{::=}{}\<[18E]%
\>[23]{}\bullet \mid \Gamma ,\mathit{x}\colonT (\tau ,\Varid{n})\mid \Gamma ,\Varid{a}\mid \Gamma , \ghl{\Varid{ev}\colonT\!\!}(\Conid{C},\Varid{n}){}\<[127]%
\>[127]{}\quad \;{}\<[132]%
\>[132]{}\mbox{\onelinecomment  type environment}{}\<[E]%
\\
\>[B]{}\Conid{P}{}\<[18]%
\>[18]{}\mathbin{::=}{}\<[18E]%
\>[23]{}\bullet \mid \Conid{P}, \ghl{\Varid{ev}\colonT\!\!}(\forall a.\overline{\Conid{C}}\Rightarrow \Conid{C})\mid \Conid{P},\Varid{k}\colonT\sigma {}\<[127]%
\>[127]{}\quad {}\<[132]%
\>[132]{}\mbox{\onelinecomment  program environment}{}\<[E]%
\ColumnHook
\end{hscode}\resethooks
\endgroup
The judgement \ensuremath{\Conid{P};\Gamma  \vdashn{\Varid{n}} \Varid{e} : \tau\ghl{\leadsto\Varid{t}~\vert~\Conid{TSP}}} states
that in program theory \ensuremath{\Conid{P}} and environment \ensuremath{\Gamma }, the expression~\ensuremath{\Varid{e}} has type~\ensuremath{\tau}
at level~\ensuremath{\Varid{n}} and elaborates to term \ensuremath{\Varid{t}} whilst producing splices \ensuremath{\Conid{TSP}}.

\subsection{Splice Elaboration}

The \ensuremath{\Conid{TSP}} is a function from a level to a splice environment \ensuremath{\Conid{SP}}.
In many rules, we perform level-pointwise union of produced splices,
written $\cup$.

\nw{Do we need a more formal description in a figure or as an equation? Perhaps not.}
During elaboration, all splices are initially added to the \ensuremath{\Conid{TSP}} at
the level of their contents, so a splice that occurs at level~\ensuremath{\Varid{n}} is added
at level~\ensuremath{\Varid{n}\mathbin{-}\mathrm{1}} by means of the operation~$\cup_{n-1}$ as in rule~\ESplice.

What happens with the splices contained in the~\ensuremath{\Conid{TSP}} depends on their
level. If they occur at a positive level, they will be bound by a surrounding
quotation in rule~\EQuote. The notation \ensuremath{\Conid{TSP}_{\Varid{n}}} denotes the
projection of the splices contained in~\ensuremath{\Conid{TSP}} at level~\ensuremath{\Varid{n}}. They become part
of the splice environment associated with the quotation. Via \ensuremath{\lfloor\Conid{TSP}\rfloor^{\Varid{n}}},
we then truncate~\ensuremath{\Conid{TSP}} so that it is empty at level~\ensuremath{\Varid{n}} and above.


If a splice
occurs at non-positive level, it is a top-level splice and will become a top-level
splice definition in rules~\Def or~\Inst, in such a way that the splice definitions
are made prior to the value definition which yielded them.
The splice definitions are created by the \ensuremath{\Varid{collapse}} judgement (Figure~\ref{fig:collapse-definition}), which
takes a splice environment returned by the expression elaboration judgement and
creates top-level splice declarations for each negative splice which appeared inside a term.
To guarantee a stage-correct
execution, the splices are inserted in order of their level.

We maintain the invariant on the \ensuremath{\Conid{TSP}} where it only contains
splices of levels prior to the current level of the judgement. Therefore,
when the judgement level decreases as in \EQuote, the splices for that level
are removed from the splice environment and bound at the quotation.
As the top-level of the program is at level~\ensuremath{\mathrm{0}}, the splice environment
returned by an elaboration judgement at level~\ensuremath{\mathrm{0}} will only contain splices at negative
levels, which is why the appeal to the \ensuremath{\Varid{collapse}} function starts from level~\ensuremath{\mathbin{-}\mathrm{1}}.

\subsection{Constraint Elaboration}
\matt{example}
The second point of interest are the constraint elaboration rules given in
Figure~\ref{fig:constraint-elaboration}. Since in our language type classes have just
a single method, we use the function corresponding to the method itself
as evidence for a class instance in rule~\Inst.


Constraints of the new constraint form
\ensuremath{\Conid{CodeC}\;\Conid{C}} are elaborated into values of type \ensuremath{\Conid{Code}\;\tau}. Therefore
inspecting the entailment elaboration form \CDecr and \CIncr must be understood
in terms of quotation. The \CDecr entailment rule is implemented by a simple
quotation and thus similar to \EQuote. The \CIncr rule is conceptually implemented using a splice, but
as the core language does not contain splices it is understood by adding a new definition
to the splice environment, which mirrors \ESplice. These rules explains the necessity of
level-indexing constraints in the source language; the elaboration would not
be well-staged if the stage discipline was not enforced.

The remainder of the elaboration semantics which elaborate the simple terms and
constraint forms are fairly routine.

%

\subsection{Example}

As an example of elaboration, let us consider the expression \ensuremath{\$(\Varid{c}_{1}')}
where \ensuremath{\Varid{c}_{1}'\mathrel{=}\qq{\Varid{show}}} from Example~\ref{ex:c1}. There are two points of interest
here: there is a top-level splice which will be floated to a top-level splice definition,
and the \ensuremath{\Conid{CodeC}\;(\Conid{Show}\;\Varid{a})} constraint of \ensuremath{\Varid{c}_{1}'} must be elaborated into quoted evidence
using rule \CDecr.

We assume that the program environment contains \ensuremath{\Varid{c}_{1}'}:
\begin{hscode}\SaveRestoreHook
\column{B}{@{}>{\hspre}l<{\hspost}@{}}\ColumnHook
\column{E}{@{}>{\hspre}l<{\hspost}@{}}\ColumnHook
\>[B]{}\Conid{P}\mathrel{=}\bullet ,\Varid{c}_{1}'\colonT\forall \Varid{a}.\Conid{CodeC}\;(\Conid{Show}~\Varid{a})\Rightarrow \Conid{Code}\;(\Varid{a}\to \Conid{String}){}\<[E]%
\ColumnHook
\end{hscode}\resethooks
As our program comprises a single main expression, we have to use rule~\PExpr. This rule
requires us to first elaborate the expression itself at level~\ensuremath{\mathrm{0}} in an empty type environment. We obtain
\begin{hscode}\SaveRestoreHook
\column{B}{@{}>{\hspre}l<{\hspost}@{}}\ColumnHook
\column{E}{@{}>{\hspre}l<{\hspost}@{}}\ColumnHook
\>[B]{}\Conid{P};\bullet  \vdashn{\mathrm{0}} \$(\Varid{c}_{1}') : \forall \Varid{a}.\Conid{Show}~\Varid{a}\Rightarrow \Varid{a}\to \Conid{String}{}\<[E]%
\\
\>[B]{}\quad\ghl{\leadsto \Lambda \Varid{a}.\lambda \Varid{ev}\colonT \Varid{a}\to \Conid{String}.\Varid{sp}~\vert~\{\Delta\vdash\Varid{sp}\colonT\Varid{a}\to \Conid{String}=\Varid{c}_{1}'~\langle \Varid{a}\rangle ~\qq{\Varid{ev}}_{\bullet }\} \cup_{\mathbin{-}\mathrm{1}}\bullet }{}\<[E]%
\ColumnHook
\end{hscode}\resethooks
where
\begin{hscode}\SaveRestoreHook
\column{B}{@{}>{\hspre}l<{\hspost}@{}}\ColumnHook
\column{E}{@{}>{\hspre}l<{\hspost}@{}}\ColumnHook
\>[B]{}\Delta\mathrel{=}\bullet ,\Varid{a},\Varid{ev}\colonT (\Varid{a}\to \Conid{String},\mathrm{0}){}\<[E]%
\ColumnHook
\end{hscode}\resethooks
The splice point \ensuremath{\Varid{sp}} has been introduced for the splice via rule \ESplice.
Evidence for the \ensuremath{\Conid{Show}} constraint is introduced into the type environment at level~\ensuremath{\mathrm{0}}
via rule \ECAbs. It is captured in \ensuremath{\Delta} for use in the splice. Because the use of
the evidence occurs at level~\ensuremath{\mathbin{-}\mathrm{1}} and the required constraint is \ensuremath{\Conid{CodeC}\;\Conid{Show}}, rule \CDecr
is used to quote the evidence. The splice environment attached to the quote is empty,
because there are no further splices.

Back to rule \PExpr, we furthermore obtain that the resulting main expression is of the form
\begin{hscode}\SaveRestoreHook
\column{B}{@{}>{\hspre}l<{\hspost}@{}}\ColumnHook
\column{E}{@{}>{\hspre}l<{\hspost}@{}}\ColumnHook
\>[B]{}\Varid{p}\mathrel{=}\Lambda \Varid{a}.\lambda \Varid{ev}\colonT \Varid{a}\to \Conid{String}.\Varid{sp}\colonT \forall \Varid{a}.(\Varid{a}\to \Conid{String})\to \Varid{a}\to \Conid{String}{}\<[E]%
\ColumnHook
\end{hscode}\resethooks
The type of \ensuremath{\Varid{p}} results from elaborating the original \ensuremath{\Conid{Show}\;\Varid{a}} constraint to the type of its method \ensuremath{\Varid{a}\to \Conid{String}}
via the rule \CTC. We can now look at \ensuremath{\Varid{collapse}} and observe that it will extract the one splice
at level~\ensuremath{\mathbin{-}\mathrm{1}} passed to it into a top-level splice definition that ends up before~\ensuremath{\Varid{d}}, the result
being
\begin{hscode}\SaveRestoreHook
\column{B}{@{}>{\hspre}l<{\hspost}@{}}\ColumnHook
\column{E}{@{}>{\hspre}l<{\hspost}@{}}\ColumnHook
\>[B]{}\mathbf{spdef}~\Delta \vdashn{\mathbin{-}\mathrm{1}}\Varid{sp} \colonT \Varid{a}\to \Conid{String}\mathrel{=}\Varid{c}_{1}'~\langle \Varid{a}\rangle ~\qq{\Varid{ev}}_{\bullet };\Varid{p}{}\<[E]%
\ColumnHook
\end{hscode}\resethooks


\noindent
The operational semantics for the language evaluates the definition of \ensuremath{\Varid{sp}}
first to a quotation before the quoted expression is substituted into the remainder
of the program so that evaluation can continue.






\begin{figure*}
\inlinemathhs
\ruletype{%
\begin{hscode}\SaveRestoreHook
\column{B}{@{}>{\hspre}l<{\hspost}@{}}\ColumnHook
\column{E}{@{}>{\hspre}l<{\hspost}@{}}\ColumnHook
\>[B]{}\Varid{collapse}(\Varid{n},\Conid{TSP},\Varid{pgm})\mathrel{=}\Varid{ds}{}\<[E]%
\ColumnHook
\end{hscode}\resethooks
}
\begin{mathpar}
\labelledinferrule{ }{%
\begin{hscode}\SaveRestoreHook
\column{B}{@{}>{\hspre}l<{\hspost}@{}}\ColumnHook
\column{E}{@{}>{\hspre}l<{\hspost}@{}}\ColumnHook
\>[B]{}\Varid{collapse}(\Varid{n},\bullet ,\Varid{pgm})\mathrel{=}\Varid{pgm}{}\<[E]%
\ColumnHook
\end{hscode}\resethooks
}
{\CEmpty}
          \and
\labelledinferrule{%
\begin{hscode}\SaveRestoreHook
\column{B}{@{}>{\hspre}l<{\hspost}@{}}\ColumnHook
\column{E}{@{}>{\hspre}l<{\hspost}@{}}\ColumnHook
\>[B]{}\Varid{ds}\mathrel{=}\Varid{collapse}(\Varid{n}\mathbin{-}\mathrm{1},\lfloor\Conid{TSP}\rfloor^{\Varid{n}},\overline{\Varid{spdef\char95 i}};\Varid{pgm}){}\<[E]%
\ColumnHook
\end{hscode}\resethooks
\and
\begin{hscode}\SaveRestoreHook
\column{B}{@{}>{\hspre}l<{\hspost}@{}}\ColumnHook
\column{E}{@{}>{\hspre}l<{\hspost}@{}}\ColumnHook
\>[B]{}\overline{\Varid{spdef\char95 i}\mathrel{=}\mathbf{spdef}~\Delta_i \vdashn{\Varid{n}}\Varid{sp}_{\Varid{i}} \colonT \tau_{\Varid{i}}\mathrel{=}\Varid{e}_{\Varid{i}}}{}\<[E]%
\ColumnHook
\end{hscode}\resethooks
\\
\and
\begin{hscode}\SaveRestoreHook
\column{B}{@{}>{\hspre}l<{\hspost}@{}}\ColumnHook
\column{E}{@{}>{\hspre}l<{\hspost}@{}}\ColumnHook
\>[B]{}\Conid{TSP}_{\Varid{n}}\mathrel{=}\overline{\Delta_{\Varid{i}} \vdash \Varid{sp}_{\Varid{i}} \colonT \tau_{\Varid{i}}\mathrel{=}\Varid{e}_{\Varid{i}}}{}\<[E]%
\ColumnHook
\end{hscode}\resethooks
}
{%
\begin{hscode}\SaveRestoreHook
\column{B}{@{}>{\hspre}l<{\hspost}@{}}\ColumnHook
\column{E}{@{}>{\hspre}l<{\hspost}@{}}\ColumnHook
\>[B]{}\Varid{collapse}(\Varid{n},\Conid{TSP},\Varid{pgm})\mathrel{=}\Varid{ds}{}\<[E]%
\ColumnHook
\end{hscode}\resethooks
}
{\CStrip}
\end{mathpar}
\caption{Definition of $collapse$}
\label{fig:collapse-definition}
\end{figure*}
\begingroup
\def\ghl#1{#1}\let\!\relax





\begin{figure*}
\inlinemathhs
\ruletype{%
\begin{hscode}\SaveRestoreHook
\column{B}{@{}>{\hspre}l<{\hspost}@{}}\ColumnHook
\column{E}{@{}>{\hspre}l<{\hspost}@{}}\ColumnHook
\>[B]{}\ghl{\Gamma \leadsto\Delta}{}\<[E]%
\ColumnHook
\end{hscode}\resethooks
}
\begin{mathpar}
\labelledinferrule{ }
{%
\begin{hscode}\SaveRestoreHook
\column{B}{@{}>{\hspre}l<{\hspost}@{}}\ColumnHook
\column{E}{@{}>{\hspre}l<{\hspost}@{}}\ColumnHook
\>[B]{}\ghl{\bullet \leadsto\bullet }{}\<[E]%
\ColumnHook
\end{hscode}\resethooks
}{TE\_Empty}
\and
\labelledinferrule{%
\begin{hscode}\SaveRestoreHook
\column{B}{@{}>{\hspre}l<{\hspost}@{}}\ColumnHook
\column{E}{@{}>{\hspre}l<{\hspost}@{}}\ColumnHook
\>[B]{}\ghl{\Gamma \leadsto\Delta}{}\<[E]%
\ColumnHook
\end{hscode}\resethooks
\and
\begin{hscode}\SaveRestoreHook
\column{B}{@{}>{\hspre}l<{\hspost}@{}}\ColumnHook
\column{E}{@{}>{\hspre}l<{\hspost}@{}}\ColumnHook
\>[B]{}\Gamma \vdash_{\mathsf{ty}}\tau\ghl{\leadsto\tau^\prime}{}\<[E]%
\ColumnHook
\end{hscode}\resethooks
}{%
\begin{hscode}\SaveRestoreHook
\column{B}{@{}>{\hspre}l<{\hspost}@{}}\ColumnHook
\column{E}{@{}>{\hspre}l<{\hspost}@{}}\ColumnHook
\>[B]{}\ghl{\Gamma ,\mathit{x}\colonT (\tau,\Varid{n})\leadsto\Delta,\mathit{x}\colonT (\tau^\prime,\Varid{n})}{}\<[E]%
\ColumnHook
\end{hscode}\resethooks
}{TE\_Var}
\and
\labelledinferrule{%
\begin{hscode}\SaveRestoreHook
\column{B}{@{}>{\hspre}l<{\hspost}@{}}\ColumnHook
\column{E}{@{}>{\hspre}l<{\hspost}@{}}\ColumnHook
\>[B]{}\ghl{\Gamma \leadsto\Delta}{}\<[E]%
\ColumnHook
\end{hscode}\resethooks
}{%
\begin{hscode}\SaveRestoreHook
\column{B}{@{}>{\hspre}l<{\hspost}@{}}\ColumnHook
\column{E}{@{}>{\hspre}l<{\hspost}@{}}\ColumnHook
\>[B]{}\ghl{\Gamma ,\Varid{a}\leadsto\Delta,\Varid{a}}{}\<[E]%
\ColumnHook
\end{hscode}\resethooks
}{TE\_TyVar}
\and
\labelledinferrule{%
\begin{hscode}\SaveRestoreHook
\column{B}{@{}>{\hspre}l<{\hspost}@{}}\ColumnHook
\column{E}{@{}>{\hspre}l<{\hspost}@{}}\ColumnHook
\>[B]{}\ghl{\Gamma \leadsto\Delta}{}\<[E]%
\ColumnHook
\end{hscode}\resethooks
\and
\begin{hscode}\SaveRestoreHook
\column{B}{@{}>{\hspre}l<{\hspost}@{}}\ColumnHook
\column{E}{@{}>{\hspre}l<{\hspost}@{}}\ColumnHook
\>[B]{}\Gamma \vdash_{\mathsf{ct}}\Conid{C}\ghl{\leadsto\tau}{}\<[E]%
\ColumnHook
\end{hscode}\resethooks
}{%
\begin{hscode}\SaveRestoreHook
\column{B}{@{}>{\hspre}l<{\hspost}@{}}\ColumnHook
\column{E}{@{}>{\hspre}l<{\hspost}@{}}\ColumnHook
\>[B]{}\ghl{\Gamma , \ghl{\Varid{ev}\colonT\!\!}(\Conid{C},\Varid{n})\leadsto\Delta,\Varid{ev}\colonT (\tau,\Varid{n})}{}\<[E]%
\ColumnHook
\end{hscode}\resethooks
}{TE\_Ctx}
\end{mathpar}
\caption{\ghl{\text{Elaboration of Type Environments}}}
\end{figure*}
\endgroup

\section{Pragmatic Considerations}
\label{sec:future}

Now that the formal developments are complete and the relationship between the
source language and core language has been established by the elaboration
semantics, it is time to consider how our formalism interacts with
other language extensions, and to discuss implementation issues.

\subsection{Integration into GHC}
\label{ref:implementation}

The implementation and integration of this specification into GHC's
sophisticated architecture naturally requires certain design decisions to be
made, which we now discuss.

\nw{I have removed the para talking about the fact that typed quotations and the frontend need to be modified. I think that's quite obvious to anybody reading this paper!}

\subsubsection{Type Inference}
\label{ref:inference}

Type inference for our new constraint form should be straightforward to integrate
into the constraint solving algorithm used in GHC~\cite{vytiniotis2011outsidein}.
The key modification is to keep track of the level of constraints and only solve goals with
evidence at the right level. If there is no evidence available for a constraint at
the correct level, then either the \CIncr or \CDecr rule can be invoked
in order to correct the level of necessary evidence.

\al{I do not understand the following sentence.}
\nw{me neither}
\matt{Are we advocating the removal of this sentence or rewording? The point
is supposed to be that if you need to generalise at level 0 and require a constraint at
level 1 you should generate a \ensuremath{\Conid{CodeC}\;\Conid{C}} constraint rather than \ensuremath{\Conid{C}}}
\matt{commented out}

\subsubsection{Substituting Types}
Substituting inside quotations poses some implementation challenges depending on the
quotation representation.
In previous implementations which used low-level
representations to represent quotations~\cite{pickering2019multi, roubinchtein2015irmeta}
the solution was to maintain a separate environment for free variables which can
be substituted into without having to implement substitution in terms of the low-level
representation. The idea is then that by the time the low-level representation is evaluated
all the variables bound in the environment are already added to the environment and
so computation can proceed as normal when it encounters what was previously a free variable.

Therefore we should treat every free type variable in a quotation as a splice
point, create an environment to attach to the quotation which maps the splice
point to the variable of the name and then when the quotation is interpreted
substitute the type into the quotation.

\subsubsection{Erasing Types}

The operational semantics of System F do not depend on the type information and
so the types can be erased before
evaluation. It is this observation that leads us to accept Example~\ref{ex:tv1}.
However, there is another
point in the design space -- we could have elaborated to an erased version of System
F where types were replaced by placeholders. This would
save us the complications of having to substitute types inside quotations.
However, the option of maintaining the type information is
important for practical purposes. GHC has an optional internal typechecking phase
called core lint which verifies the correctness of code generation and optimisation, it
would be a shame to lose this pass in any program which used metaprogramming.
In a language where the type information dwarfs the runtime content of terms,
it would be desirable to also explore the option to store erased or partially erased
terms~\cite{brady2003inductive,jay2008scrap}.

\subsubsection{Cross-Stage Persistence}

Earlier work on cross-stage persistence~\citep{pickering2019multi} has
suggested
that implementing cross-stage persistence for instance dictionaries should
be possible because a dictionary was ultimately a collection of top-level functions
so some special logic could be implemented in the compiler to lift a dictionary.

\al{Or perhaps even more drastically, a dictionary embedded into a GADT?}
\matt{Yes, but not sure how to weave that in}
Additional complexity arises when functions with
local constraints are passed an arbitrary dictionary which could have been
constructed from other dictionaries, which in turn come from dictionaries. At
the point the dictionary is passed, the required information about its
structure has been lost so it is
impossible to interpret into a future stage.
A solution to this could be to use a different evidence form which passes the
derivation tree for a constraint to a function as evidence before the function
constructs the required dictionary at the required stage. This would increase
the runtime overhead of using type classes and would not be practical to implement.

\subsection{Interaction with Existing Features}

So far we have considered how metaprogramming interacts with qualified types,
but of course there are other features that are specific to GHC that need to be considered.

\subsubsection{GADTs}
Local constraints can be introduced by pattern matching on a GADT.
For simplicity our calculus did not include GADTs or local constraints but
they require similar treatment to other constraints introduced locally. The
constraint solver needs to keep track of the level that a GADT pattern
match introduces a constraint and ensure that the constraint is only used at
that level.

Note that this notion of a ``level'' is the stage of program execution where the
constraint is introduced and not the same idea of a level the constraint solver
uses to prevent existentially quantified type variables escaping their scope.
Each nested implication constraint increases the level so type variables
introduced in an inner scope are forbidden from unifying with type variables
which are introduced at a previous level.
\matt{Cite? http://okmij.org/ftp/ML/generalization.html}

\subsubsection{Quantified Constraints}

The quantified constraints extension~\cite{bottu2017quantified}
relaxes the form of the constraint schemes
allowed in method contexts to also allow the quantified and implication forms
which in our calculus are restricted to top-level axioms. Under this restriction
there are some questions about how the \ensuremath{\Conid{CodeC}} constraint form should
interact especially with implication constraints. In particular, whether
constraint entailment should deduce that \ensuremath{\Conid{CodeC}\;(\Conid{C}_{\mathrm{1}}\Rightarrow \Conid{C}_{\mathrm{2}})} entails
\ensuremath{\Conid{CodeC}\;\Conid{C}_{\mathrm{1}}\Rightarrow \Conid{CodeC}\;\Conid{C}_{\mathrm{2}}} or the inverse and what consequences this has for
type inference involving these more complicated constraint forms.

\subsection{Interaction with Future Features}

GHC is constantly being improved and extended to encompass new and ambitious
features,
and so we consider how metaprogramming should interact with features that are
currently in the pipeline.

\subsubsection{Impredicativity}
\label{ref:impredicativity}

For a number of the examples that we have discussed in this paper,
an alternative would be to use impredicative instantiation to more
precisely express the binding position of a constraint.
For instance, the function \ensuremath{\Varid{c}_{1}\mathbin{::}\Conid{Show}\;\Varid{a}\Rightarrow \Conid{Code}\;(\Varid{a}\to \Conid{String})} from Example~\ref{ex:c1} might instead have been expressed in the following manner:
\begin{hscode}\SaveRestoreHook
\column{B}{@{}>{\hspre}l<{\hspost}@{}}\ColumnHook
\column{E}{@{}>{\hspre}l<{\hspost}@{}}\ColumnHook
\>[B]{}\Varid{c}_{1}'\mathbin{::}\Conid{Code}\;(\forall \Varid{a}\hsforall .\Conid{Show}\;\Varid{a}\Rightarrow \Varid{a}\to \Conid{String}){}\<[E]%
\\
\>[B]{}\Varid{c}_{1}'\mathrel{=}\qq{\Varid{show}}{}\<[E]%
\ColumnHook
\end{hscode}\resethooks
The type of \ensuremath{\Varid{c}_{1}'} now binds and uses the \ensuremath{\Conid{Show}\;\Varid{a}} constraint at level $1$ without the use of \ensuremath{\Conid{CodeC}}.

However, despite many attempts~\cite{jones2007practical}, GHC has never properly
supported impredicative instantiation due to complications with type
inference.
Recent work \citep{serrano2018impred, serrano2020quicklook} has
proposed the inclusion of restricted impredicative instantiations,
and these would accept \ensuremath{\Varid{c}_{1}'}.
In any case, impredicativity is not a silver bullet, and there is still a need for
the \ensuremath{\Conid{CodeC}} constraint form. Without the \ensuremath{\Conid{CodeC}} constraint
form there is no way to manipulate ``open'' constraints. That is, constraints which
elaborate to quotations containing free variables.

\al{I'm not sure if I concretely get what you mean by ``open'' constraints.
In my mind, the primary need for \ensuremath{\Conid{CodeC}} arises from anything that abstracts
over constraints. E.g. classes might need \ensuremath{\Conid{CodeC}} as instance context. Or
\ensuremath{\Conid{All}} from generics-sop.}

Experience has taught us that writing code generators which manipulate open
terms is a lot more convenient than working with only closed terms. We predict
that the same will be true of working with the delayed constraint form as well.
In particular, features such as super classes, instance contexts and type families
can be used naturally with \ensuremath{\Conid{CodeC}} constraints.
So whilst relaxing the impredicativity restriction will have some positive
consequences to the users of Typed Template Haskell, it does not supersede our
design but rather acts as a supplement.

\nw{Would be good to show an example of this specifically here}

\subsubsection{Dependent Haskell}

Our treatment of type variables is inspired by the in-built phase distinction
of System F. As Haskell barrels at an impressive rate to a dependently
typed language~\cite{gundry2013type, eisenberg2016dependent},
the guarantees of the phase distinction will be lost in some
cases. At this point it will be necessary to revise the specification in order
to account for the richer phase structure.

The specification for Dependent Haskell~\cite{weirich2017specification} introduces
the so-called relevance quantifier in order to distinguish between relevant and
irrelevant variables. The irrelevant quantifier is intended to model a form
of parametric polymorphism like the $\forall$ in System F, the relevant quantifier
is written as $\Pi$ after the dependent quantifier from dependent type theory.

Perhaps it is sound to modify their system in order to enforce the stage discipline
for relevant type variables not irrelevant ones. It may be that the
concept of relevance should be framed in terms of stages,
where the irrelevant stage
proceeds all relevant and computation stages -- in which case it might be
desirable to separate the irrelevant stage itself into multiple stages
which can be evaluated in turn.
The exact nature of this interaction is left as a question for future work.

\section{Related Work}
\label{sec:related}





Multi-stage languages with explicit staging annotations were first suggested
by \citet{taha1997multi, taha2000metaml}. Since then there has been a reasonable amount of theoretical
interest in the topic which has renewed in recent years. There are several practical
implementations of multi-stage constructs in mainstream programming languages
such as BER MetaOCaml~\cite{kiselyov2014design}, Typed Template Haskell and
Dotty~\cite{stucki2018practical}.

\matt{reference other formalisms of MetaML? \cite{moggi1999idealized} \cite{taha2003environment}}

At a first glance there is surprisingly little work which attempts to reconcile
multi-stage programming with language features which include polymorphism. Most
presented multi-staged calculi are simply typed despite the fact that
all the languages which practically implement these features support polymorphism.
The closest formalism is by \citet{kokaji2011polymorphic} who consider
a language with polymorphism and control effects. Their calculus is presented
without explicit type abstraction and application. There
is no discussion about cross-stage type variable references or qualified types
and their primary concern, similar to \citet{kiselyov2017polylet} is the interaction of the
value restriction and staging. Our calculus in contrast, as it models
Haskell, does not contain any effects so we have concentrated on qualified types.
\citet{calcagno2003implementing} present a similar ML-like language with
let-generalisation.

Combining together dependent types and multi-stage features is a more common
combination. The phase distinction is lost in most dependently typed languages
as the typechecking phase involves evaluating expressions. Therefore in order
to ensure a staged evaluation, type variables must also obey the same
stage discipline as value variables. This is the approach taken by \citet{kawata2019dependently}.
\citet{pasalic2004role} defines the dependently-typed multi-stage language Meta-D
but doesn't consider constraints or parametric polymorphism. Concoqtion~\cite{fogarty2007concoqtion} is an
extension to MetaOCaml where Coq terms appear in types. The
languge is based on $\lambda_{H\bigcirc}$~\cite{pasalic2002tagless} which includes
dependent types but is motivated by removing tags in the generated program.
\citet{brady2006verified} observe similarly that it is worthwhile to combine together
depedent types and multi-stage programming to turn a well-typed interpreter
into a verified compiler. The language presented does not consider parametric
polymorphism nor constraints.

We are not aware of any prior work which considers the implications of relevant
implicit arguments formally, although there is an informal characterization by~\citet{pickering2019multi}.

\paragraph{Formalising Template Haskell}
With regards to formalisms of Template Haskell, a brief description of
Untyped Template Haskell is given by~\citet{sheard2002template}. The language
is simply-typed and does not account for multiple levels. The language has also
diverged since their formalism as untyped quotations are no longer typechecked
before being converted into their representation. Their formalism does
account for the \ensuremath{\Conid{Q}}~monad which provides operations that allow a
programmer to ``reify'' types, declarations and so on in order to inspect the internal
structure. These are typically used in untyped Template Haskell in order to
generically define instances or other operations.

It is less common to use the reification functions in Typed Template Haskell programs and
so we have avoided their inclusion in our formalism for the sake of simplicity.
In our calculus we wanted to precisely understand the basic interaction between
constraints and quotations, and the existence of the \ensuremath{\Conid{Q}}~monad is orthogonal to this.

Code generators are typically effectful in order to support operations such as
let insertion or report errors so it is an important question to define a
calculus with effects. From GHC 8.12, the type of quotations will be generalised~\cite{overloaded2019pickering}
from \ensuremath{\Conid{Q}\;(\Conid{TExp}\;\Varid{a})} to a minimal interface \ensuremath{\forall \Varid{m}\hsforall .\Conid{Quote}\;\Varid{m}\Rightarrow \Varid{m}\;(\Conid{TExp}\;\Varid{a})} so a user will
have more control over which effects they are allowed to use in their code
generators. We leave formalising this extension open to future work.

\paragraph{Metaprogramming In GHC}

GHC implements many different forms of metaprogramming from the principled
to the ad-hoc. At the principled end of the spectrum in a similar vein
as Typed Template Haskell there is Cloud Haskell~\cite{epstein2011cloud}, which
implements a modality for distributed computing. Another popular principled style
is generic programming~\cite{rodriguez2008comparing, magalhaes2010generic} which allows the representation of datatypes to be
inspected and interpreted at runtime. Untyped Template Haskell
is used for untyped code generation in the combinator style. As well as generating
expressions it can be used to generate patterns, declarations and types. Programs
are typechecked after being generated rather than typechecking the generators
as in Typed Template Haskell. Untyped Template Haskell has a limited interface
into the typechecker but Source Plugins~\cite{pickering2019source} allow
unfettered access to the internal state and operations of the typechecker and other
compiler phases.

\paragraph{Modal Type Systems}

Type systems motivated by modal logics have more commonly
contemplated the interaction of modal operators and polymorphism. In particular
attention has turned recently to investigating dependent modal type
theories and the complex interaction of modal operators in such theories~\cite{gratzer2020multimodal}. It
seems probable that ideas from this line of research can give a formal account of the interaction
of the code modality~\cite{davies2001modal} and the parametric quantification from System F which
can also be regarded as a modality~\cite{pfenning2001intensionality, nuyts2018degrees}.

In recent times, Fitch-Style Modal calculi~\cite{clouston2018fitch, gratzer2019implementing} have become a popular way of specifying
a modal type system due to their good computational properties. It would
be interesting future work to attempt to modify our core calculus to a Fitch-Style
system which was not level indexed. In particular the calculus for Simple RaTT~\cite{bahr2019simply}
looks like a good starting point.

In short, we are sure there is a lot to learn from the vast amount of literature
on modal type systems but we are not experts in this field and the literature
does not deal with our practical concerns regarding implementing and writing
programs in staged programming languages. The goal of this paper was not to uncover a
logically inspired programming language but to give a practical and understandable
practical specification which can be understood by people without extensive
background in modal type theories.


\section{Conclusion}
\label{sec:conclusion}





Now that we have presented the first formalism of Typed Template Haskell,
the way is clear for future researchers to understand and extend the basic system.
We envisage that the system will be useful for two different communities.
Firstly, researchers into extensions such as Linear Haskell~\cite{bernardy2017linear}
and Dependent Haskell~\cite{weirich2017specification}
now have the possibility to consider how their features interact with stages so that the language
remains sound with respect to staged evaluation.
Secondly, users interested in
multi-stage programming now have a firmer foundation to base further extensions to
the multi-stage features in GHC.

In the long-term our hope is to make multi-stage programming a more popular
and accessible
paradigm for functional programmers. Even with the present implementation of
Typed Template Haskell, \citet{yallop2018partially}
have shown how different features implemented in GHC can be used together in
order to express elegant code generators.
We anticipate that a firm foundation for Typed Template Haskell that supports
finer control over the type of qualified constraints will
enable more practitioners to explore this unexplored territory,
and begin to reach for staging as a useful tool in their toolbox of techniques
to enhance the predictability and performance of their code.

\bibliography{bibliography}
\clearpage

\end{document}